\documentclass[a4paper,11pt]{article}
\pdfoutput=1 

\usepackage{jcappub} 

\usepackage[varg]{txfonts}
\usepackage[utf8]{inputenc}

\usepackage{longtable}
\usepackage{epsfig,caption}

\usepackage{lineno} 

\usepackage[load-configurations=astronomy]{siunitx}
\DeclareSIUnit\parsec{pc}
\DeclareSIUnit\hubble{\ensuremath{h}}
\newcommand\ion[2]{#1$\;${\scshape{#2}}}
\newcommand{\lya}{Ly\ensuremath{\alpha}}%

\title{\bf  Constraint on neutrino masses from SDSS-III/BOSS Ly$\alpha$ forest and other cosmological probes}

\author[a,b]{Nathalie Palanque-Delabrouille,}
\author[a]{Christophe Y\`eche,}
\author[c,d,e]{Julien Lesgourgues,}
\author[a,f]{Graziano Rossi,}
\author[a]{Arnaud Borde,}
\author[g,h]{Matteo Viel,}
\author[i]{Eric Aubourg,}
\author[j]{David Kirkby,}
\author[a]{Jean-Marc LeGoff,}
\author[a]{James Rich,}
\author[b]{Natalie Roe,}
\author[k]{Nicholas P. Ross,}
\author[l,m]{Donald~P.~Schneider,}
\author[n]{David Weinberg}

\emailAdd{nathalie.palanque-delabrouille@cea.fr, christophe.yeche@cea.fr, Julien.Lesgourgues@cern.ch}

\affiliation[a]{CEA, Centre de Saclay, IRFU/SPP,  F-91191 Gif-sur-Yvette, France}
\affiliation[b]{Lawrence Berkeley National Laboratory, Berkeley, CA 94720, USA}
\affiliation[c]{Institut de Th\'eorie des Ph\'enom\`enes Physiques, \'Ecole Polytechnique F\'ed\'erale de Lausanne, CH-1015, Lausanne, Switzerland}
\affiliation[d]{CERN, Theory Division, CH-1211 Geneva 23, Switzerland}
\affiliation[e]{LAPTh, Univ. de Savoie, CNRS, B.P.110, Annecy-le-Vieux F-74941, France}
\affiliation[f]{Department of Astronomy and Space Science, Sejong University, Seoul, 143-747, Korea}
\affiliation[g]{INAF, Osservatorio Astronomico di Trieste, Via G. B. Tiepolo 11, 34131 Trieste, Italy}
\affiliation[h]{INFN/National Institute for Nuclear Physics, Via Valerio 2, I-34127 Trieste, Italy}
\affiliation[i]{APC, Universit\'{e} Paris Diderot-Paris 7, CNRS/IN2P3, CEA, Observatoire de Paris, 10, rue A. Domon \& L. Duquet,  Paris, France}
\affiliation[j]{Department of Physics and Astronomy, University of California, Irvine, CA 92697, USA}
\affiliation[k]{Department of Physics, Drexel University, 3141 Chestnut Street, Philadelphia, PA 19104, USA}
\affiliation[l]{Department of Astronomy and Astrophysics, The Pennsylvania State University, University Park, PA 16802, USA}
\affiliation[m]{Institute for Gravitation and the Cosmos, The Pennsylvania State University, University Park, PA 16802, USA}
\affiliation[n]{Department of Physics and Center for Cosmology and Astro-Particle Physics, Ohio State University, Columbus, OH 43210, USA}

\date{Received xx; accepted xx}

\abstract{ 
We present constraints on the parameters of the $\Lambda$CDM cosmological model in the presence of massive neutrinos, using the one-dimensional Ly$\alpha$ forest power spectrum obtained with the Baryon Oscillation Spectroscopic Survey (BOSS) of the Sloan Digital Sky Survey (SDSS) by  \citet{Palanque-Delabrouille2013}, complemented by additional cosmological probes.  
The interpretation of the measured Ly$\alpha$  spectrum is done using a  second-order Taylor expansion of the simulated power spectrum.   BOSS
 Ly$\alpha$ data alone provide better bounds than previous Ly$\alpha$ results, but are still poorly constraining, especially for the sum of neutrino masses $\sum m_\nu$, for which we obtain an upper bound of 1.1~eV (95\% CL), including systematics for both  data and   simulations. Ly$\alpha$ constraints on $\Lambda$CDM parameters and neutrino masses are  compatible with CMB bounds from the {\rm Planck} collaboration~\citep{PlanckCollaboration2013}. Interestingly, the combination of Ly$\alpha$ with CMB data reduces the uncertainties  significantly, due to very different directions of degeneracy in parameter space, leading to the strongest cosmological bound to date on the total neutrino mass, $\sum m_\nu < 0.15$~eV at 95\% CL (with a best-fit in zero). Adding recent BAO results further tightens this constraint to $\sum m_\nu < 0.14$~eV at 95\% CL. This bound is nearly independent of the statistical approach used, and of the different combinations of CMB and BAO data sets considered in this paper in addition to Ly$\alpha$. Given the measured values of the two squared mass differences $\Delta m^2$, this result tends to favor the normal hierarchy scenario against the inverted hierarchy scenario for the masses of the active neutrino species.

}

\begin{document}
\maketitle
\flushbottom

\section{Introduction}
\label{sec:intro}

The three active neutrinos are the most elusive particles in the standard model of particle physics. 
Despite their tiny cross section, experiments have determined the
number of active families and measured two of the squared
mass differences $\Delta m^2$, all three mixing angles, and one of the complex
phases. 
However, the most difficult measurement is that of the absolute neutrino mass scale, which remains poorly constrained by laboratory experiments. Current bounds from $\beta$-decay experiments translate into a constraint on the total neutrino mass (summed over the three families) of $\sum m_\nu \leq 7$~eV at the 95\% Confidence Level (C.L.) \cite{Beringer:1900zz}. The tritium decay experiment KATRIN is expected to reduce this bound by one order of magnitude \cite{Osipowicz:2001sq}. This value is still far from the lower bound inferred from neutrino oscillation experiments, $\sum m_\nu \geq 0.06$~eV (95\%C.L.) \cite{Capozzi:2013csa}.

If the total neutrino mass was measured, the individual mass of each neutrino mass eigenstate could be reconstructed from the knowledge of the two measured $\Delta m^2$’s, up to an uncertainty on the ordering of the three masses. In the so-called Normal Hierarchy (NH) scenario, one eigenstate is much heavier and the lower bound is indeed $\sum m_\nu \geq 0.06$~eV (95\%C.L.). In the Inverted Hierarchy scenario, the two heaviest neutrinos are nearly degenerate, and the lower bound is actually closer to $\sum m_\nu \geq 0.10$~eV (95\%C.L.). The knowledge of the total mass and of the type of hierarchy is crucial to complete our understanding of the neutrino sector, and to provide a better knowledge of other issues in particle physics like leptogenesis, baryogenesis, the right-handed neutrino sector, etc.

The most sensitive probes of the total neutrino mass are cosmological observations \cite{Lesgourgues:2006nd,Hannestad:2010kz,lesgourgues2013neutrino}. There are so many neutrinos in the universe that they accounted for 40\% of its total energy density during the radiation-dominated epoch, and  despite their tiny mass, they  still  account for at least 0.5\% of the total density today (for $\sum m_\nu=0.06$~eV). The transition from the relativistic to non-relativistic regime of the heaviest neutrinos may have left a small signature in the CMB temperature anisotropy spectrum, on intermediate scales, through the early integrated Sachs-Wolfe effect~\cite{Lesgourgues:2012uu,Hou:2012xq,lesgourgues2013neutrino}. Once they become non-relativistic, neutrinos contribute to a small fraction of Dark Matter, but due to their large velocity dispersion they constitute a sub-dominant Hot Dark Matter component. By contributing to the expansion of the universe, but not to the clustering of dark matter on small scales, massive neutrinos tend to slow down the structure formation rate of the dominant (Cold) Dark matter component. Hence, neutrino masses induce a characteristic scale-dependent growth factor in the structure formation history~\cite{Bond:1980ha}. At the level of linear perturbation theory, this behavior causes a step-like suppression in the matter power spectrum, with a redshift-dependent amplitude~\cite{Hu:1997mj,Lesgourgues:2006nd,lesgourgues2013neutrino}. On non-linear scales, this effect is further enhanced by the fact that massive neutrinos delay the time of non-linear collapse~\cite{Brandbyge:2009ce,Viel2010,Bird:2011rb,Wagner:2012sw}. Today, the small-scale suppression in the matter power spectrum is predicted to be at least  5\% (for $\sum m_\nu=0.06$~eV).

These effects have been used to constrain $\sum m_\nu$ with several types of cosmological observations. The CMB temperature anisotropy spectrum probes neutrino masses through the integrated Sachs-Wolf effect, and also through CMB lensing effects, which are related to the matter power spectrum in the recent universe, and hence to the growth factor. None of the available CMB data sets has detected neutrino mass effects so far. Results from the first Planck satellite release, combined with other CMB data, provide a bound $\sum m_\nu < 0.66$~eV (95\%C.L.)~\cite{PlanckCollaboration2013}. The combination of CMB data with probes of the expansion history allows one  to remove degeneracies and to improve this bound. For instance, the combination of CMB data and Baryon Acoustic Oscillation (BAO) scale data available in 2013 yields $\sum m_\nu < 0.23$~eV (95\%C.L.)~\cite{PlanckCollaboration2013}. 

The situation on the side of Large Scale Structure observations is more controversial. Several results from galaxy redshift surveys, weak lensing surveys, or previous Ly$\alpha$ surveys are compatible with the above CMB+BAO bound~\cite{Zhao12}. There are, however, a number of experiments which are best fitted by assuming a sizeable total neutrino mass, of the order of 0.3 or 0.4~eV. These studies include: the lensing power spectrum extracted from the Planck temperature map from 2013~\cite{Ade:2013tyw}; the cluster abundance extracted from the Planck SZ catalogue~\cite{Ade:2013lmv}; the galaxy lensing power spectrum measured by the CFHTLens survey~\cite{Heymans:2013fya}; and, more recently, the redshift space distorsions measured by BOSS~\cite{Beutler:2014yhv}. There are on-going debates on the efforts to reconcile these experiments, perhaps by better modeling systematic effects. Fortunately, current and future experiments tend to be  more sensitive to the scale-dependence and redshift-dependence of the matter power spectrum. Since we understand how to compute theoretically the scale- and redshift-dependence induced by hot dark matter, new data sets will allow us to better break the degeneracy between neutrino mass effects and possible systematics that affect the overall amplitude of the measured power spectrum. 

The analysis of the Lyman-$\alpha$ (Ly$\alpha$) forest in quasar spectra allows the measurement of the one-dimensional flux power spectrum, which is related to the underlying three-dimensional matter power spectrum. The impact of neutrino masses on the flux power spectrum has been computed with increasing precision over the years, and a comparison of these predictions with observational data has led to several Ly$\alpha$ constraints on neutrino masses~\cite{Seljak:2004xh,Seljak2006,Viel2010,2014arXiv1407.8338C}. With various hypotheses and combinations of Ly$\alpha$ and CMB data sets, these limits range from 0.17~eV to 0.9~eV for the sum of the three neutrino masses, at 95\% CL .

In~\cite{Palanque-Delabrouille2013}, the BOSS collaboration presented a new reconstruction of the flux power spectrum in several redshift bins in the range $2.2 \leq z \leq 4.4$. The new data have such small error bars that they can be used for cosmological parameter extraction only if theoretical predictions are also highly accurate. A  large suite of dedicated hydrodynamical simulations has been developed for that purpose, following the grid approach of~\cite{Viel2006} but improving on several aspects, such as size and mass resolution of the simulations, and considering additional simulations in order to compute a second-order likelihood of  the 1D flux power spectrum~\cite{Borde2014}. Massive neutrinos  have been included in ~\cite{Rossi2014},  and the dependence of the flux power spectrum on several cosmological and astrophysical parameters plus $\sum m_\nu$ has been precisely estimated. The goal of this paper is to compare the results of these simulations with the data of~\cite{Palanque-Delabrouille2013} to derive new bounds on cosmological parameters and neutrino masses.


\section{Cosmological probes}
\label{sec:probes}
We describe in this section the different probes that we use to constrain the sum of neutrino masses. We first present the 
one-dimensional Ly$\alpha$ forest (both the data and the hydrodynamical simulations we produced to interpret them), then the cosmic microwave background data, and finally the additional probes that contribute to obtaining better constraints. 

\subsection{One-dimensional Ly$\alpha$ forest power spectrum} \label{sec:Lya}

\subsubsection{ Description of data and measurement of  power spectrum}

As our large-scale structure probe, we use the 1D Ly$\alpha$ flux power spectrum measurement from the  first release of  BOSS quasar data~\citep{Palanque-Delabrouille2013}. The data consist of a sample of $13~821$  spectra selected from the larger sample of about $60~000$ quasar spectra of the SDSS-III/BOSS DR9 \cite{Ahn2012, Dawson2012, Eisenstein2011, Gunn2006, Ross2012,Smee2013} on the basis of their high quality, high signal-to-noise ratio and good spectral resolution ($<85\,\rm km.s^{-1}$ on average over a quasar forest). The aim of this tight selection was to reduce the systematic uncertainties on the measurement to a level comparable to the statistical uncertainties. We also remove all quasars that have broad absorption line (BAL) features  in their spectra (about 12\% of the data),  damped Ly$\alpha$ or detectable Lyman limit  systems (LLS) in their forest (close to 20\% of the data).  
The measured flux power spectrum is obtained in twelve redshift bins from $\langle z\rangle = 2.2$ to $4.4$, each covering a range $\Delta z=0.2$, and thirty-five wavenumbers from $k=0.001~\rm(km/s)^{-1}$ to $0.02~\rm(km/s)^{-1}$. 

The Ly$\alpha$ forest region, defined by the rest-frame wavelength interval  $1050<\lambda_{\rm RF}<1180\,$\AA, was divided into up to three distinct $z$-sectors, where each $z$-sector has  a redshift extension of at most $0.2$, thus strongly reducing  correlations between consecutive $z$-bins  in the power spectrum measurement.  In the following, we neglect any  correlation between different $z$-bins. In a given redshift bin, however, weak correlations are present  between different scales, with typical values of at most 15\% in neighboring bins on large scales,  rapidly dropping to $<5\%$ after a few bins. 
The redshift distribution of BOSS quasars is such that most Ly$\alpha$ absorbers have redshifts that lie in the range $2.1<z<3.1$, with a significant reduction in number of measured forests with increasing redshifts (cf. table~\ref{tab:distrib_z}).  The last two redshift bins have a statistical uncertainty that exceeds 10\% on all scales. The statistical uncertainty is largest on small scales where the window function from the resolution of the spectrograph  reduces the power by up to a factor $\sim 20$. 
\begin{table}[htdp]
\begin{center}
\begin{tabular}{|c|r|ccc|ccc|}
\hline
Mean $z$ & $\sharp$ forests & \multicolumn{3}{c|} {$\sigma_{\rm stat}/P_k$ (\%)} & \multicolumn{3}{c|} {$\sigma_{\rm syst}/P_k$ (\%)}\\
\hline
2.2& 7053 &  2 & 4 & 12 &        1  &  2  &  4\\
2.4 & 8737 &  1  &  2  &  8 & 1  &  1  &  4\\
2.6 & 7440 & 1  &  2  &  6 & 1  &  1  &  3\\
2.8 & 5683 & 1  &  2  &  5 & .  &  1  &  2\\
3.0 & 3368 & 2  &  2  &  6 & .  &  .  &  2\\
3.2 & 1794 & 2  &  3  &  5 & .  &  .  &  2\\
3.4 & 1092 &  3  &  4  &  7 & .  &  .  &  2\\
3.6 &1092 & 5  &  5  &  8 & 2  &  2  &  3\\
3.8 & 535 & 7  &  7  &  10 & 1  &  2  &  2\\
4.0 & 260 &  9  &  11  &  15 & 1  &  1  &  2\\
4.2 & 127 & 11  &  14  &  18 & 2  &  3  &  5\\
4.4 & 81 &  16  &  20  &  25 & 1  &  1  &  2\\
\hline
\end{tabular}
\end{center}
\caption{\it Redshift distribution of the Ly$\alpha$ $z$-sectors in the BOSS DR9 data. The second column lists the number of quasars with forest absorption in the redshift bin.  The last two columns provide the $15^{\rm th}$-percentile (5$^{\rm th}$ lowest value out of 35, with one power spectrum value for each $k$-mode), median and $85^{\rm th}$-percentile (30$^{\rm th}$ lowest out of 35)  fractional statistical and systematic uncertainties per $k$-bin relative to the power spectrum value, in \%, for each redshift bin. }
\label{tab:distrib_z}     
\end{table}

Careful treatments were applied to the data to correct the measured power spectrum for the contributions of noise and spectrograph resolution. The remaining uncertainty in the determination of these two quantities led to systematic uncertainties. We  include nuisance parameters in the fit for residual imperfections in their modeling.  Additional systematics arose from the procedure for masking of the sky lines, for continuum fitting, for noise subtraction, and estimation of the contamination by all species other than atomic hydrogen (referred to as `metals' in this paper). The ranges of  statistical and  total systematic uncertainties present in each redshift bin are indicated in Table~\ref{tab:distrib_z}. As expected from the data selection and work described above,  systematic uncertainties have indeed been reduced to a  lower  level than  statistical uncertainties for all redshift bins.

\begin{figure}[htbp]
\begin{center}
\epsfig{figure= 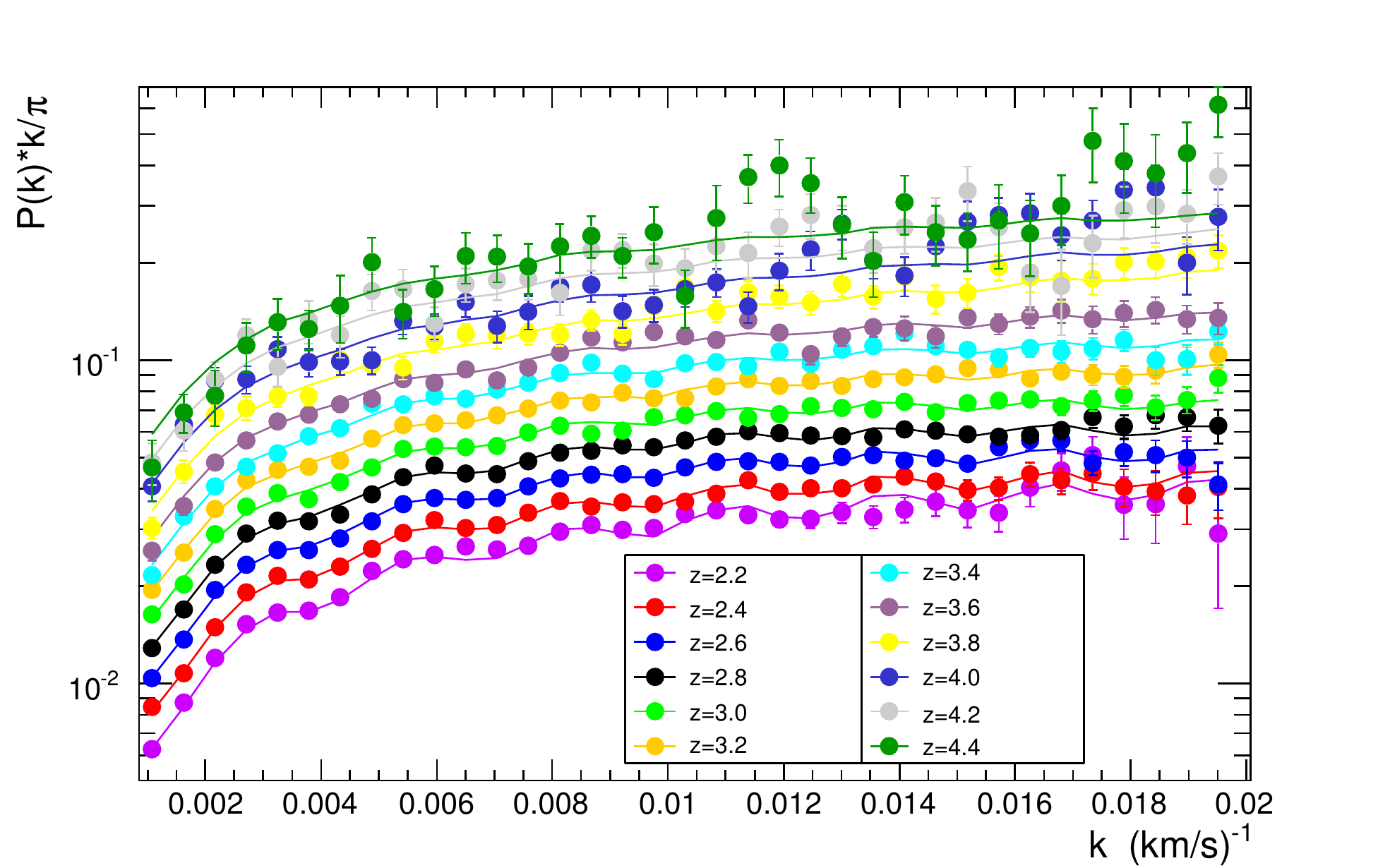,width = \linewidth} 
\caption{\it 1D Ly$\alpha$ forest power spectrum from the SDSS-III/BOSS DR9 data. The solid curves show the best-fit model obtained in section~\ref{sec:lya_alone} when considering Ly$\alpha$ data alone.} 
\label{fig:P1D}
\end{center}
\end{figure}

The measured 1D flux power spectrum is illustrated in figure~\ref{fig:P1D}, along with the best cosmological fit for Ly$\alpha$ alone described in section~\ref{sec:lya_alone}. The wiggles that appear on the power spectrum at all redshifts are due to the \ion{Si}{iii} - Ly$\alpha$ cross-correlation. The absorption wavelengths of these two metals are $9\, \AA$ apart, resulting in an oscillating pattern with peak separation of $\Delta k = 0.0028 {\rm \, (km/s)^{-1}}$.

\subsubsection{Simulations}\label{sec:1Dsim}

The cosmological interpretation of the power spectrum measurement is obtained by comparison to a set of  full hydrodynamical cosmological simulations that were produced specifically for that purpose. The methodology and technical framework for these simulations are presented in~\cite{Borde2014}, while all issues concerning the inclusion of neutrinos in the pipeline and their impact  on the power spectrum are described in detail in \cite{Rossi2014}. We summarize below the aspects of the simulation procedure that are most relevant to this work.\\

{\noindent\bf Hydrodynamical simulations with neutrinos}\\
The neutrinos, considered as three degenerate species, are globally introduced as a third particle type, in addition to cold dark matter and baryons. The simulations were run using \texttt{CAMB}\footnote{\url{http://camb.info}} \citep{Lewis2000} to compute the transfer functions and linear power spectra at $z=30$, then \texttt{2LPT}\footnote{\url{http://cosmo.nyu.edu/roman/2LPT/}}  (second-order Lagrangian Perturbation Theory) to compute the initial displacement of the particles, and finally  \texttt{GADGET-3}~\citep{Springel2001,Springel2005}  for the hydrodynamical processing.  The choice of  a second-order rather than first-order (or Zel'dovich approximation) code for the initial conditions was motivated by our initial redshift for  \texttt{GADGET-3} of $z=30$, itself motivated by the inclusion of neutrinos. 
Indeed, because of their high velocity, neutrinos require initial conditions taken at rather low redshift in order to reduce Poisson noise. \\
    
{\noindent\bf Splicing procedure}\\
The 1D power spectrum is computed for each set of cosmological parameters on the grid using a set of three simulations. The large-scale power is   derived from a simulation with $768^3$ particles per species (baryons, dark matter and neutrinos) in a box  of 100~$h^{-1}$~Mpc on a side, where $h$ is the  value of the Hubble constant today normalized  to 100~$\rm km~s^{-1}~Mpc^{-1}$. This size is chosen  to cover a maximum scale equivalent to the Ly$\alpha$ forest $z$-sectors defined in the analysis of the BOSS data. The small-scale resolution is achieved with simulations of $768^3$ particles per species in a box of 25~$h^{-1}$~Mpc on a side. A splicing technique  \cite{McDonald2003} is then  used to combine the large-scale simulation and the high-resolution one, using a transition simulation with $192^3$ particles per species in a 25~$h^{-1}$~Mpc box, i.e., a simulation with the same resolution as the large-box simulation and the same box size as the high-resolution one. As a result of this procedure, the spliced power spectrum is expected to be equivalent to  the one measured  from  a single simulation spanning a total volume of $(100~h^{-1}~{\rm Mpc})^3$ with $3072^3$ particles per species, i.e., an equivalent mean mass  per gas particle of  $1.2\times10^{5}~h^{-1}~M_\odot$. 

The precision of the splicing technique was studied on the simulation set produced with all parameters having their central value. The precision was determined by comparing the  power spectrum spliced from a simulation with $256^3$ particles per species in a 100~$h^{-1}$~Mpc box and another with $256^3$ particles per species in a 25~$h^{-1}$~Mpc box, 
to the `exact' power spectrum measured from a full simulation spanning a volume of $(100~{h^{-1}~\rm Mpc})^3$ with $1024^3$ particles per species. 
The ratio of the spliced to the exact power spectrum exhibited a small residual. We modeled this residual as a shift  varying  linearly with $k$-mode to reach 5\% at $k=0.02\;h\;{\rm Mpc^{-1}}$, according to the equation $P_{\rm cor}(k) = P_{\rm spliced}(k)/[1.01-2.5 \cdot  k]$, where $P_{\rm spliced}$ is the spliced power spectrum from section 5 of \cite{Borde2014}. The standard deviation of the residuals to this fit are at the level of $1.0\%$.  In Sec.~\ref{sec:likelihood} we describe how we correct the simulations for the  observed $k$-dependent but redshift-independent offset, and we present in section~\ref{sec:systsimus} an estimate of the systematic uncertainty coming from  the application of the splicing technique. 
\\

{\noindent\bf Taylor expansion}\\
These simulations explore the impact on the 1D flux power spectrum of four cosmological parameters  ($n_s$, $\sigma_8$, $\Omega_m$ and $H_0$), two astrophysical parameters ($T_0$ and $\gamma$)  related to the heating rate of the intergalactic medium, and  the sum $\sum m_\nu$ of the masses of the three active neutrino species. Using these simulations, we compute the 1D flux power spectrum at 14 predefined redshifts equally spaced by $\Delta z = 0.2$ from $z=2.2$ to 4.6, corresponding to the central redshift of each of the redshift bins of the  analysis of the DR9 BOSS data. 
The photo-ionization rate (or equivalently the UV flux) of each simulation is fixed by requiring the effective optical depth at each redshift to follow the empirical law $\tau_{\rm eff}(z) =  A^\tau \times (1+z) ^{\eta^\tau}$, where $A^\tau=0.0025$ and $\eta^\tau=3.7$~\citep{Meiksin2009}. The rescaling coefficients are determined independently for each redshift using all the line-of-sight pixels; these coefficients typically lie between 0.8 and 1.2. As justified in \cite{Theuns2005}, this normalization is done a posteriori, and thus does not require running additional simulations. 
We  modify  the   parameters $A^\tau$ and $\eta^\tau$ to probe the impact of  different  mean flux  normalizations and evolutions with redshift. 

By varying the input parameters (cosmological and astrophysical parameters, or total neutrino mass) around a central model chosen to be in agreement with the latest Planck results~\citep{PlanckCollaboration2013}, the simulations were used to derive a  second-order Taylor expansion, including cross-terms, around the central model. To derive the first and second-order derivatives, we ran simulations for a total of 36 different combinations of the input parameters, and three different values of $A^\tau$ or $\eta^\tau$. We chose to distribute the  values on a regularly-spaced grid whose central and step values for each parameter are listed in table~ \ref{table:grid_parameters}.  For all the simulations of the grid, $\Omega_b h^2$ is fixed at 0.0 221. 

\begin{table}[h]
\begin{center}
\begin{tabular}{|lll|}
\hline
Parameter & Central value & Range\\
\hline
$n_s$\dotfill & $0.96$ & $\pm\,0.05$\\
$\sigma_8$\dotfill & $0.83$ & $\pm\,0.05$\\
$\Omega_m$\dotfill & $0.31$ & $\pm\,0.05$\\
$H_0$\dotfill & $67.5$ & $\pm\,5$\\
$T_0(z=3)$\dotfill & $14000$ & $\pm\,7000$\\
$\gamma(z=3)$\dotfill & $1.3$ & $\pm\,0.3$\\
$A^\tau$ \dotfill & 0.0025 & $\pm\,0.0020$\\
$\eta^\tau$ \dotfill & 3.7& $\pm\,0.4$\\
$\sum m_\nu$ (eV)\dotfill & 0.0 & $0.4, 0.8$\\
\hline
\end{tabular}
\end{center}
\caption{Central values and variation ranges of the cosmological and astrophysical parameters  for our simulation grid.  The neutrino derivatives, unlike the others, are one-sided and not symmetric around the `central' value. The cross-derivatives including a non-zero neutrino mass are computed with $\sum m_\nu = 0.8\;{\rm  eV}$. }\label{table:grid_parameters}
\end{table}

Tests were performed to assess the precision of the  Taylor expansion resulting from this  grid of simulations. In particular, we produced simulations for several sets of parameters with values different from those used in the grid, and compared the power spectrum derived to the one predicted by our Taylor expansion. All showed excellent agreement as long as the tested parameters remained within the range of values used to compute the expansion  (cf.  table~ \ref{table:grid_parameters}). 

Other approaches could have been considered to model  the 1D flux power spectrum. For instance, an interesting alternative is the emulation of cosmological simulations as introduced originally by~\citet{Heitmann06} and  revisited in several papers such as \cite{Habib07, Schneider08, Heitmann2009,Heitmann10,Lawrence09}. The simulation parameters are selected on a Latin hypercube instead of on a regularly spaced grid. In a recent development,  \citet{Schneider11} have updated the technique by drawing the parameters on a hypersphere, thus reducing the  parameter space volume and better concentrating the simulations in the highest probability region. The power spectrum is then constructed from Principal Component Analysis combined with Gaussian Process modeling. While the authors themselves show that the much-simpler-to-implement polynomial interpolation gives comparable accuracy as the Gaussian Process, an optimal coverage of the parameter space, in contrast, can provide significant improvements over random sampling for instance. 
The main drawback of this approach is the computational cost. \citet{Schneider11}  used a total of 100 simulations to obtain 1\% accuracy on the predicted CMB temperature power spectrum. While this is possible for relatively fast Boltzmann solvers, it is prohibitive for orders-of-magnitude slower  hydrodynamical codes. In the non-linear regime, \citet{Heitmann2009} have obtained percent-level  matter power spectra to wavenumbers $k\sim 1~h~{\rm Mpc}^{-1}$ using a suite of 38 N-body simulations spanning  5 cosmological parameters. Though promising, the cosmic emulator approach has not yet been extended  to the emulation of the transmitted flux power spectrum that requires fully hydro-simulations and additional parameters to describe the IGM. Such a study is beyond the scope of the present work. To emulate the transmitted flux in the Ly-$\alpha$ forest, we therefore use a traditional Taylor expansion as in previous analyses~\cite{Viel2006, Wang2013}. We derive it from a suite of 36 simulations (and 3 UV-flux normalizations) that span 9 parameters,  and we verify the achieved accuracy in the parameter range that fits the data.
\\
{\noindent\bf Statistical uncertainties}\\
For each simulation, i.e., for  each  set of parameters, $100~000$ skewers were drawn with random origin and direction, and the 1D power spectrum was computed. The final 1D power spectrum is taken as the average over the $100~000$ lines of sight, and the statistical uncertainty on the power spectrum  as their $rms$ divided by $\sqrt{100~000}$. The statistical uncertainty on the simulation is largest at  small $k$, but always remains much smaller than the data statistical uncertainty, i.e., by a factor  5 to 150. As explained in~\cite{Borde2014}, we have checked that we do not oversample our simulation volume by the following test: we varied the number of lines of sight drawn from $n=5~000$ to $100~000$ and verified that the uncertainty on each point  of the power spectrum scaled as $\sqrt{n}$ at better than the percent level. \\

\begin{figure}[hhtbp]
\begin{center}
\epsfig{figure= 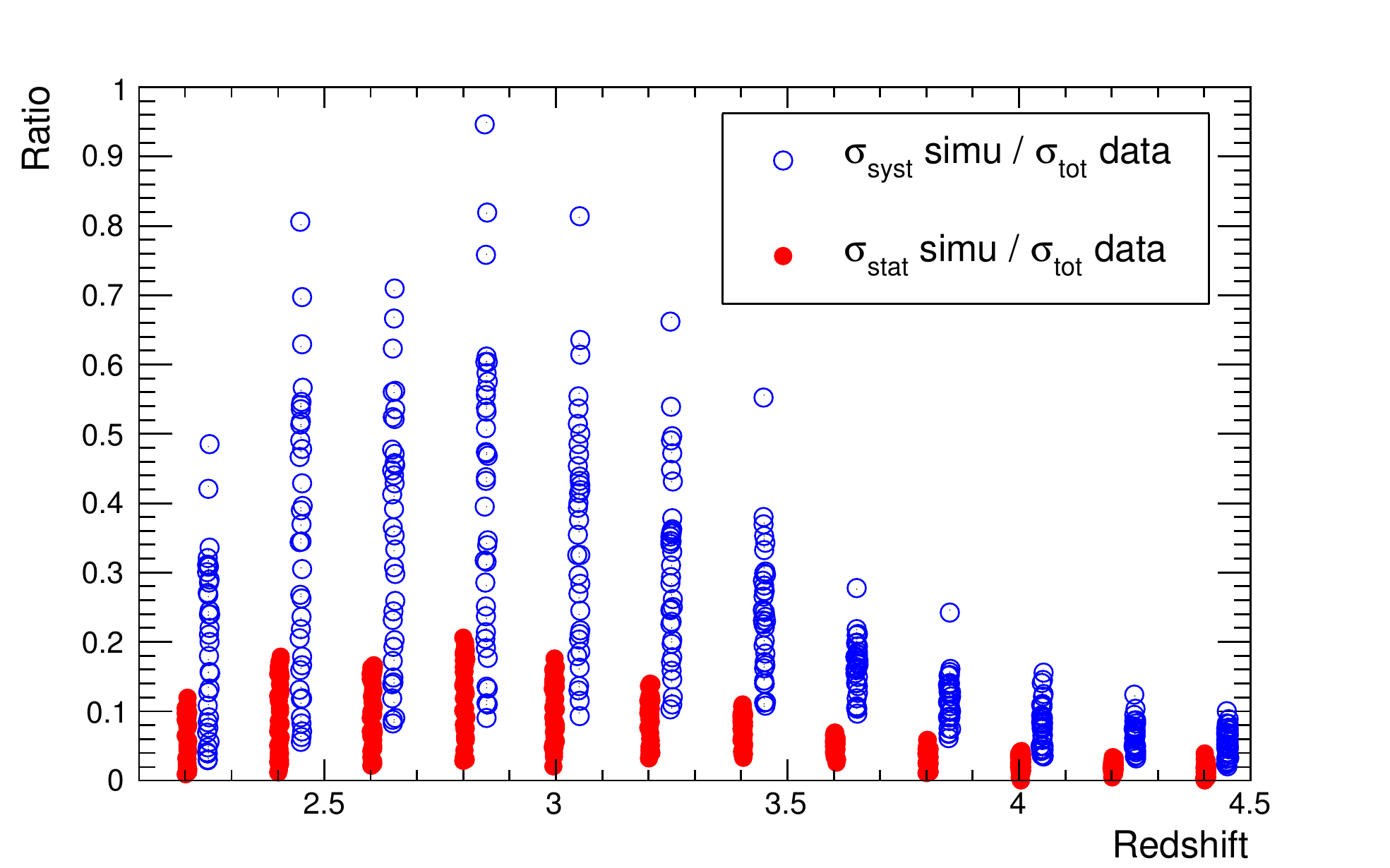,width = \linewidth} 
\caption{\it Simulation uncertainties (statistical and systematic)  relative to  data total uncertainty for each redshift bin. The highest systematic uncertainties in the simulation at all redshifts occur on the largest scales ($k<0.002\;({\rm km/s})^{-1}$) where  sample variance is large. Systematic uncertainties are shifted in redshift by 0.05  for better clarity.} 
\label{fig:syst}
\end{center}
\end{figure}

{\noindent\bf Sample variance}\\
A sample variance effect is expected on large scales since the size of the simulation volume is similar to the largest modes measured. The uncertainty related to this effect was determined by comparing  simulations with different random seeds to initiate the distribution of particles. This exercise led to two different realizations of the Universe, whose 1D power spectra were in excellent agreement  on small scales but differed on large scales by up to $\sim3\sigma$. This test was performed for our central $\Lambda$CDM model, and we assumed the result to be valid for all models. We assigned a simulation sample-variance systematics equal to 30\% of the difference between the power spectra measured from these two  simulations with  different random seeds. \\

{\noindent\bf Simulation total uncertainties}\\
In the fitting procedure, the simulations are assigned a diagonal covariance matrix, with total uncertainty equal to the quadratic sum of  the statistical and systematic uncertainties listed above. The statistical and total systematic uncertainties of the hydrodynamical simulations are presented in figure~\ref{fig:syst} compared to the total uncertainties in the data measurement; the simulation uncertainties remain smaller than data uncertainties at all redshifts and all scales. \\

\begin{figure}[htbp]
\begin{center}
\epsfig{figure= 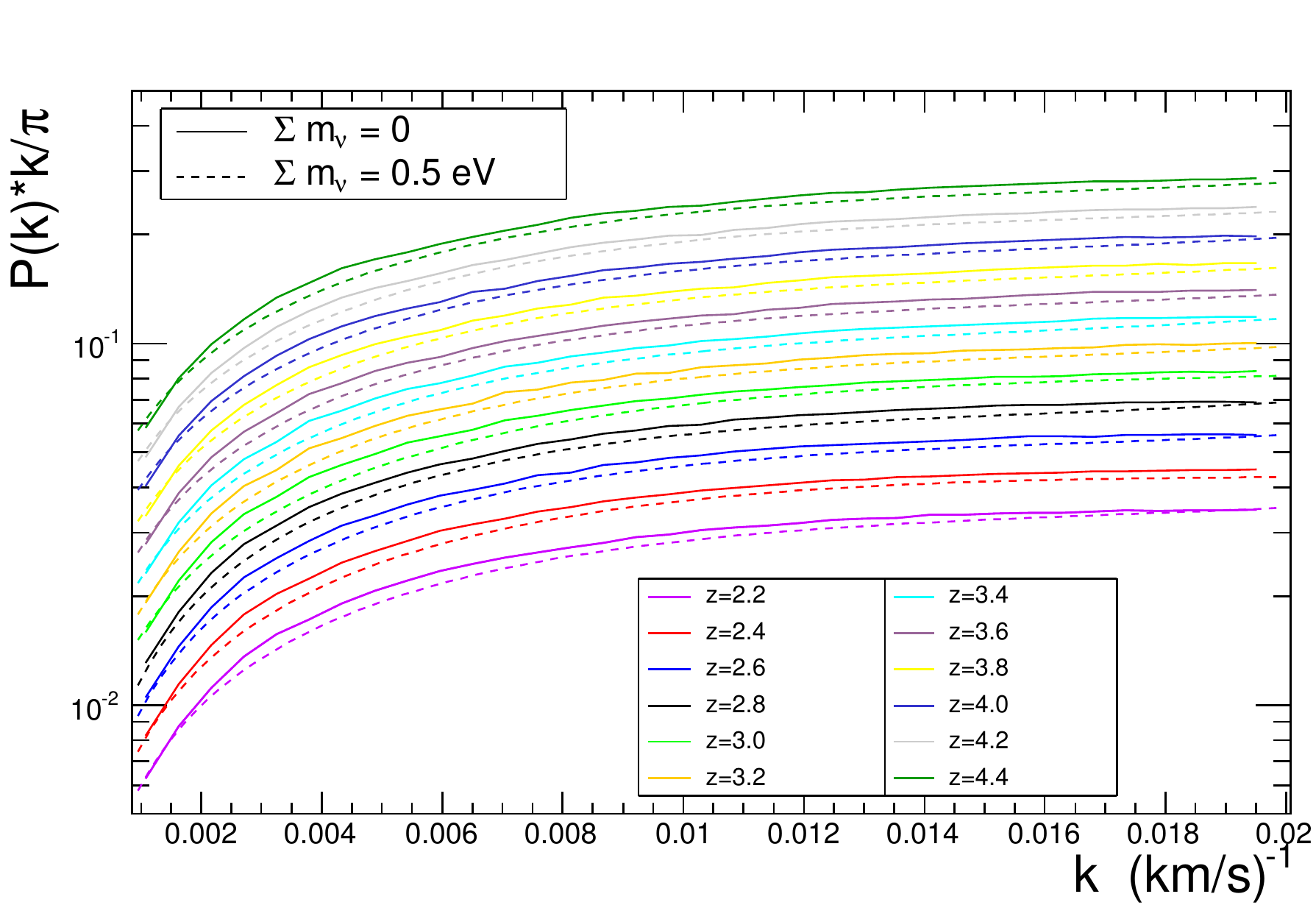,width = 7.5cm} 
\epsfig{figure= 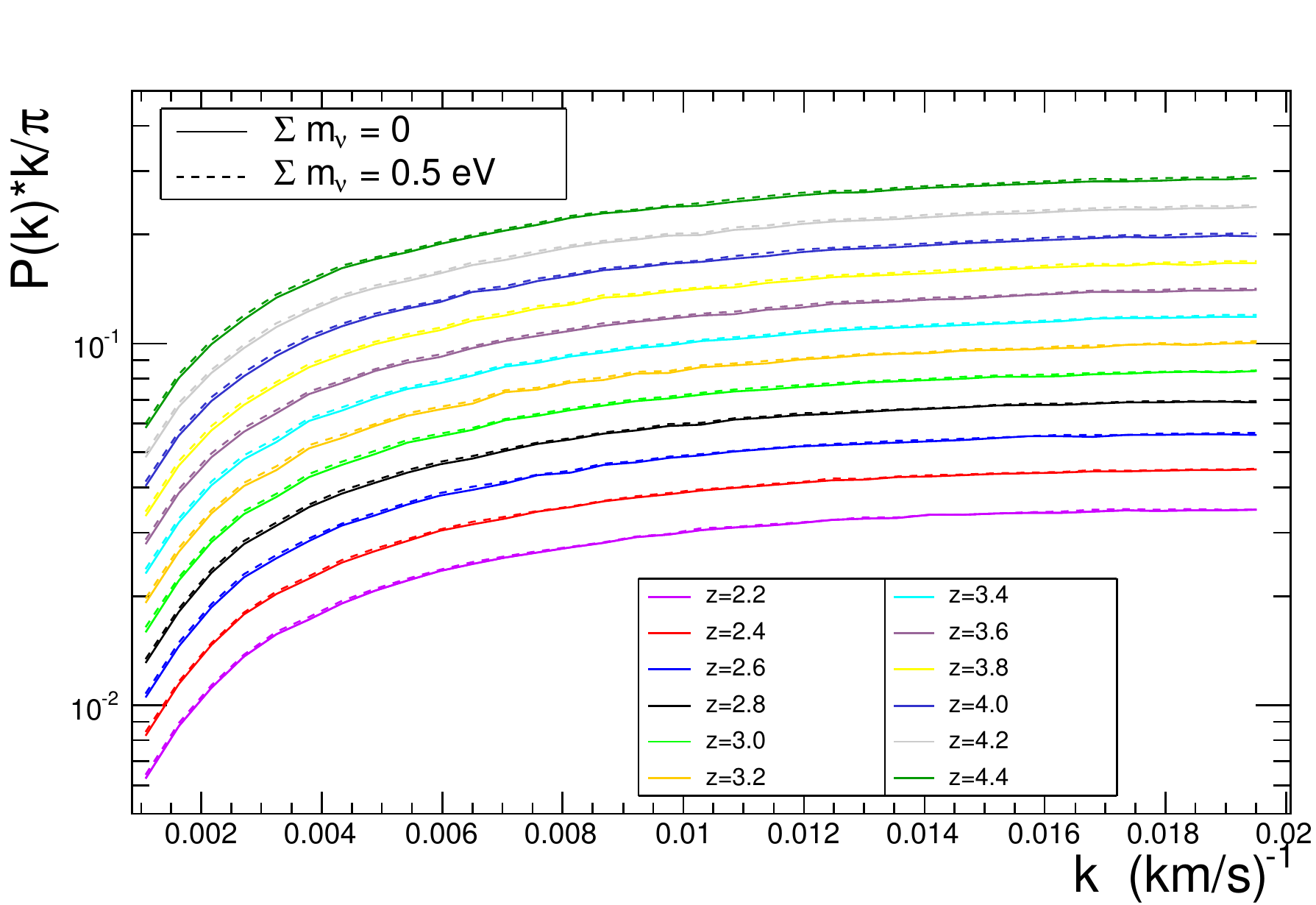,width = 7.5cm}
\caption{\it 1D Ly$\alpha$ forest power spectrum obtained with the full hydrodynamical simulations. The solid and dotted lines are respectively obtained for simulations with  massless neutrinos and  massive neutrinos ($\sum m_\nu = 0.5$ eV).  On the left panel, the two simulations were produced with the same primordial spectrum amplitude $A_s$, thus they correspond to two different values of $\sigma_8(z=0)$. On the right panel, in the massive neutrino simulation, the primordial spectrum amplitude used for generating the initial conditions has been renormalized to return the same value of $\sigma_8(z=0)$ as in the massless neutrino simulation. } 
\label{fig:PkNumass}
\end{center}
\end{figure}

 {\noindent\bf Illustration of simulated 1D flux power spectra}\\
 Fig.~\ref{fig:PkNumass} displays the impact of a small total neutrino mass ($\sum m_\nu = 0.5$~eV) on the 1D flux power spectrum, when either the primordial fluctuation amplitude $A_s$ (left panel) or the current fluctuation amplitude $\sigma_8$ (right panel) is held fixed (while $H_0$, $\Omega_m$, $n_s$ and all astrophysical parameters are also  fixed).

The right panel of Fig.~\ref{fig:PkNumass} corresponds to the genuine impact on the power spectrum of varying $\sum m_\nu$ while keeping fixed all other variables of the cosmological grid.  Despite the small value of $\sum m_\nu$,  the tiny effect of neutrino masses is both  scale and redshift dependent. The redshift dependence is a direct consequence of the reduction of the growth rate of fluctuations in presence of a hot dark matter component. This physical effect can only be observed in $N$-body simulations with separate species accounting for cold and hot dark matter. The non-trivial scale and redshift dependence explains why, in section \ref{sec:results}, we will find that the effects of ($\sum m_\nu$, $\sigma_8$, $n_s$) are not fully degenerate with one another, even when using  Ly$\alpha$ data only. However, these effects (visible in the right panel) are roughly ten times smaller than the overall suppression of power (seen in the left panel). This result indicates that neutrino mass bounds can be improved by  about one order of magnitude when Ly$\alpha$ data are combined with other cosmological probes, such as CMB data, that can constrain $\sigma_8$ or $n_s$ independently.

\subsubsection{Likelihood}\label{sec:likelihood}

We model the  likelihood with three categories of parameters which are floated in the maximization procedure. The first category describes the cosmological model in the simplest case of $\Lambda$CDM assuming a flat Universe. The second category models the astrophysics within the IGM, and the relationship between the gas temperature and its density.  
The  purpose of the third category is  to describe the imperfections of our measurement of the 1D power spectrum. By fitting the parameters of the latter category, we improve significantly the goodness of the fit, at the expense, however, of a reduction in the sensitivity on the other parameters. The complete list of the parameters used in the fit is described below. The first two categories correspond to the nine cosmological  and astrophysical parameters of  our grid of simulations  (see table~\ref{table:grid_parameters}), which are all included in the Taylor expansion from which we derive directly the  theoretical power spectrum $P^{th}(k_i,z_j)$ for  scale $k_i$ and redshift $z_j$. The parameters of the third category are introduced in the likelihood as explained below. 

\begin{itemize}
\item {\bf Cosmological parameters:} We model the cosmological dependance by five parameters: the fluctuation amplitude of the matter power spectrum $\sigma_8$ taken at $z=0$, the spectral index of primordial density fluctuations, $n_s$, the matter density $\Omega_m$, the Hubble constant $H_0$ and the sum of neutrino mass,  $\sum m_\nu$  .

\item {\bf Astrophysical parameters:} Two parameters ($A^\tau$ and $\eta^\tau$) taken at $z=3$ describe the effective optical depth assuming a power law evolution, $\tau_{\rm eff}(z) =  A^\tau\times [(1+z)/4 ]^{\eta^\tau}$. Two parameters ($T_0$ and $\gamma$)  are related to the heating rate of the intergalactic medium. To account for the effect of the correlated \ion{Si}{iii}  absorption, we  introduce a multiplicative  term to $P^{th}(k_i,z_j)$: $1+a^2+2a\cos(vk)$ with $a = f_ {\rm{Si\,III}}/(1-{\bar F}(z))$ as  in~\cite{McDonald2006}. The parameter $f_ {\rm{Si\,III}}$ is allowed to vary in the fit and $v$ is fixed at 2271~km~$\rm s^{-1}$. Similarly, we model a possible correlation with \ion{Si}{ii} by floating a parameter  $f_ {\rm{Si\,II}}$ with $v$  fixed at 5577~km~$\rm s^{-1}$.

\item {\bf Nuisance parameters:} \\
- We  take into account the impact on $P^{th}$ of  an imperfect resolution estimate, in the analysis described in \citet{Palanque-Delabrouille2013},   using a multiplicative correction  term: $ {\cal C}_{reso}(k) =  \exp( - k^2\cdot \alpha_{reso})$.  The nuisance parameter $\alpha_{reso}$ is constrained to stay in a physical range by adding a Gausssian constraint in the total likelihood, of mean 0 and width  $(5 \rm{km/s})^2$ .\\
- We allow for imperfection in the noise estimate of  the 1D power spectrum through  twelve additive correction terms (one per redshift bin $z_i$): $ {\cal C}_{noise}(k,z_i) =  P_{noise}(k_i,z_j) \cdot \alpha_{noise}^i$.  The noise power spectrum  $P_{noise}(k_i,z_j)$ is described in \citet{Palanque-Delabrouille2013}. The nuisance parameters are allowed to vary around 0 with a Gaussian constraint of $\sigma =10\%$.\\
- We introduce a multiplicative correction ${\cal C}_{splicing}(k)$ to model the uncertainty of the splicing technique evaluated in~\citep{Borde2014}. The corrective term depends linearly on the $k$-mode as   ${\cal C}_{splicing}(k) = 1.01 - \alpha_{splicing} \cdot k$, where $\alpha_{splicing}$ is allowed to vary in the fit. This form reproduces the correction to the splicing residuals discussed in section~\ref{sec:1Dsim}. To maintain the nuisance parameter $\alpha_{splicing}$  within a reasonable range, we impose  a Gaussian constraint on $\alpha_{splicing}$ of mean 0 and width  $\sim 3\%$ at the $k$ maximal value of 0.02  $(\rm{km/s})^{-1}$.

\end{itemize}

 For a given cosmological model defined by the $n$ cosmological, astrophysical and nuisance parameters $\Theta=(\theta_{1},\ldots,\theta_{n})$, and for a data set of power spectra $P(k_i,z_j)$ measured with the covariance matrix $C$, the likelihood function can therefore be written as:

\begin{equation}
{\cal L}\bigl(P;\Theta\bigr)  = \frac{1}{(2\pi)^{N_k \cdot N_z/2} \sqrt{\det(C)}}\exp \left(-\frac{\Delta^T C_{\rm tot}^{-1} \Delta}{2} \right) \cdot  
{\cal L}_{prior}\bigl( \alpha \bigr) 
\end{equation}
where the $N_k \times N_z$ elements of the  vector $\Delta$  are defined as $\Delta(k_i,z_j)= P(k_i,z_j) - P^{th}(k_i,z_j)$, and $C_{\rm tot}$ is the sum of the data and the simulation covariance matrices. The power spectrum $P(k_i,z_j)$ is measured for $N_k$ bins in $k$ and $N_z$ bins in redshift. The theoretical power spectrum $P^{th}(k_i,z_j)$ is the  predicted value of the power spectrum for the bin $k_i$ and redshift $z_j$ knowing $\Theta$. The last term of the likelihood corresponds to the constraint on the nuisance parameters $\alpha=(\alpha_{reso},\alpha_{noise}^{1},\ldots,\theta_{noise}^{12}, \alpha_{splicing})$, which are a subset of the parameters $\Theta$.

\subsection{Cosmic Microwave Background}
\label{sec:cmb}

We include in our analysis the  CMB data sets described in the March 2013 Planck papers (see e.g. \cite{PlanckCollaboration2013}). The `Planck’ data set mentioned in tables and figures hereafter includes the low-$l$ and high-$l$ CMB temperature likelihoods from the first Planck release. The `CMB’ data set includes `Planck’ plus a low-$l$ WMAP+Planck polarisation likelihood (called `WP’ in \cite{PlanckCollaboration2013}), and the high-$l$ likelihoods from the Atacama Cosmology Telescope (ACT)~\cite{Das:2013zf} and the South Pole Telescope (SPT)~\cite{Reichardt:2011yv} ground-based experiments (called `highL’ in \cite{PlanckCollaboration2013}).

\subsection{Other probes}
\label{sec:other}

In order to further remove degeneracies between cosmological parameters, we will occasionally combine CMB data with the measurement of the Baryon Acoustic Oscillation (BAO) scale by the BOSS collaboration, as presented in \cite{Anderson:2013zyy}. This measurement provides independent information on the BAO scale at $z=0.57$ along and perpendicular to the line of sight. We include the full correlation between these two measurements, as computed by the authors of \cite{Anderson:2013zyy} and implemented in the {\sc Monte Python} parameter extraction code\footnote{\tt http://www.montepython.net} (likelihood called {\tt bao\_boss\_aniso} in the code).


\section{Interpretation methodology}
\label{sec:methodology}

The latest Ly$\alpha$ analyses used Markov Chain Monte Carlo simulations \citep{Viel2010} with Bayesian inference.  Even more recently, the Planck collaboration~\cite{PlanckCollaboration2014Freq} used a frequentist approach to interpret the CMB data and eventually to derive constraints on  the neutrino masses.  The study demonstrates the interest of conducting in parallel the two interpretations. The
debate between the Bayesian and the frequentist statistical approaches is beyond the scope of this paper.  The conceptual difference between the two methods should not generally lead, in the end, to major discrepancies in the determination of physical parameters and their confidence intervals when the model parameters can all be constrained by the data (see~\cite{Yeche2006}). This is expected to be the case in the present work.   We decided to perform both approaches to check the robustness of the results.

\subsection{Frequentist Interpretation}

 Our determination of the coverage intervals of unknown cosmological parameters is based on the `frequentist' (or `classical') confidence level method originally defined by \citet{neyman1937}. This approach avoids any potential bias due to the choice of priors.  We start with the likelihood ${\cal L}\bigl(x,\sigma_x;\Theta)$, for a given cosmological model defined by the $n$ cosmological, astrophysical and nuisance parameters $\Theta=(\theta_{1},\ldots,\theta_{n})$, and for data measurements $x$ with Gaussian experimental errors $\sigma_{x}$.  In the rest of this paper, we adopt a $\chi^2$ notation, which means that the following quantity is minimized:

\begin{equation}
\chi^2(x,\sigma_x;\Theta) = -2 \ln ({\cal L}(x,\sigma_x;\Theta))\,.
\label{eq:chi2}
\end{equation}

We first determine the minimum $\chi^2_0$ of $\chi^2(x,\sigma_{x};\Theta)$ leaving  all the cosmological parameters free. Then, to set a confidence level (CL) on any individual cosmological parameter $\theta_i$, we scan the variable $\theta_i$: for each fixed value of $\theta_i$, we minimize again $\chi^2(x,\sigma_{x};\Theta)$ but with $n-1$ free parameters. The $\chi^2$ difference, $\Delta \chi^2(\theta_i)$, between the new minimum and  $\chi^2_0$, allows us to compute the CL on the variable, assuming that the experimental errors are Gaussian,
\begin{equation}
{\rm CL}(\theta_i) = 1-\int_{\Delta \chi^2(\theta_i)}^{\infty}  f_{\chi^2}(t;N_{dof}) dt,
\label{Eq:CL}
\end{equation}
with
 \begin{equation}
 f_{\chi^2}(t;N_{dof})=\frac{e^{-t/2}t^{N_{dof}/2 -  1}}{\sqrt{2^{N_{dof}}} \Gamma(N_{dof}/2)}   \label{Eq:chi2}
\end{equation}
where $\Gamma$ is the Gamma function and the number of degrees of freedom $N_{dof}$
is equal to 1.
This method can be easily extended to two variables. In this case, the minimizations are
performed for $n-2$ free parameters and the confidence level ${\rm CL}(\theta_i,\theta_j)$ is
derived from Eq.~\ref{Eq:CL} with $N_{dof}=2$.

By definition, this frequentist approach does not require any marginalization to determine the sensitivity on a single individual
cosmological parameter.  Moreover, in contrast with Bayesian treatment, no prior on the cosmological parameters is needed.  With this approach, the correlations between the variables are naturally taken into account and the minimization fit can explore the entire phase space of the cosmological, astrophysics and nuisance parameters.

In this study, we have to deal with a very specific case, when a variable is bounded to a physical region. Even if the likelihood for Ly$\alpha$ is technically defined in the unphysical region corresponding to $\sum m_\nu<0$, the definition of the coverage intervals are more difficult to derive from the original construction of \citet{neyman1937}.  In particular, for the specific variable, $\sum m_\nu$, we cross-check our definition of the upper limit with the elegant approach proposed in~\cite{FC98},  which unifies the treatment of upper limits for null results and two-sided confidence intervals for non-null results.

\subsection{Bayesian Interpretation}

The widely used `Bayesian’ approach relies on Bayes theorem. For a given data set, a given model, a set of priors (the probability assigned to each model parameter before fitting the experiment) and a likelihood (the probability of the data given the model), this theorem describes the computation of the posterior distribution, i.e., the probability of each particular value of model parameters. The full posterior can be marginalized over some of the model parameters in order to provide a posterior on the remaining parameters: this allows the computation of credible intervals for each single parameters, joint confidence contours on pairs of parameters, etc. Several numerically-friendly implementation of this calculation, all relying on a Monte Carlo exploration of the parameter space, have been described in the literature. 

For our Bayesian analysis, we use the parameter extraction code {\sc Monte Python}\footnote{\tt http://www.montepython.net} \cite{Audren:2012wb}, which implements several of these methods. In this work, we only need to use the Metropolis-Hastings Monte Carlo Markov Chain (MCMC) method described in \cite{Lewis:2002ah}, which scales  well with the number of free parameters. This feature is crucial for this study, given the large number of nuisance parameters associated to the Planck, ACT/SPT and Ly$\alpha$ likelihoods. To speed up the MCMC exploration, when we combine Ly$\alpha$ with CMB data, we use a Cholesky decomposition of the proposal density matrix \cite{Lewis:2013hha}, with oversampling factors set to 1 for cosmological parameters, 2 for CMB nuisance parameters and 4 for Ly$\alpha$ parameters. We verified that our results are stable under a change in the oversampling factors. Our choice of priors will be specified for each case in the next section.


\section{Results}
\label{sec:results}

\subsection{ Constraints on cosmological parameters using Ly$\alpha$ alone} \label{sec:lya_alone}
\label{sec:resLya}
\subsubsection{Frequentist approach}
\label{sec:resLya_F}

We checked explicitly that  Ly$\alpha$ data can constrain independently the five cosmological parameters used in this analysis ($\sigma_8$, $n_s$, $\Omega_m$, $H_0$, $\sum m_\nu$) and the 19 astrophysical and nuisance parameters introduced in section~\ref{sec:likelihood}. This check was performed by running simulations with off-the-grid values of the various parameters, and verifying that the fitted values agreed with the simulated ones; details can be found in \cite{Borde2014}.
The parameters have indeed distinct effects, especially on the amplitude, shape (slope, curvature, etc.) and redshift dependence of the flux power spectrum. For instance, the three parameters $n_s$, $H_0$ and $\sum m_\nu$ all affect the shape of the spectrum in different ways, as can be confirmed visually on the simulations presented in \cite{Borde2014, Rossi2014},  as well as in Figure~\ref{fig:PkNumass} (see section~\ref{sec:sensitivity} for further detail).

There still remains, however, a large degeneracy between $H_0$ and $n_s$, with a correlation of about $70\%$. Indeed, the impact of these two parameters on the spectrum shape is {\it nearly} the same, unless  considering  extreme values of $H_0$. Hence, in order to present interesting results, it is useful to impose a constraint on $H_0$ forcing this parameter to remain in an  observationally favored region. In most of this section, we impose a Gaussian constraint  with mean value and standard deviation $H_0 = 67.4 \pm 1.4 \;{\rm km~s^{-1}~Mpc^{-1}}$, corresponding to the constraint derived from the Planck 2013 temperature data assuming the ``base model’’, i.e., a $\Lambda$CDM model with fixed $\sum m_\nu=0.06$~eV \cite{PlanckCollaboration2013}. We will investigate the consequence of choosing a more conservative constraint on $H_0$ in subsection~\ref{lya_only_B}.

\begin{figure}[htb]
\begin{center}
\epsfig{figure= 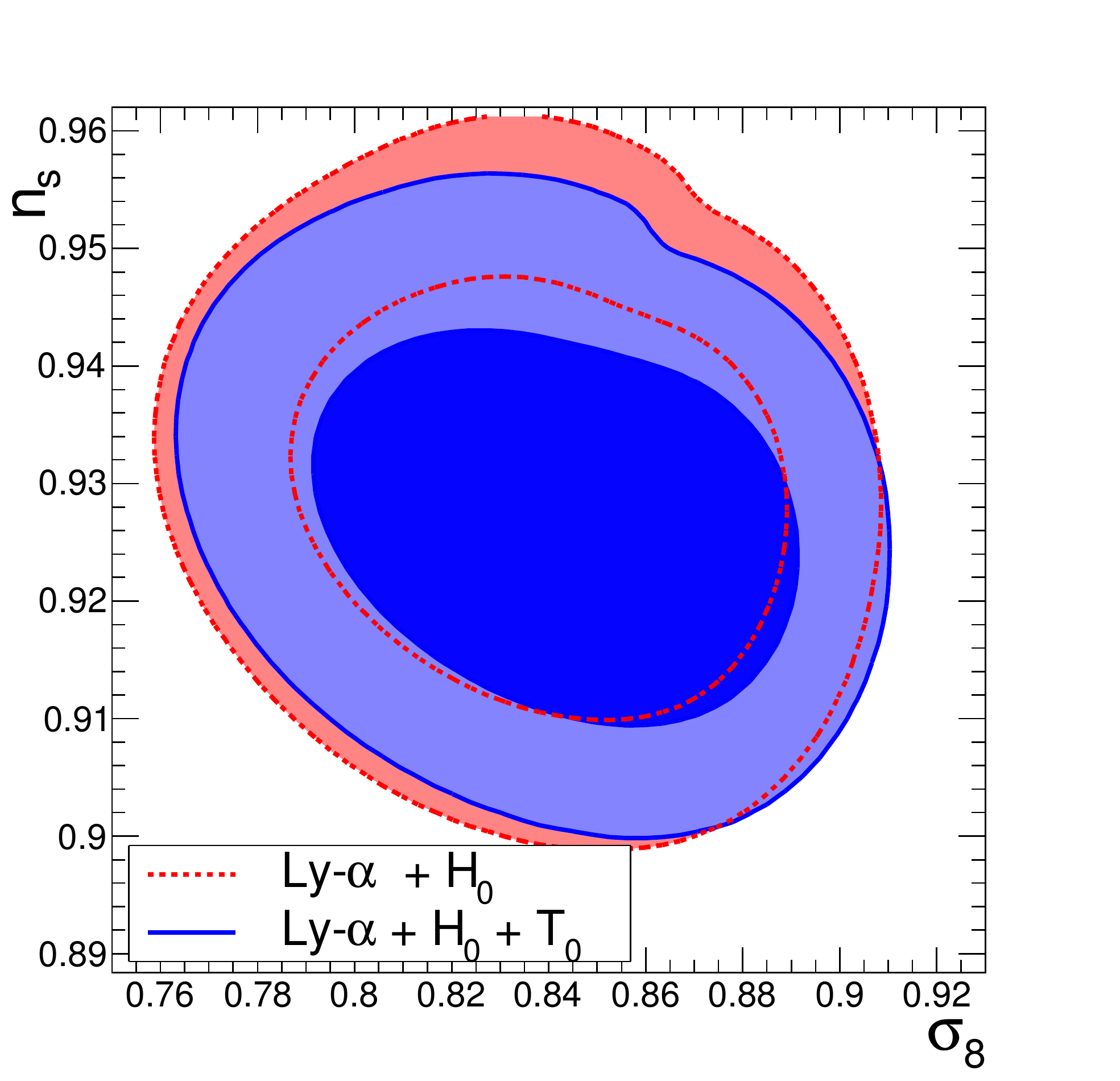,width = 7.5cm} \epsfig{figure= 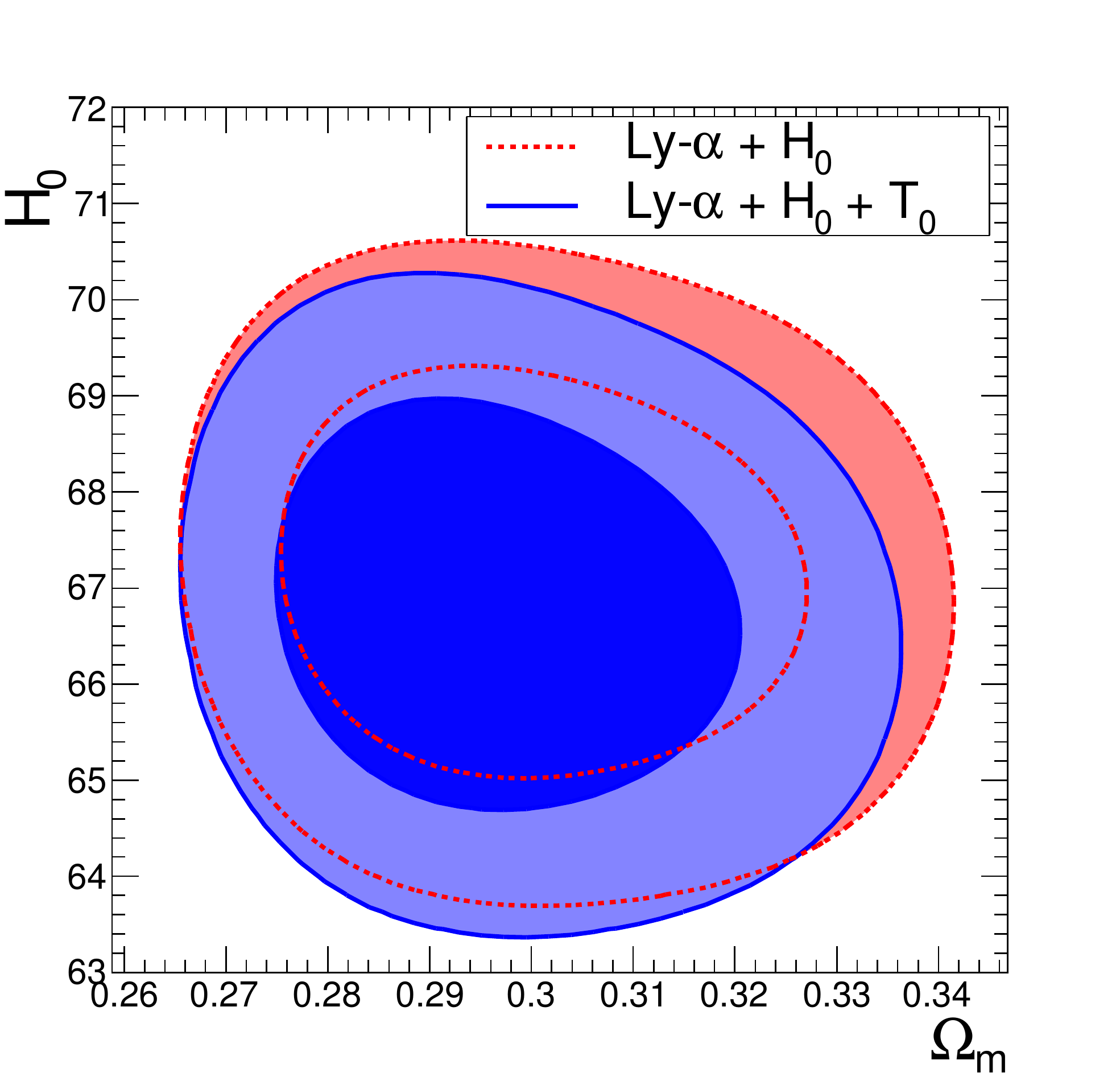,width = 7.5cm}
\epsfig{figure= 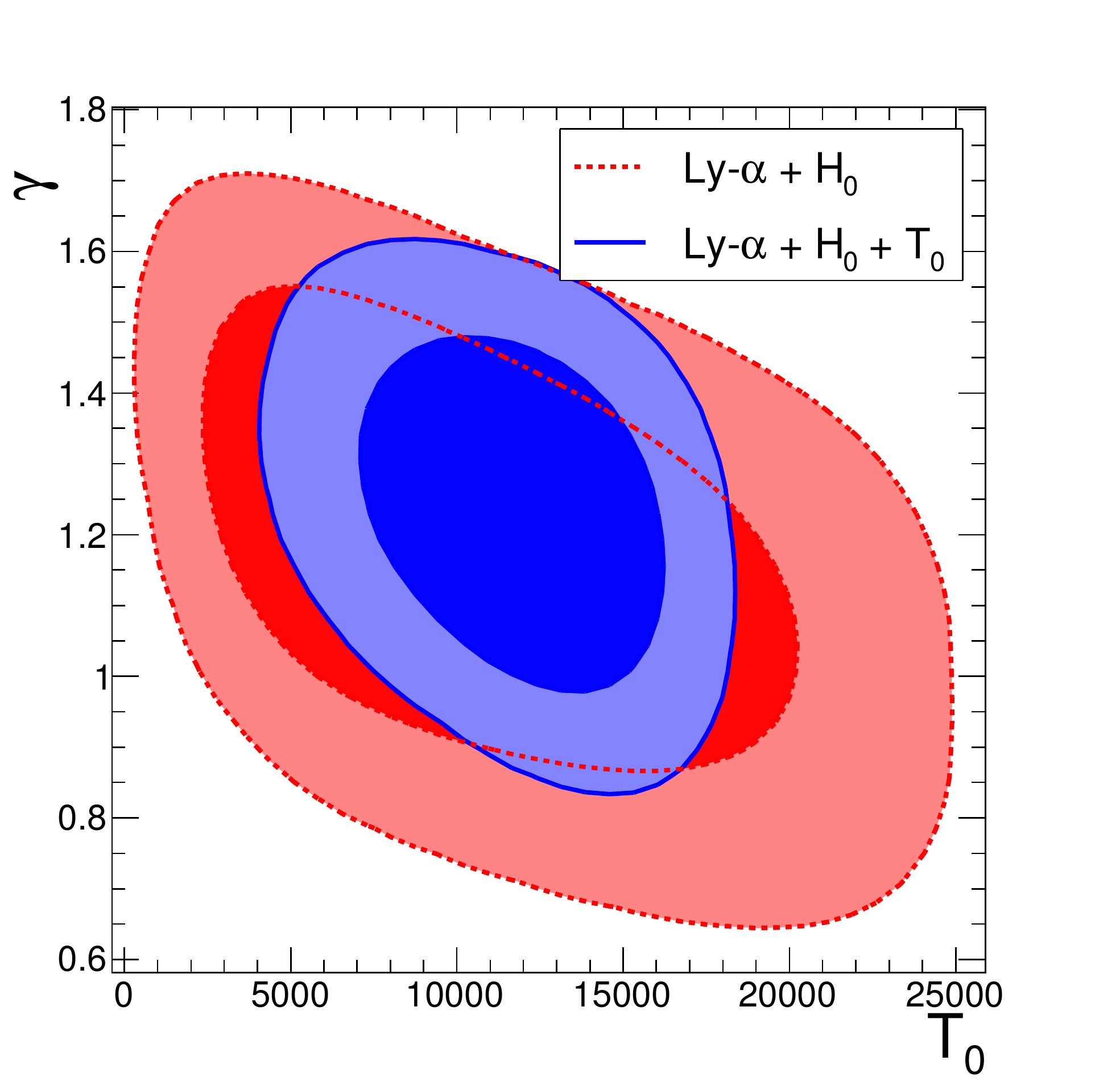,width = 7.5cm} \epsfig{figure= 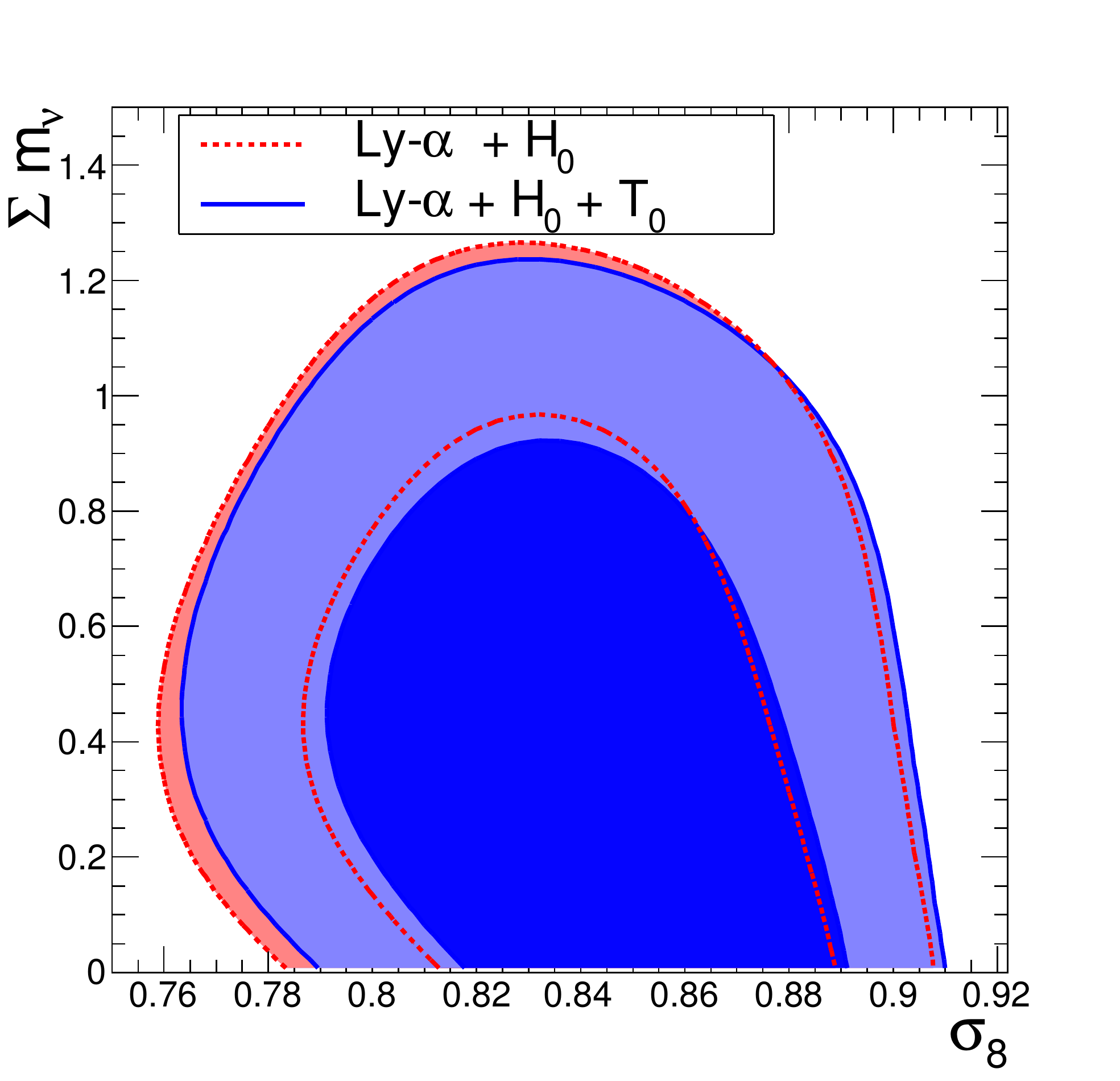,width = 7.5cm}
\caption{\it Two-dimensional probability contours for  the astrophysical and cosmological parameters $(\sigma_8 , n_s)$,  $(\Omega_m ,  H_0)$, $(T_0 ,  \gamma)$,  and  $(\sigma_8 , \sum m_\nu)$, at the 68\% and 95\% confidence levels, following a frequentist interpretation ($\sum m_\nu$ is in units of eV and $H_0$ in ${\rm km~s^{-1}~Mpc^{-1}}$).  The contours are obtained with  BOSS Ly$\alpha$ data  using a Gaussian constraint $H_0 = 67.4 \pm 1.4 \;{\rm km~s^{-1}~Mpc^{-1}}$. The solid blue lines and dotted red lines correspond respectively to a configuration with 12 redshift bins $z=[2.1-4.5]$  and an external constraint on $T_0$, or 10 redshift bins $z=[2.1-4.1]$ with $T_0$ let free. }
\label{fig:ContoursLyaAlone}
\end{center}
\end{figure}
The heating rate parameters $T_0$ and $\gamma$ of the IGM are difficult to determine with high precision  using the medium-resolution  data of BOSS (cf., however, the recent work of~\cite{2014arXiv1405.1072L} for a measurement of $\gamma$ in BOSS). They are usually measured from high-resolution data such as in the recent work of \citep{Bolton2014}.
When left free in the fits, $T_0$ and $\gamma$ are strongly correlated to the amplitude  $A^\tau$ and slope $\eta^\tau$ that  describe the effective optical depth $\tau_{\rm eff} (z)= A^\tau \times (1+z)^{\eta^\tau}$ of the IGM. Since we do not aim to determine IGM properties, we can constrain $T_0$ to a   reasonable range ($T_0 = 14000 \pm 3000$ at $1\sigma$ using a Gaussian constraint) in agreement with most recent measurements~\citep{Becker2011,Garzilli2012, Lidz2010, Schaye2000} .  We verified that this restriction had no impact on the best-fit value of the cosmological parameters (see Fig.~\ref{fig:ContoursLyaAlone}). We also consider  configurations where we limit the data to the better-measured 10 lowest redshift bins.  The slope $\gamma$ is left free in all cases.

We performed four fits, with or without neutrino masses and including 12 or 10 redshift bins (with or without a constraint on $T_0$, respectively). The results of these four fits (all based on the frequentist approach) are given in table~\ref{tab:fit_Lya}.

\begin{table}[htb]
\caption{\it Best-fit value and frequentist confidence limits of the cosmological and astrophysical parameters of the model fitted to the flux power spectrum $P(k_i,z_j)$ measured with the BOSS Ly$\alpha$ data. We consider either 12 or 10 redshift bins, covering respectively   $z=[2.1-4.5]$ or  $[2.1-4.1]$, and  $\sum m_\nu$ fixed at 0~eV or let free. We use a Gaussian constraint $H_0 = 67.4 \pm 1.4 \;{\rm km~s^{-1}~Mpc^{-1}}$, and a constraint on $T_0$ for the 10 redshift-bin case  as explained in the text. $T_0$ in K and $\sum m_\nu$ in eV.  For each parameter, we quote the 68\% confidence levels, except for the total neutrino mass, for which we quote the 95\% upper bound.}
\begin{center}
\begin{tabular}{lcccc}
\hline
Parameter & $2.1<z<4.5$ & $2.1<z<4.1$ & $2.1<z<4.5$& $2.1<z<4.1$ \\
\hline \\[-10pt]
$\sigma_8$ &  $0.858\pm0.025$ & $0.855\pm0.025$& $0.850\pm0.036$& $0.846\pm0.039$ \\[5pt]
$n_s$ &  $0.929\pm0.011$&  $0.928\pm0.012$&  $0.925\pm0.011$&  $0.928\pm0.012$ \\[5pt]
$\Omega_m$ &  $0.292\pm0.013$ &  $0.294\pm0.014$&  $0.293\pm0.014$&  $0.296\pm0.017$ \\[5pt]
$H_0$ &  $66.8\pm1.4$&  $67.2\pm1.4$&  $66.8\pm1.4$&  $67.2\pm1.4$ \\[5pt]
$\sum m_\nu$ & fixed at 0& fixed at 0&$<1.1$ {\scriptsize (95\%)}&$<1.1$ {\scriptsize (95\%)} \\[5pt]
\hline \\[-10pt]
$f_ {\rm{Si\,III}}$ &  $0.0061\pm0.0004$ &  $0.0061\pm0.0004$ &  $0.0061\pm0.0004$ &  $0.0061\pm0.0004$ \\[5pt]
$f_ {\rm{Si\,II}}$ &  $0.0007\pm0.0005$ &  $0.0007\pm0.0005$ &  $0.0007\pm0.0005$ & $0.0007\pm0.0005$ \\[5pt]
$T_0$& $12100^{+2800}_{-3200}$  &   $8400^{+8000}_{-4400}$  &  $12100^{+2800}_{-3200}$ &  $8200^{+8000}_{-4400}$ \\[5pt]
$\gamma$& $1.1\pm 0.2$& $1.1\pm 0.2$& $1.2\pm 0.2$&$1.2\pm 0.2$ \\[5pt]
$A^\tau$ & $0.0026\pm0.0001$ & $0.0028\pm0.0002$ & $0.0026\pm0.0001$  & $0.0028\pm0.0002$ \\[5pt]
$\eta^\tau$ &$3.67\pm0.02$ & $3.67\pm0.02$ & $3.67\pm0.02$ & $3.67\pm0.02$ \\[5pt]
\hline \\[-10pt]
$\chi^2 / dof$ & $416/ 398$& $339/ 330$& $416/397$ & $339/329$ \\[5pt]
 p-value & $ 22\%$& $24\%$& $21\%$ & $24\%$ \\
\hline
\end{tabular}
\end{center}
\label{tab:fit_Lya}
\end{table}
\newpage

\begin{figure}[htb]
\begin{center}
\epsfig{figure= 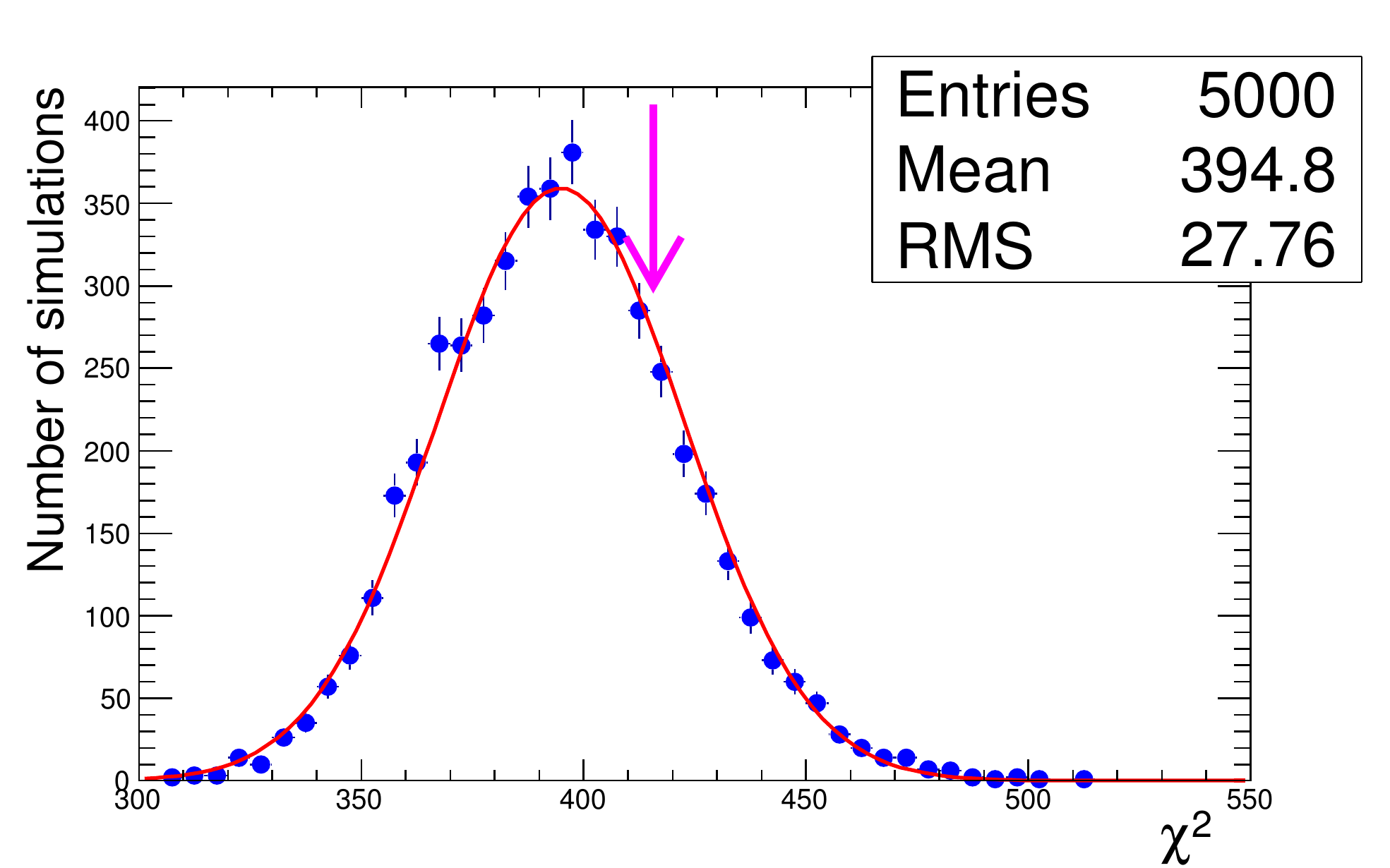,width = 7.5cm} \epsfig{figure= 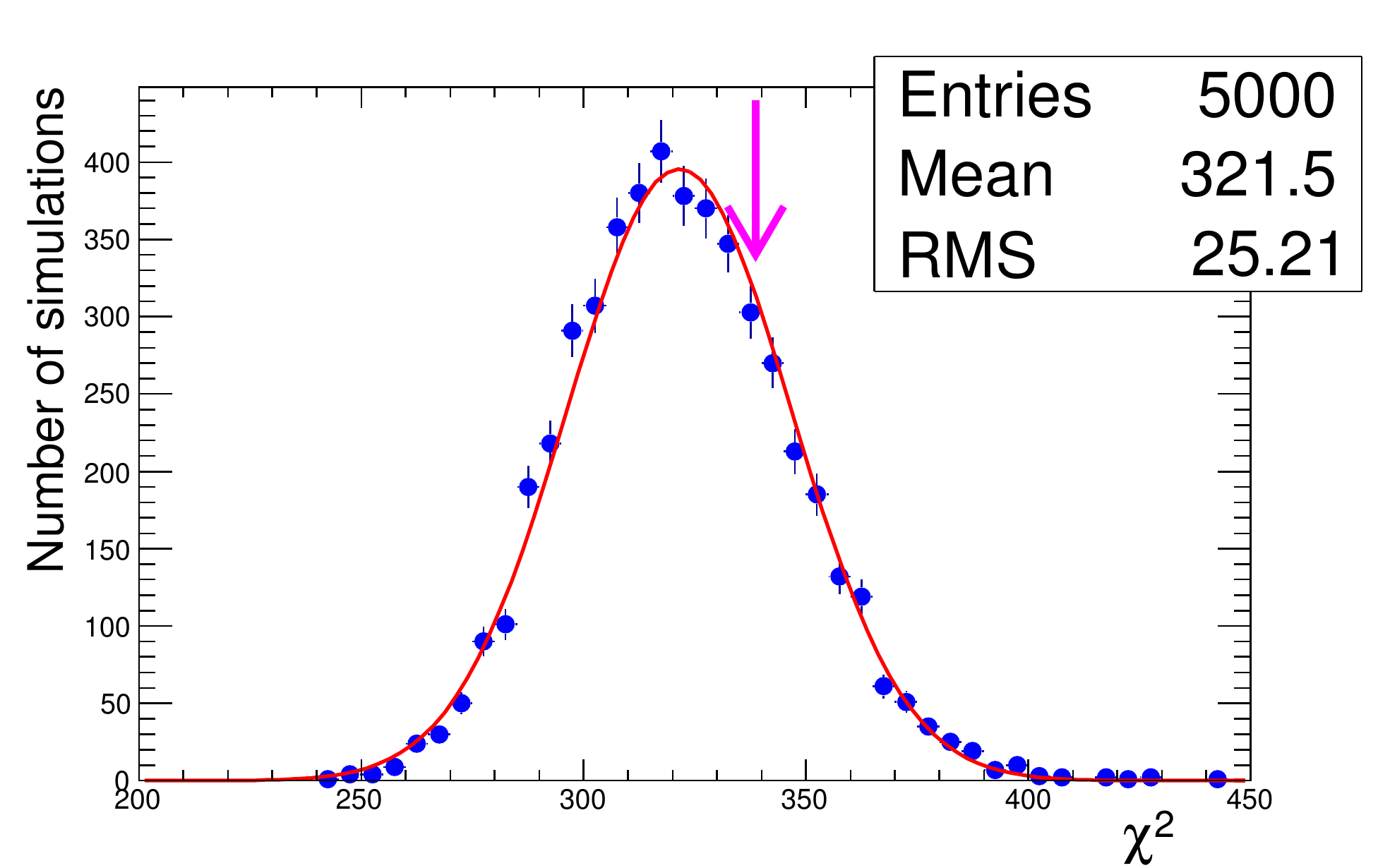,width = 7.5cm}
\caption[]{\it The distribution of the $\chi^2$ obtained for 5000 simulations of the 1D power spectrum measurement by BOSS. The left and right panels show, respectively, the  $\chi^2$ distribution for a  12 redshift bin $z=[2.1-4.5]$ configuration with an external constraint on $T_0$ and a 10 redshift bin $z=[2.1-4.1]$ configuration.  The arrows represent the values obtained for the $\chi^2$ minimum with the fit to the data.  The p-value $\chi^2$  are  respectively 22\% and 24\% for the two configurations. The deviation from the mean value of the $\chi^2$ distribution is less than $1\sigma$.}
\label{fig:GOF}
\end{center}
\end{figure}

The fits are extremely stable. Allowing the neutrino mass float does not modify any of the cosmological parameters by more than $1\sigma$.  Restricting the fit to the lowest 10 redshift bins  also does not  cause any significant change in the fitted parameters. Removing the constraint on the IGM temperature decreases the best-fit value of $T_0$ by about $5000$~K when fitting all 12 redshift bins and by about $2000$~K when restricting to the lowest 10, but does not affect any of the other parameters; no shift is observed beyond the $0.5\sigma$ level. We also verified that removing or doubling the value of the simulation uncertainties  modified by less than $0.5\sigma$ the best-fit value of any of the parameters, the greatest effect being seen on  $A_\tau$. Removing the simulation uncertainties increases the best-fit $\chi^2$ by 10\%.

As the actual number of degrees of freedom of our problem is affected in a non-trivial manner by correlations between parameters, we must determine the goodness of fit to the data with an alternative method. We compare the  absolute value of $\chi^2_0$ obtained with the data to the $\chi^2$ distribution. We perform 5000 simulations of the measured power spectrum and  repeat the fit for each simulation. The distribution of the $\chi^2$ of Fig.~\ref{fig:GOF} can be used to derive the goodness of fit. The fractions of simulations having a $\chi^2$ value larger than  $\chi^2_0$, the p-value,  are  respectively 22\% and 24\% for a  12 redshift bin $z=[2.1-4.5]$ configuration with an external constraint on $T_0$, and a 10 redshift bin $z=[2.1-4.1]$ configuration without constraint on $T_0$.   The measured $\chi^2$ is less than a $1\sigma$ away from the most probable  $\chi^2$ for the two configurations, thus demonstrating  the consistency of the data with the model. 

The upper limit  obtained on the neutrino mass from Ly$\alpha$ alone is $\sum m_\nu < 1.1$~eV at 95\% CL, whether for  the fit using all data to $z=4.5$ or when restricting to data with redshift $z<4.1$. The $\chi^2$ minima are at $\sum m_\nu\sim 0.3$ and 0.2~eV for the two configurations respectively, and are consistent with 0. The main correlation  observed is between $\sum m_\nu$ and $\sigma_8$, where the correlation coefficient reaches $\sim 70\%$ (see Fig.~\ref{fig:ContoursLyaAlone}).
In descending order of importance,  $\sum m_\nu$ has a 45\% correlation with $\Omega_m$ and a 20\% correlation  with $n_s$.
Despite the significant improvement in data quality compared to previous measurements, this upper limit on neutrino mass  is similar to the one found by~\cite{Viel2010} using the 1D power spectrum measured with SDSS data. This result  is a consequence  of our better accounting of systematics in the simulations, but also our improved modeling of the dependence of the power spectrum on the various relevant parameters.  In particular, our analysis includes all terms in the second-order Taylor expansion of the power spectrum, while previous studies neglected cross derivatives.

As we have explained above, we obtain robust results with a better  $\chi^2 / dof$  and p-value when restricting the BOSS Ly$\alpha$ data to the lowest 10 redshift bins ($z=[2.1-4.1]$). In the rest of this paper, we will therefore consider this configuration as a baseline for the Ly$\alpha$ likelihood.

\subsubsection{Bayesian approach}\label{lya_only_B}
\label{sec:resLya_B}

For comparison,  table~\ref{tab:fit_Lya_B} presents the results of a Bayesian analysis, with 10 redshift bins and a free neutrino mass. The first column reports our results when flat priors are assumed for all parameters except $H_0$ (which is assigned a Gaussian prior), and can be directly compared with the frequentist results in the last column of table~\ref{tab:fit_Lya}, obtained under the same assumptions. Frequentist and Bayesian confidence limits should agree perfectly in the ideal case of a multivariate Gaussian posterior distribution. Here, the actual posterior deviates only slightly from a multivariate Gaussian, except along the direction of $\sum m_\nu$ (constrained to be positive) and $T_0$ (the marginalized posterior on $T_0$ is found to be highly skewed). Hence we find only minor differences in the final confidence limits (for all parameters except $T_0$, for which the frequentist confidence limits and Bayesian credible intervals are shifted by nearly one sigma). The 95\%~C.L. upper bound on $\sum m_\nu$ differs by only 12\% in the two analyses. We conclude that our bounds on cosmological parameters are robust  against a change of statistical approach.

\begin{table}[htb]
\caption{\it Bayesian mean values and credible intervals of the cosmological and astrophysical parameters of the model fitted to the flux power spectrum $P(k_i,z_j)$ measured with the BOSS Ly$\alpha$ data (presented in section~\ref{sec:Lya}). $H_0$ is in units of ${\rm km~s^{-1}~Mpc^{-1}}$, $T_0$ in K and $\sum m_\nu$ in eV. For each parameter, we quote the 68\% confidence levels, except for the total neutrino mass, for which we quote the 95\% upper bound. The  columns correspond to different choices of priors on $H_0$ and other parameters, as explained in the text. We do not show the results for the 13 nuisance parameters varied simultaneously with these 11 parameters.}
\begin{center}
\begin{tabular}{lccc}
\hline
& Ly$\alpha$ + $H_{0}^\mathrm{Gaussian}$ & Ly$\alpha$ + $H_{0}^\mathrm{Gaussian}$ & Ly$\alpha$ + $H_{0}^\mathrm{tophat}$ \\
& {\scriptsize ($H_0 = 67.4 \pm 1.4$)} & {\scriptsize ($H_0 = 67.4 \pm 1.4$)} & {\scriptsize ($62.5 \leq H_0 < 72.5$)} \\
Parameter & flat $(\Omega_\mathrm{m}, \sigma_8)$ prior & flat $(A_s, \omega_m)$ prior & flat $(A_s, \omega_m)$ prior \\
\hline \\[-10pt]
$\sigma_8$         & $0.83_{-0.03}^{+0.03}$       & $0.83 \pm 0.03$         & $0.84 \pm 0.03$ \\[5pt]
$n_s$              & $0.928 \pm 0.012$          & $0.928_{-0.012}^{+0.011}$ & $0.931_{-0.012}^{+0.012}$ \\[5pt]
$\Omega_m$         & $0.303 \pm 0.015$          & $0.315_{-0.021}^{+0.017}$ & $0.316_{-0.021}^{+0.018}$ \\[5pt]
$H_0$              &  $67.0 \pm 1.4$            & $67.0 \pm 1.4$          & $<70.9$ {\scriptsize (95\%)} \\[5pt]
$\sum m_\nu$       & $<0.95$~~{\scriptsize (95\%)} & $<0.98$ {\scriptsize (95\%)} & $<0.98$ {\scriptsize (95\%)} \\[5pt]
\hline \\[-10pt]
$f_ {\rm{Si\,III}}$ & $0.0061 \pm 0.0004$         & $0.0061 \pm 0.0004$     & $0.0061 \pm 0.0004$ \\[5pt]
$f_ {\rm{Si\,II}}$  & $0.0008 \pm 0.0005$         & $0.0008 \pm 0.0005$     & $0.0008 \pm 0.0005$ \\[5pt]
$T_0$              & $14000_{-9000}^{+3000}$      & $14000_{-9000}^{+3000}$   & $14000_{-9000}^{+3000}$ \\[5pt]
$\gamma$           & $1.2 \pm 0.2$              & $1.2 \pm 0.2$           & $1.2 \pm 0.2$ \\[5pt]
$A^\tau$           & $0.00272_{-0.00022}^{+0.00015}$ & $0.00271_{-0.00020}^{+0.00014}$ & $0.00270_{-0.00021}^{+0.00014}$ \\[5pt]
$\eta^\tau$        & $3.66 \pm 0.02$             & $3.66 \pm 0.02$         & $3.66 \pm 0.02$ \\[5pt]
\hline
\end{tabular}
\end{center}
\label{tab:fit_Lya_B}
\end{table}

In   section~\ref{sec:combined}, when combining Ly$\alpha$ and CMB data, we will assume flat priors on the primordial amplitude $A_s$ and physical density $\omega_m$, instead of flat priors on $\sigma_8$ and on the fractional density $\Omega_\mathrm{m}$. In order to show that this difference in the priors has negligible consequences, we repeated the same analysis with flat priors on $(A_s, \omega_\mathrm{m})$, using the Boltzmann code CLASS\footnote{\tt http://class-code.net} to compute $\sigma_8$ in each point. The results, shown in the second column of table~\ref{tab:fit_Lya_B}, are left invariant by the different choice of prior, except for a small shift in the central value of $\Omega_\mathrm{m}$ by  $\sim 1\sigma$.

Finally, instead of a restrictive prior $H_0 = 67.4 \pm 1.4~{\rm km~s^{-1}~Mpc^{-1}}$ (motivated by Planck data alone and for a flat $\Lambda$CDM model), we fit the data with a more conservative top-hat prior $62.5 \leq H_0 < 72.5~{\rm km~s^{-1}~Mpc^{-1}}$, overlapping with most recent distance-ladder measurement of the Hubble rate. The results, shown in the last column of table~\ref{tab:fit_Lya_B}, are  stable within 0.3$\sigma$. Hence the results of this section are robust for any reasonable assumption on $H_0$.

\vspace{0.3cm}

The final conclusion of section~\ref{sec:resLya} is that under the assumption of a minimal $\Lambda$CDM cosmology, the BOSS Ly$\alpha$ data provides a reliable upper bound on the total neutrino mass of 1.1~eV at the 95\% Confidence Level (retaining the frequentist bounds, which are less restrictive than the Bayesian ones by about 15\%).

\subsection{ Combined constraints on cosmological parameters}
\label{sec:combined}

\subsubsection{ Frequentist approach}
\label{sec:FreqCombined}
We now combine the  $\chi^2$ derived from the Ly$\alpha$ likelihood with that of CMB experiments.
 In this section (and unlike in the Bayesian analysis of the next section), we do not directly use  the CMB likelihoods. Instead,
we use the central values and the covariance matrices available in the official WMAP\footnote{http://lambda.gsfc.nasa.gov/product/map/dr5/parameters.cfm} or
{\it Planck}~\cite{PlanckExplanatorySupplement2013} repositories for the five cosmological parameters  ($\sigma_8$, $n_s$ ,  $\Omega_m$,  $H_0$, and $\sum m_\nu$). For each parameter, we assume a Gaussian CMB likelihood with asymmetric $1\sigma$ errors that we estimate on either side of the central value from the $1\sigma$ lower and upper limits, except for $\sigma_8$ where we use half of the $2\sigma$ excursion of the parameter. This approach is adopted  to better account for the narrowing of the {\it Planck} likelihood  for $\sigma_8$ beyond its $1\sigma$ contour, and, in general, this procedure  accounts for asymmetric contours. Because the results we use for  CMB are obtained under the assumption that the sum of the neutrino masses is positive, we similarly impose a lower bound $\sum m_\nu>0$ when computing the combined constraints. We discuss the impact of this assumption in figure~\ref{fig:Chi2Scan} and related text. This assumption was not applied when deriving constraints from Ly$\alpha$ data alone (cf. section~\ref{sec:resLya_F}).

\begin{figure}[htbp]
\begin{center}
\epsfig{figure= 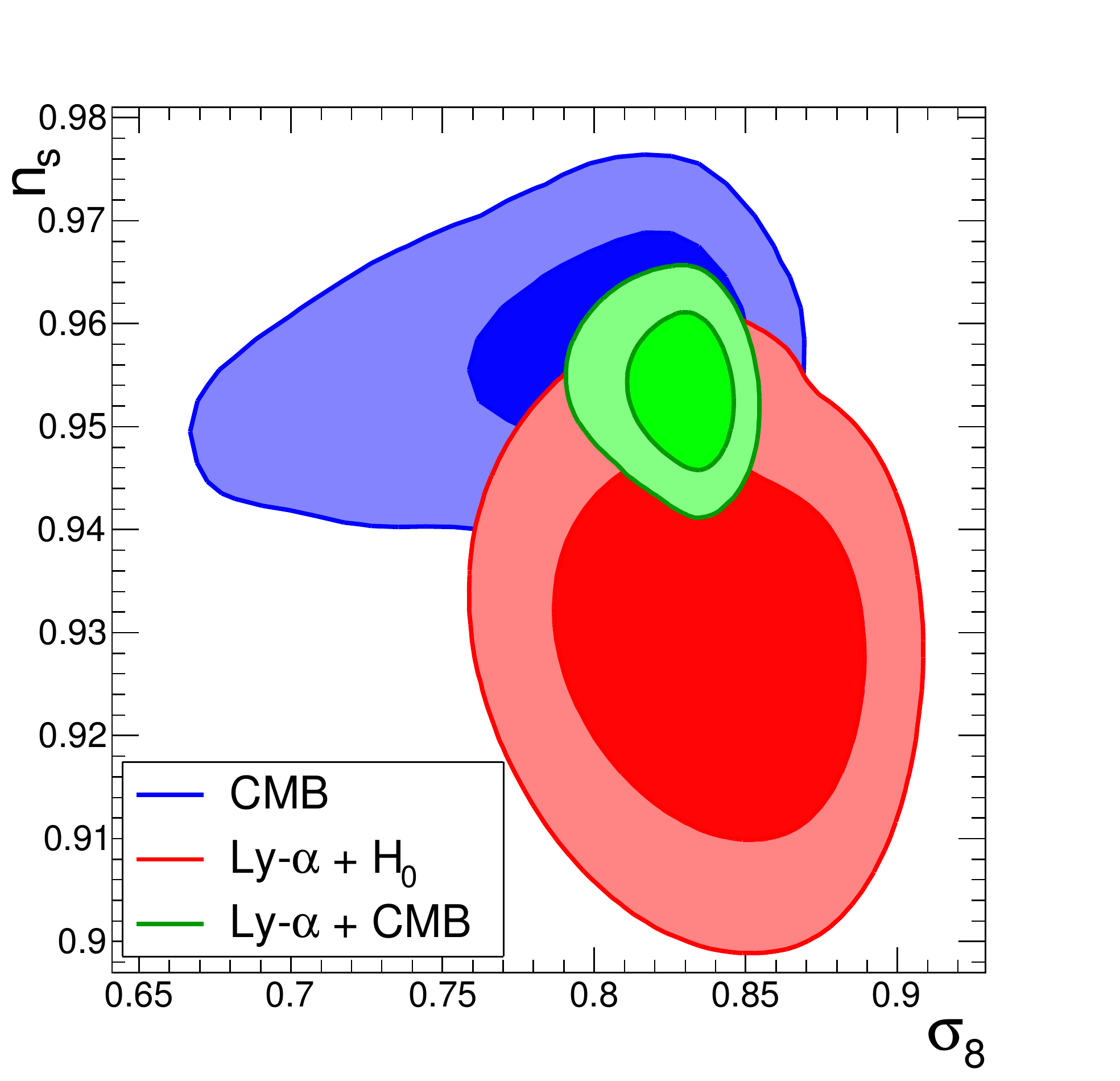,width = 5cm} \epsfig{figure= 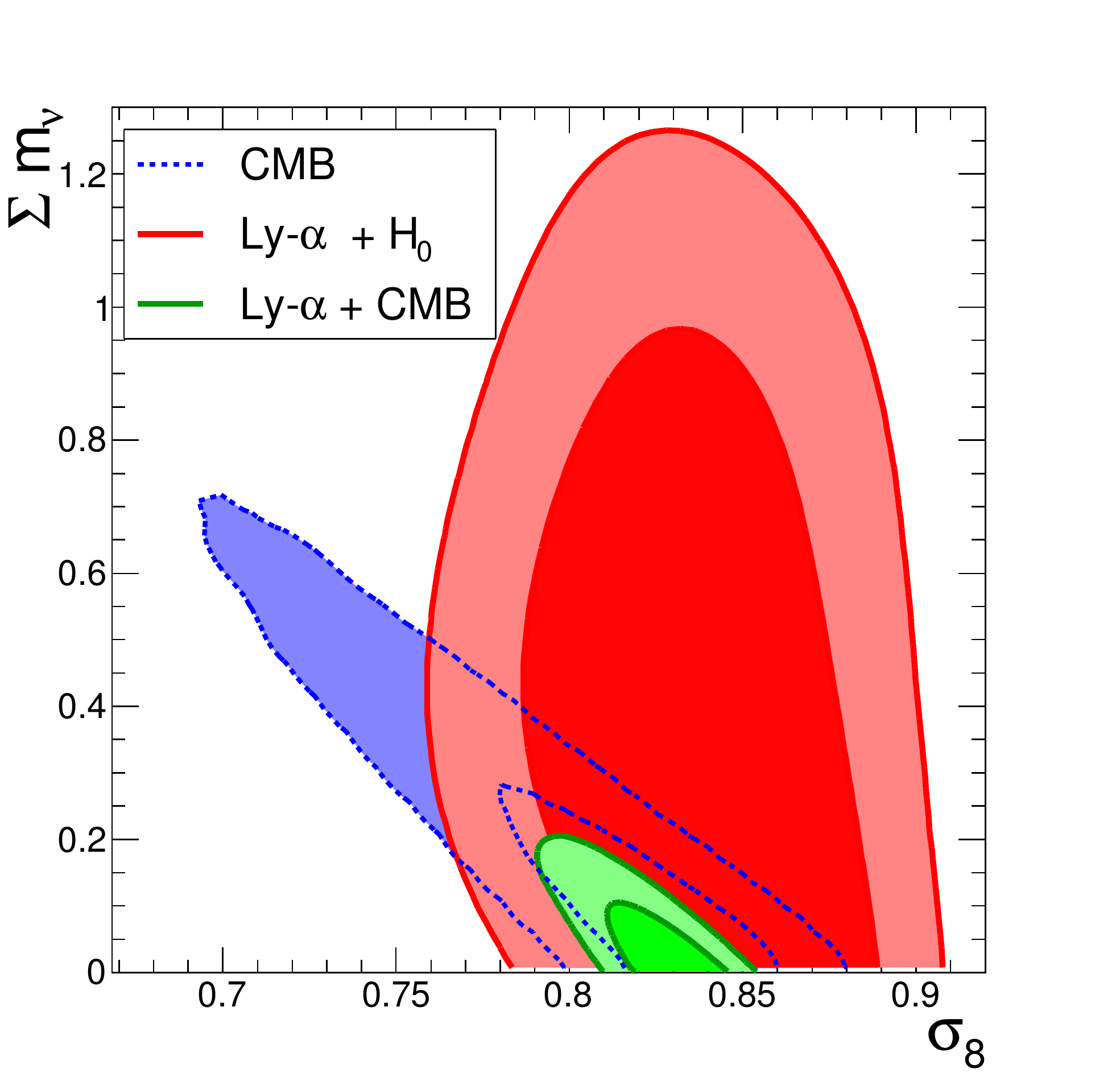,width = 5cm}\epsfig{figure= 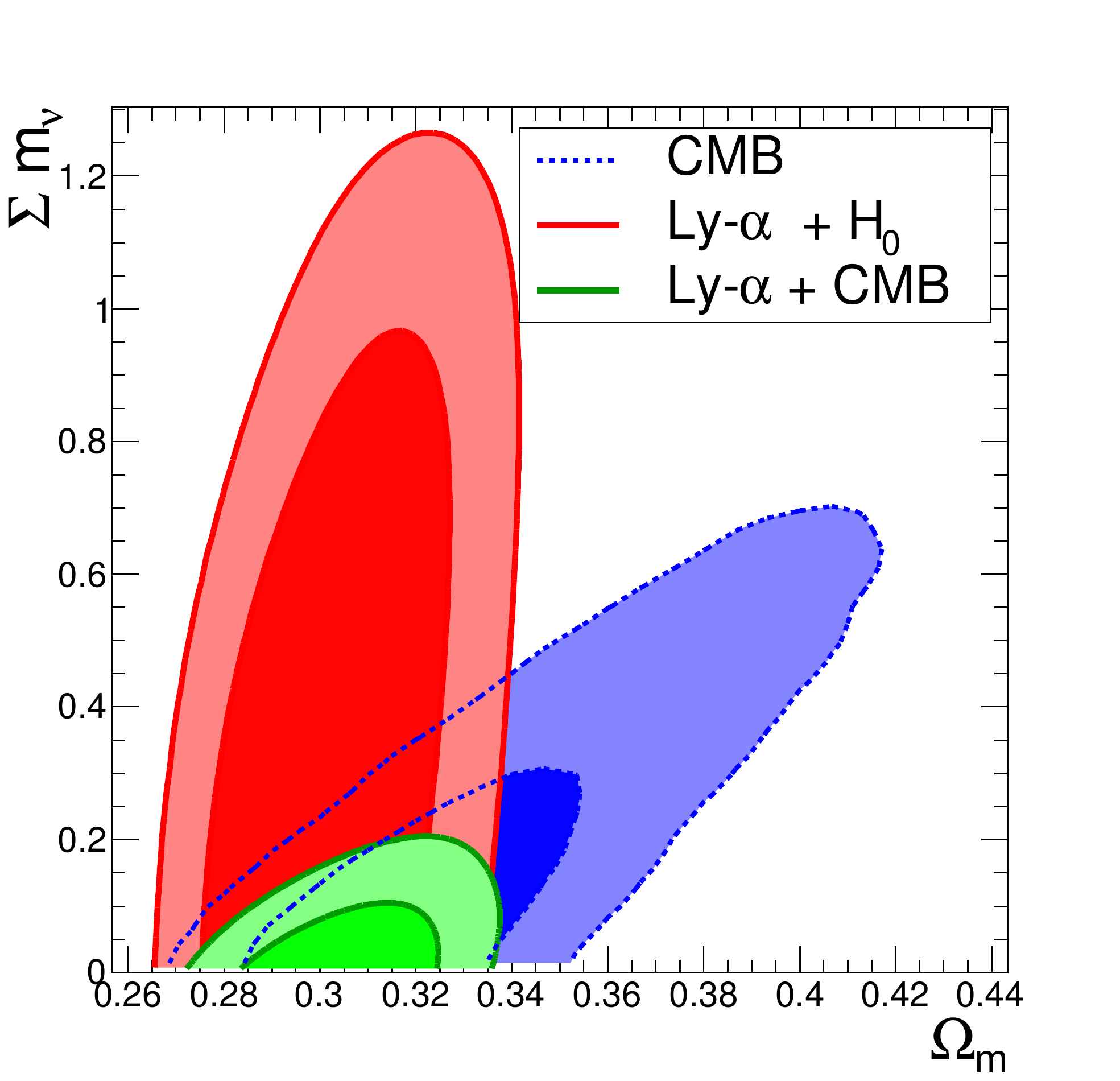,width = 5cm}
\caption{\it 2D confidence level contours for  the $(\sigma_8 , n_s)$ , $(\sigma_8 , \sum m_\nu)$ and  $(\Omega_m , \sum m_\nu)$ cosmological parameters with a frequentist interpretation.  The 68\% and 95\% confidence contours are obtained with different combinations of the BOSS Ly$\alpha$ data presented in section~\ref{sec:Lya} using the 10 redshift bins $z=[2.1-4.1]$, of the Gaussian constraint $H_0 = 67.4 \pm 1.4~{\rm km~s^{-1}~Mpc^{-1}}$ and of CMB data (Planck + WP + ACT + SPT). }
\label{fig:ContourFreqCombi}
\end{center}
\end{figure}

\begin{figure}[htbp]
\begin{center}
\epsfig{figure= 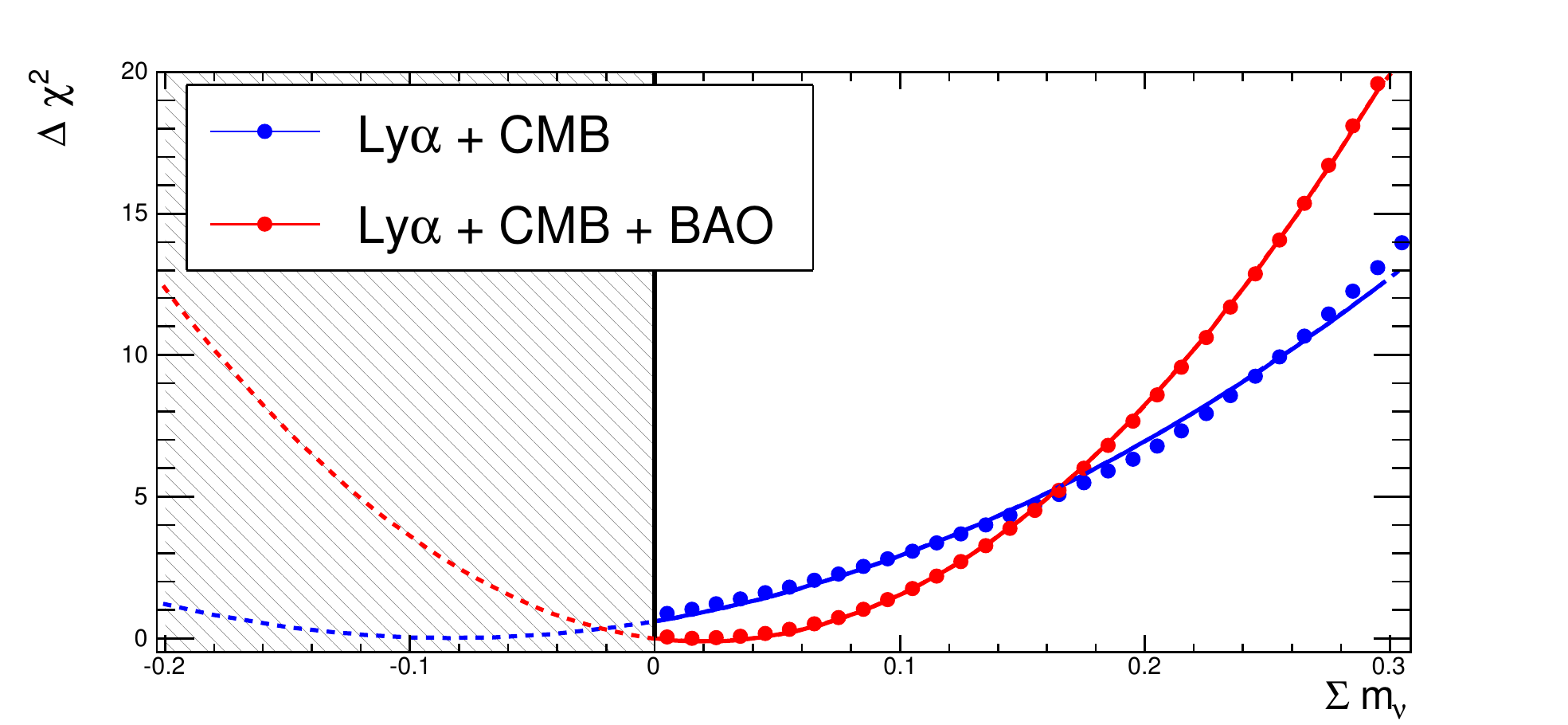,width = 15cm} 
\caption{\it  $\Delta \chi^2$ profile as a function of $\sum m_\nu$ for the two configurations (Ly$\alpha$ + CMB)  and (Ly$\alpha$ + CMB + BAO). Each point is the $\Delta \chi^2$ obtained after a maximization of the total likelihood over the other free parameters.  The points are fitted by a parabola and extrapolated into the negative region as proposed
in~\cite{PlanckCollaboration2014Freq}.}
\label{fig:Chi2Scan}
\end{center}
\end{figure}

We consider several possible combinations  of Ly$\alpha$ and {\it Planck} data. First, we include  the  results  from {\it Planck}  for  $ 2 \leq \ell \leq 2500$ (hereafter referred to as 'Planck'), and second, we combine with  a larger set of CMB data collectively referred to as `CMB' and including  {\it Planck}, WMAP polarization, ACT and SPT. Eventually, we further include  BAO constraints. The third to fifth columns in
Tab.~\ref{tab:fit_Lya_CMB} summarize the results for the three combinations. Fig.~\ref{fig:ContourFreqCombi} shows the 2D confidence level contours for  the $(\sigma_8 , n_s)$,  $(\sigma_8 , \sum m_\nu)$ and $(\Omega_m , \sum m_\nu)$  cosmological parameters. This figure demonstrates the overall compatibility (at a better than 2$\sigma$ level) between the Ly$\alpha$ and CMB data sets, under the assumption of a $\Lambda$CDM cosmology with massive neutrinos. It also leads to two interesting conclusions. First, according to the left panel in Fig.~\ref{fig:ContourFreqCombi}, there is some moderate tension (roughly at the  1.5$\sigma$ level) between the value of the spectral index inferred from the Ly$\alpha$+$H_0$ and CMB data. The results on $\sum m_\nu$, however, are not affected by this mild tension since the correlation between $\sum m_\nu$ and $n_s$ is only of order 20\%.
The other cosmological parameters are within one sigma of their preferred value from {\it Planck}. Second, the Ly$\alpha$ and CMB data sets probe different directions of degeneracy in the $(\sigma_8 , \sum m_\nu)$ plane. This explains why the bound on the total neutrino mass is much stronger when combining Ly$\alpha$ and CMB data than when using these data sets separately. Indeed, the upper limit we obtain on neutrino masses from Ly$\alpha$ alone is $\sum m_\nu < 1.1$~eV at 95\% CL, changing to $\sum m_\nu < 0.22$~eV  and $\sum m_\nu < 0.15$~eV when adding, respectively, the {\it Planck} and CMB data. The middle panel of Fig.~\ref{fig:ContourFreqCombi} is especially useful in understanding the origin of our upper limits. Increasing $\sum m_\nu$ suppresses the value of $\sigma_8(z=0)$ given the CMB normalization of the primordial fluctuation amplitude. However, this suppression only slightly varies between $z=4$ and $z=0$, so the Ly$\alpha$ forest measurement yields a $\sigma_8$ with minimal dependence on $\sum m_\nu$. The intersection of the contours is at $\sum m_\nu \simeq 0$. 

\begin{table}[htdp]

\caption{\it Best-fit value and frequentist confidence levels of the cosmological parameters of the model fitted to the flux power spectrum $P(k_i,z_j)$ measured with the BOSS Ly$\alpha$ data (presented in section~\ref{sec:Lya}), combined with several other data sets. In the third to fifth columns, we introduce the `Planck',`CMB' and 'BAO' data mentioned in section~\ref{sec:cmb} and \ref{sec:other}.  The last column features the WMAP 9-year data combined with high-$l$ ground-based experiments (ACT and SPT). For each parameter, we quote the 68\% confidence levels, except for the total neutrino mass, for which we quote the 95\% upper bound. We do not show the results for the 6 astrophysical and 14 nuisance parameters varied simultaneously with these 5 cosmological parameters. For the Ly$\alpha$ data, we include 10 redshift bins only ($z=[2.1-4.1]$), as in the last column of table~\ref{tab:fit_Lya}.}

\begin{center}

\begin{tabular}{lccccc}
\hline
Parameter &  Ly$\alpha$ +  $H_{0}^\mathrm{Gaussian}$  & Ly$\alpha$ + Planck &  Ly$\alpha$ + CMB &  Ly$\alpha$ + CMB & Ly$\alpha$ + WMAP9 \\
 & {\scriptsize ($H_0 = 67.4 \pm 1.4$)}  &  & & + BAO &+ ACT + SPT  \\
\hline \\[-10pt]
$n_s$ &  $0.928\pm0.012$  &  $0.958\pm0.006$ &  $0.954\pm0.005$ &  $0.954\pm0.005$  &  $0.950\pm0.007$ \\[5pt]
$H_0$~{\scriptsize(${\rm km~s^{-1}~Mpc^{-1}}$)}   & $67.2\pm1.4$ &   $67.9\pm1.0$  &   $68.0\pm1.0$  &   $67.8\pm0.5$ &   $67.8\pm1.1$ \\[5pt]
$\sum \! m_\nu$~{\scriptsize(eV)} &$<1.1$ {\scriptsize (95\%)} & $<0.22$ {\scriptsize (95\%)}
& $<0.15$ {\scriptsize (95\%)}  & $<0.14$ {\scriptsize (95\%)}   & $<0.31$ {\scriptsize (95\%)}   \\[5pt]
$\sigma_8$& $0.846\pm0.039$  & $0.822\pm0.018$ &  $0.832\pm0.009$ &  $0.837\pm0.011$  &  $0.789\pm0.025$  \\[5pt]
$\Omega_m$  &  $0.296\pm0.017$ &  $0.296\pm0.016$  &  $0.303\pm0.014$  &  $0.308\pm0.007$ &  $0.288\pm0.016$ \\[5pt]
\hline
\end{tabular}

\end{center}
\label{tab:fit_Lya_CMB}
\end{table}

We illustrate the $\chi^2$ profiles for two combinations of data sets  in Fig.~\ref{fig:Chi2Scan}. The minimum for  the Ly$\alpha$ + CMB configuration (blue curve) occurs for $\sum m_\nu<0$. The fact that the CMB data sets have their minimum in the unphysical (negative  $\sum m_\nu$) region was already  discussed in~\cite{PlanckCollaboration2014Freq}. In the results presented in Tab.~\ref{tab:fit_Lya_CMB},  the limit on the total neutrino mass is derived by computing the probability of $\Delta \chi^2 (\sum m_\nu) =\chi^2(\sum m_\nu) -\chi^2(\sum m_\nu=0)$ with one degree of freedom.   In~\cite{FC98},  Feldman and Cousins proposed an alternative and elegant solution  to provide confidence intervals when the $\chi^2$  minimum is not in the physical region. When  we apply their method, the limit at 95\% CL is 10\% smaller than obtained previously.  This cross-check  demonstrates the robustness of our coverage interval, which covers the upper limit given by the Feldman and Cousins  approach. Finally, as shown in   Fig.~\ref{fig:Chi2Scan} (red curve),  adding the BAO data forces the minimum  back into the physical ($\sum m_\nu>0$) region and we obtain a limit on the total neutrino mass of $\sum m_\nu < 0.14$~eV at 95\% CL (see fifth column of  Tab.~\ref{tab:fit_Lya_CMB}) .

For an insight on the impact of a change of the total neutrino mass on the 1D Ly$\alpha$ forest  power spectrum, we impose  $\sum m_\nu=0.5$~eV and  compute the best-fit value of all other cosmological, astrophysical and nuisance parameters. The $\chi^2$ changes by only $\sim 34$ for 350 data points, thus causing a change on the power spectrum at the level of one tenth of a sigma for each data point, undistinguishable by visual examination. The main effect on the parameters is a significant decrease of $\sigma_8$ from $0.832$ to $0.751$, as imposed mostly by the Planck likelihood degeneracy valley between $\sum m_\nu$ and $\sigma_8$ shown in the middle panel of figure~\ref{fig:ContourFreqCombi}.

\begin{figure}[htbp]
\begin{center}
\epsfig{figure= 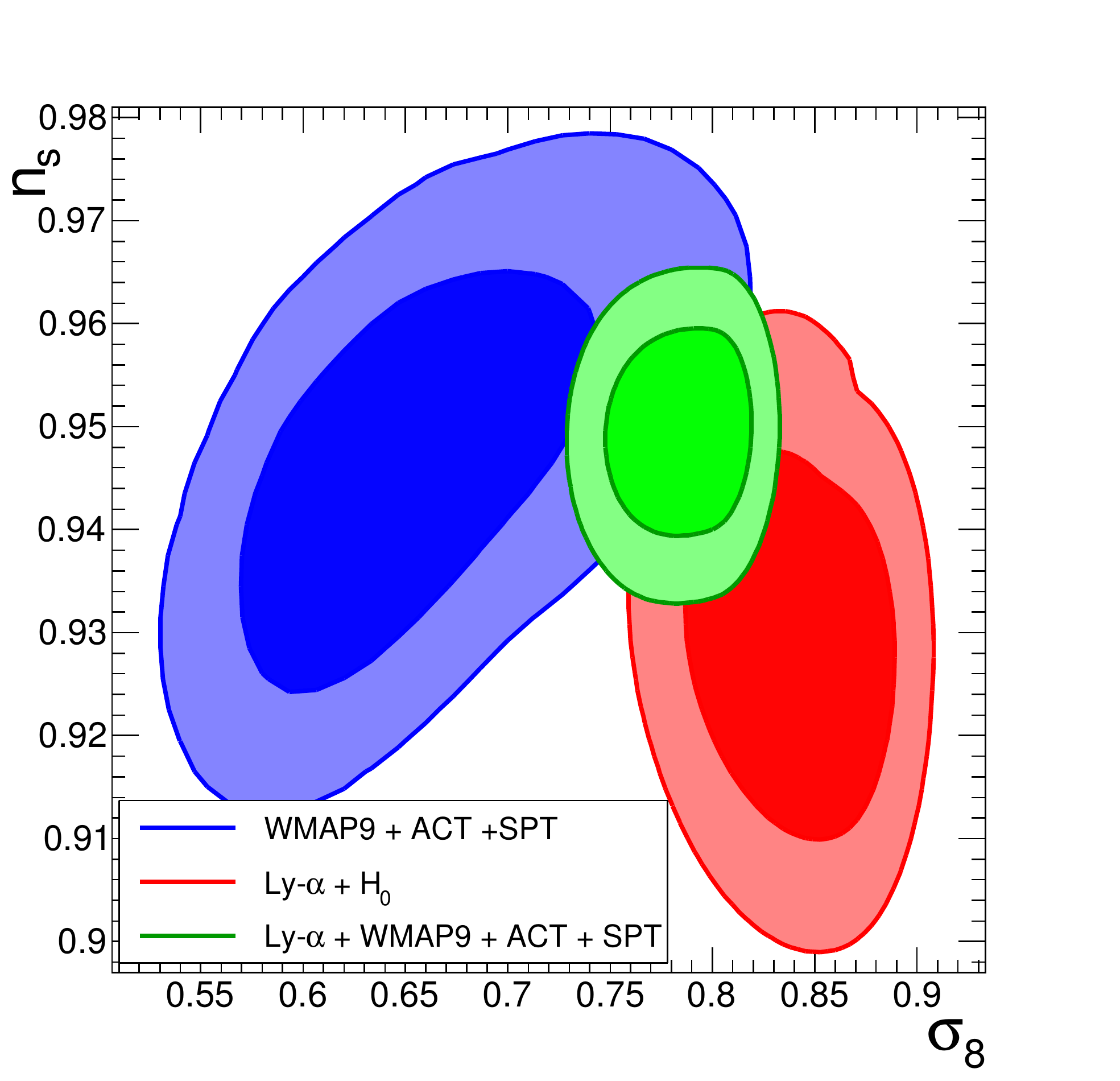,width = 5.0cm} \epsfig{figure= 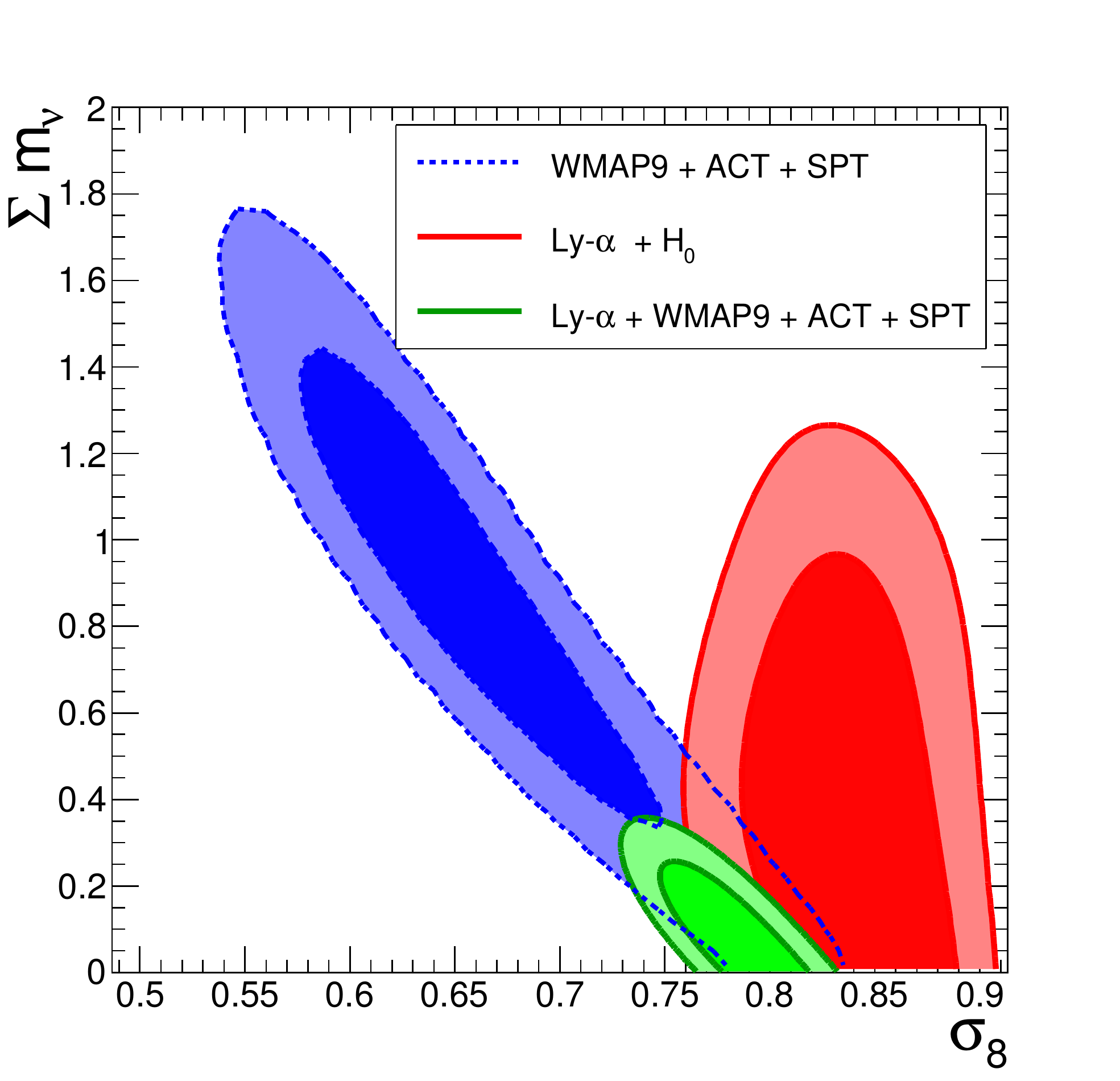,width = 5.0cm} \epsfig{figure= 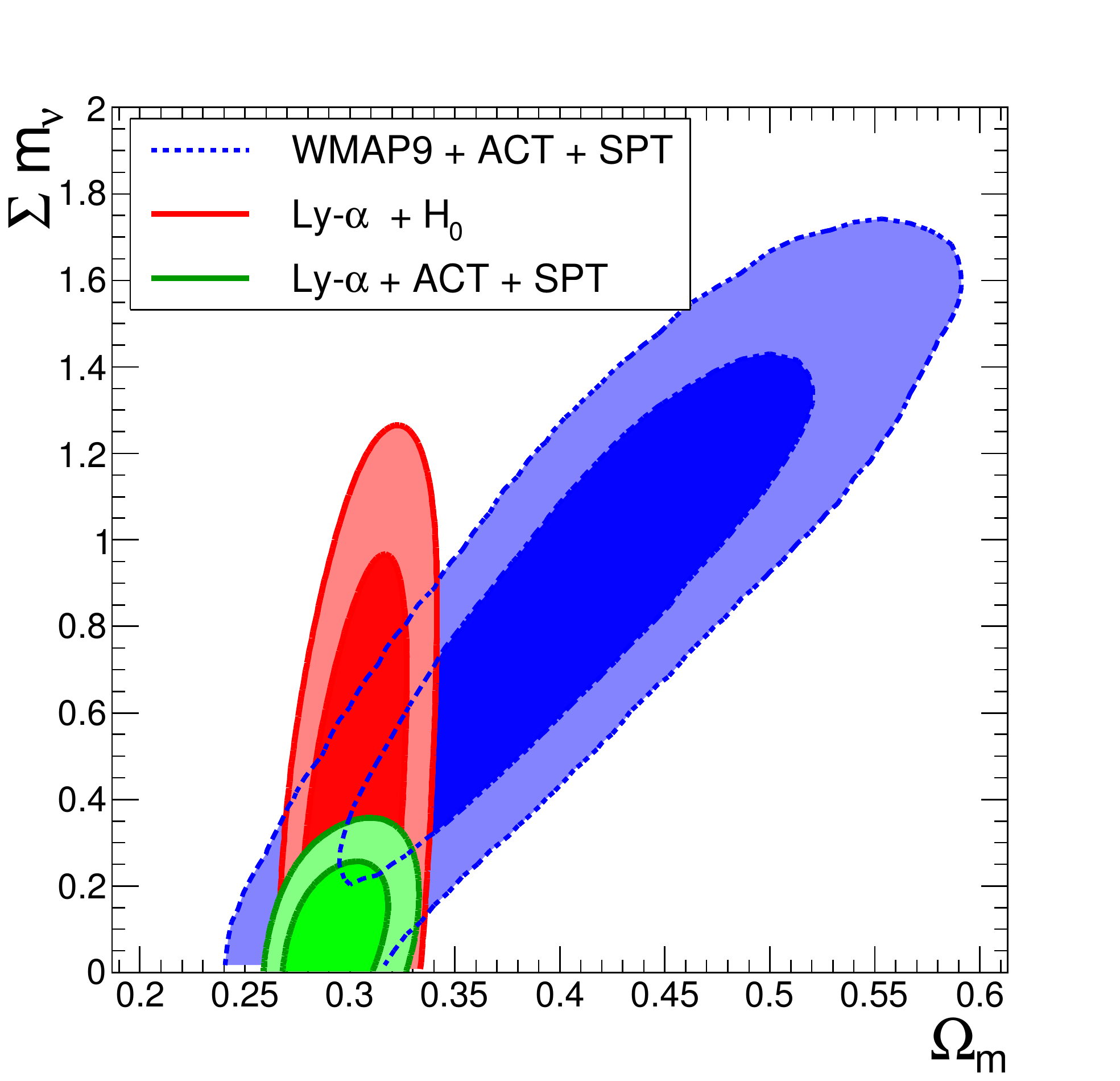,width = 5.0cm}
\caption{\it Same as figure~\ref{fig:ContourFreqCombi} except the CMB constraints  are produced  from an alternative data set (WMAP9 + ACT + SPT). }
\label{fig:ContourFreqCombiWMAP}
\end{center}
\end{figure}

We have checked the robustness of our results against different assumptions on known systematics that affect the 1D Ly$\alpha$ power spectrum. We have tested a possible systematic related to the initial conditions of our grid of simulations, i.e.,  the fact that  the simulations start at an initial redshift of $z=30$, rather than at  much higher redshift where non-linear effects are negligible. We modeled the small observed shift on the power spectrum as a  second-degree polynomial in redshift. Correcting or not the simulations for this systematic had no measurable impact on any of the cosmological parameters, even when considering Ly$\alpha$ data alone.

Finally, in order to evaluate the dependence of  our results  on Planck results as opposed to other recent CMB observations, we combine our Ly$\alpha$ data set with the results of WMAP 9-year, ACT and SPT.
The last column of Tab.~\ref{tab:fit_Lya_CMB} and the contours in Fig.~\ref{fig:ContourFreqCombiWMAP} summarize our results in that case. The Planck-based and WMAP-based results are all compatible at the 1$\sigma$ level, although this is only marginally true for $\sigma_8$. The WMAP-based data set prefers a smaller $\sigma_8$ and remains compatible with larger values of the neutrino mass. The 95\%CL bound in that case is $\sum m_\nu < 0.31~eV$ (95\% CL). A critical discussion of the difference between these two CMB data sets can be found in the appendix of Ref.~\cite{PlanckCollaboration2013}.

\vspace{0.2cm}

All these results can now be compared with those of a Bayesian analysis performed with the full CMB likelihoods.

\subsubsection{ Bayesian approach}\label{sec:combinedbayesian}

The goal of this section is to derive Bayesian results for the same models and data sets, using the full Ly$\alpha$, CMB and BAO likelihoods, and marginalizing the results explicitly over all nuisance parameters.

\begin{figure}[htbp]
\begin{center}
\epsfig{figure= 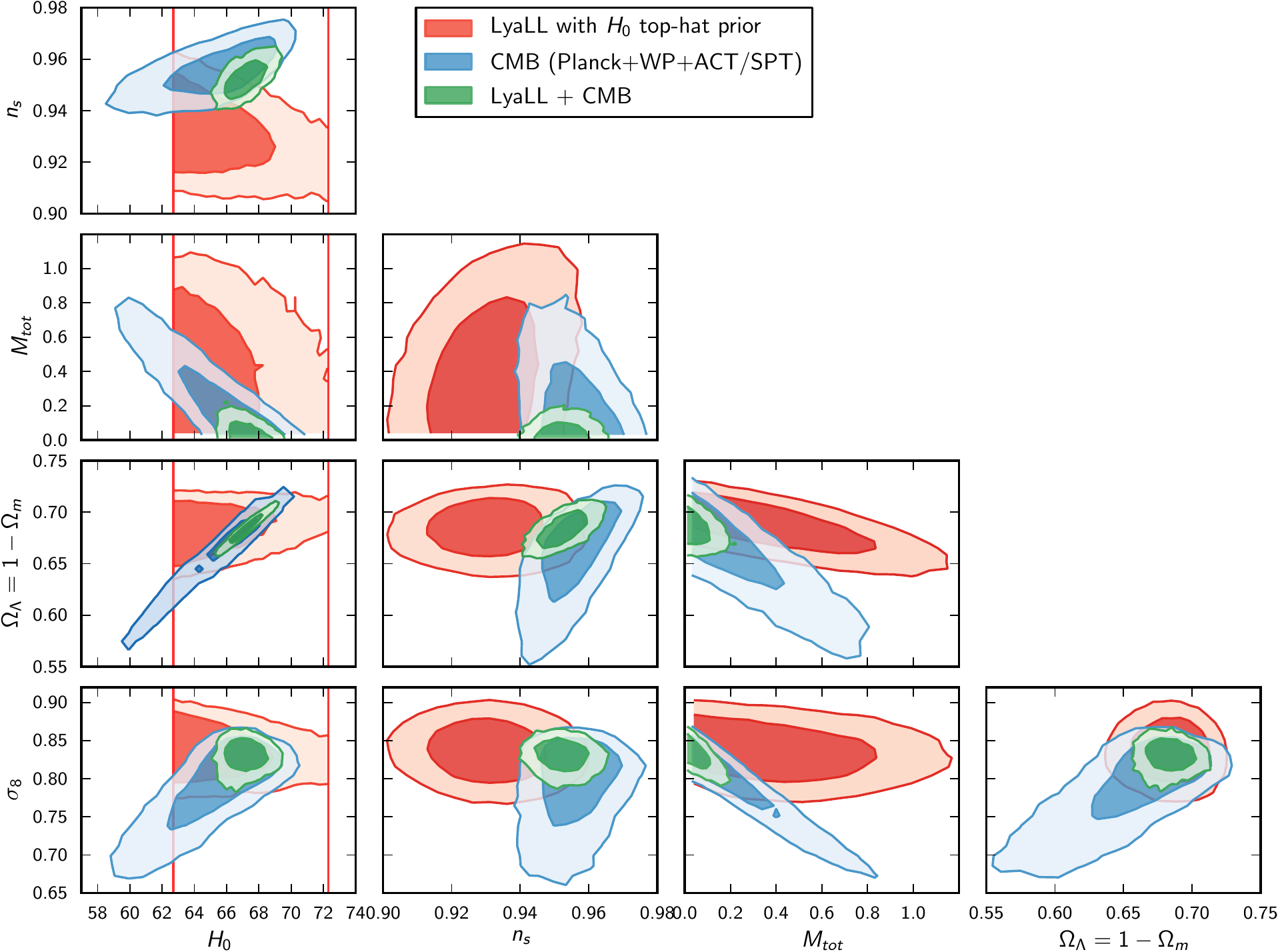,width = 16.0cm}
\caption{Joint Bayesian confidence levels for all pairs of cosmological parameters probed simultaneously by Ly$\alpha$ and CMB data. The contours are obtained with different combinations of the BOSS Ly$\alpha$ data presented in section~\ref{sec:Lya} using the 10 redshift bins $z=[2.1-4.1]$, and of the `CMB’ data set (Planck + WP + ACT + SPT).}
\label{fig:ContourBayesian}
\end{center}
\end{figure}

\begin{figure}[htbp]
\begin{center}
\epsfig{figure= 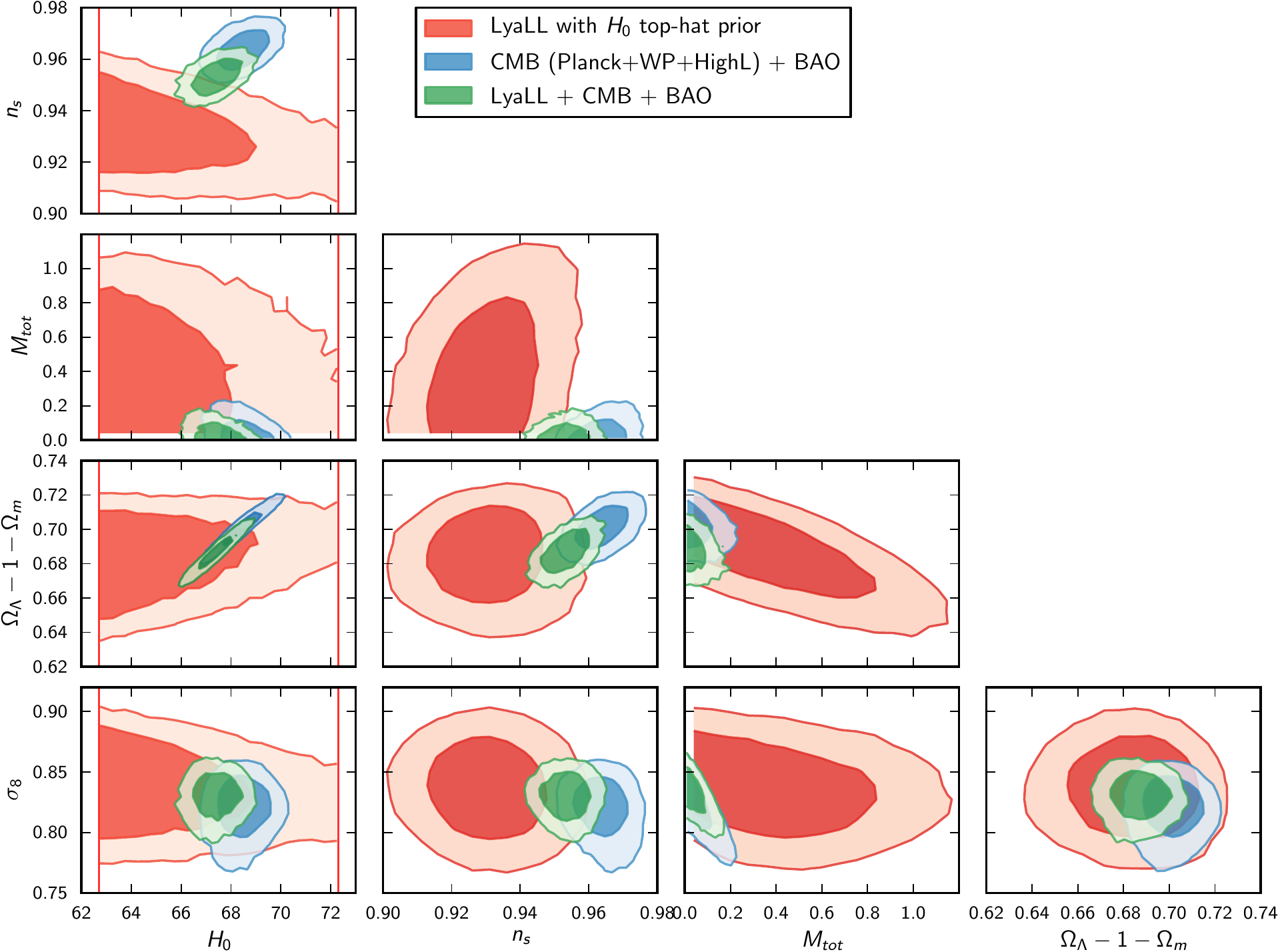,width = 16.0cm}
\caption{Joint Bayesian confidence levels for all pairs of cosmological parameters probed simultaneously by Ly$\alpha$, CMB and BAO data. The contours are obtained with different combinations of the BOSS Ly$\alpha$ data presented in section~\ref{sec:Lya} using the 10 redshift bins $z=[2.1-4.1]$, andof the `CMB+BAO' data set (Planck + WP + ACT + SPT +BAO).
}
\label{fig:ContourBayesianBAO}
\end{center}
\end{figure}

In order to use the actual CMB and BAO likelihoods, we must employ a Boltzmann solver to compute cosmological distances and CMB anisotropy spectra for each set of cosmological parameters. For this purpose, we use the CLASS\footnote{\tt http://class-code.net} solver~\cite{Lesgourgues:2011re,Blas:2011rf}, embedded in the {\sc Monte Python} code. All fixed cosmological parameters are set to the same default values as in the Planck analysis \cite{PlanckCollaboration2013}. In principle, it would be feasible to perform the MCMC exploration using flat priors on the cosmological parameters used in the Ly$\alpha$ likelihood, including $\Omega_\textrm{m}$ and $\sigma_8$. However, it is more conventional to use the primordial spectrum amplitude parameter $A_s$ (define at the pivot scale $k=0.05$~Mpc$^{-1}$) and the baryon and cold dark matter density parameters $\omega_\textrm{b}$, $\omega_\textrm{cdm}$, while treating ($\Omega_\textrm{m}$, $\sigma_8$) as derived parameters.  In the results reported in this section, flat priors have been assumed on $A_s$ and $\omega_m$. This means that $\Omega_\textrm{m}$ and $\sigma_8$ therefore have non-flat priors. However, this choice has a minor impact, as already checked  in section~\ref{sec:resLya_B}.
Our results are summarized in table~\ref{tab:fit_Lya_CMB_B} for one-dimensional credible intervals, and figs.~\ref{fig:ContourBayesian}, \ref{fig:ContourBayesianBAO} for two-dimensional confidence contours with or without BAO data.

For the Ly$\alpha$+CMB combined data sets, the frequentist and Bayesian results can be compared through the fourth column of Tab.~\ref{tab:fit_Lya_CMB} and third column of Tab.~\ref{tab:fit_Lya_CMB_B} (both labelled `Ly$\alpha$ + CMB').
Remarkably,  the constraints obtained with the frequentist and the Bayesian techniques are  similar  on all fit parameters.  For Ly$\alpha$+CMB,  both methods lead to  strong bounds on the total neutrino mass, $\sum m_\nu <0.15$ (95\% C.L.) and $\sum m_\nu <0.16$ (95\% C.L.) for the frequentist and the Bayesian approaches, respectively.
According to Fig.~\ref{fig:ContourBayesian}, this tight constraint (much stronger than from CMB or Ly$\alpha$ data taken separately) can be explained by the different degeneracy directions in the $(\sigma_8 , \sum m_\nu)$ and $(\Omega_m , \sum m_\nu)$ planes (as already described in the frequentist analysis). The fact that the values of $n_s$ preferred by the Ly$\alpha$+$H_0$ data and the CMB are in moderate tension (at about one sigma) is also confirmed, but  Fig.~\ref{fig:ContourBayesian} suggests that this situation is not contributing to the strong bound on neutrino masses, since in the $(n_s , \sum m_\nu)$ plane, the $n_s$ tension is roughly the same for small neutrino mass ($\sum m_\nu \sim 0.1$~eV) and large neutrino mass ($\sum m_\nu \sim 0.3$~eV).

\begin{table}[htdp]
\caption{\it Bayesian mean value and credible intervals of the cosmological parameters of the model fitted to the flux power spectrum $P(k_i,z_j)$ measured with the BOSS Ly$\alpha$ data (presented in section~\ref{sec:Lya}), combined with the `CMB’ and `BAO’ data sets (mentioned in section~\ref{sec:cmb}, \ref{sec:other}). For each parameter, we quote either the 68\% confidence levels or the 95\% upper bound. We assume flat priors on the seven independent cosmological parameters shown in the first rows. The last rows show the corresponding bounds for the derived parameters ($\Omega_\textrm{m}$, $\sigma_8$) with non-flat priors. We do not show the results for the 6 astrophysical and 13 nuisance parameters varied simultaneously with these 7 cosmological parameters. For the Ly$\alpha$ data, we include 10 redshift bins only ($z=[2.1-4.1]$). In the last column, we let the CMB lensing amplitude $A_L$ vary as a free parameter.}
\begin{center}
\begin{tabular}{lcccc}
\hline
Parameter &  Ly$\alpha$ + $H_{0}^\mathrm{tophat}$ & Ly$\alpha$ + CMB & Ly$\alpha$ + CMB & Ly$\alpha$ + CMB($A_L$) \\
& {\scriptsize ($62.5 \leq H_0 < 72.5$)}&&+ BAO&\\
\hline \\[-10pt]
$10^9 A_s$              & $3.2_{-0.7}^{+0.5}$           & $2.20_{-0.06}^{+0.05}$          & $2.20_{-0.06}^{+0.05}$      & $2.18_{-0.06}^{+0.05}$ \\[5pt]
$10^2 \omega_\mathrm{b}$ &{\scriptsize (fixed to $2.22$)}& $2.20 \pm 0.02$             & $2.20 \pm 0.02$            & $2.22 \pm 0.03$ \\[5pt]
$\omega_\mathrm{cdm}$    & $0.110_{-0.013}^{+0.008}$    & $0.1200_{-0.0018}^{+0.0019}$     & $0.1196_{-0.0014}^{+0.0015}$ & $0.1191 \pm 0.002$ \\[5pt]
$\tau_\mathrm{reio}$     & {\scriptsize (irrelevant)}   & $0.091_{-0.013}^{+0.012}$      & $0.091_{-0.013}^{+0.011}$    & $0.0871_{-0.013}^{+0.012}$ \\[5pt]
$n_s$                   & $0.931\pm 0.012$              & $0.953 \pm 0.005$            & $0.953 \pm 0.005$         & $0.955_{-0.006}^{+0.005}$ \\[5pt]
$H_0$                   & $<70.9$ {\scriptsize (95\%)}  & $67.2_{-0.9}^{+0.8}$          & $67.4\pm 0.7$              & $67.5_{-1.1}^{+1.0}$ \\[5pt]
$\sum \! m_\nu$~(eV)    & $<0.98$ {\scriptsize (95\%)}  & $<0.16$ {\scriptsize (95\%)} & $<0.14$ {\scriptsize (95\%)} & $<0.21$ {\scriptsize (95\%)}\\[5pt]
$A_L$ & {\scriptsize (fixed to $1$)} & {\scriptsize (fixed to $1$)} & {\scriptsize (fixed to $1$)}                 & $1.12 \pm 0.10$ \\[5pt]
\hline \\[-10pt]
$\sigma_8$              & $0.84\pm 0.03$                & $0.830_{-0.013}^{+0.017}$      & $0.830_{-0.012}^{+0.016}$   & $0.818_{-0.014}^{+0.021}$ \\[5pt]
$\Omega_\mathrm{m}$      & $0.316_{-0.021}^{+0.018}$    & $0.316 \pm 0.012$               & $0.313 \pm 0.009$         & $0.312 \pm 0.013$ \\[5pt]
\hline
\end{tabular}
\end{center}
\label{tab:fit_Lya_CMB_B}
\end{table}

Including BAO data produces a much better measurement of the background parameters $\Omega_\mathrm{m}=1-\Omega_\Lambda$ and $H_0$ than with CMB data alone. The CMB likelihood correlates strongly these background parameters with the other parameters ($n_s$, $\sigma_8$, $\sum m_\nu$) used in our analysis. As a consequence, the CMB+BAO data set is more constraining on each of these five parameters than the CMB data alone~\cite{PlanckCollaboration2013}. Figure~\ref{fig:ContourBayesianBAO} demonstrates that this tighter constraint goes in the direction of increasing the tension on $n_s$ between the Ly$\alpha$+$H_0$ and external data sets, roughly at the 2$\sigma$ level, suggesting that we should be extremely careful when combining these data sets. Assuming that all systematic errors have been properly modeled in each data set, there remains the possibility that we are not fitting the correct model to the data. In the presence of BAO data, the $n_s$ tension impacts the result for $\sum m_\nu$:  figure~\ref{fig:ContourBayesianBAO} shows that the tension in the $(n_s , \sum m_\nu)$ plane is minimal for the smallest possible values of $\sum m_\nu$. However, this impact is small: the upper bound only sharpens from 0.16~eV to 0.14~eV (95\% C.L.). Other bounds produced by the inclusion of BAO data can be found in the fourth column of Tab.~\ref{tab:fit_Lya_CMB_B}.

The Planck result paper~\cite{PlanckCollaboration2013} presents a discussion of the  degeneracies between the parameter $A_L$ and the mass of neutrinos. By definition, the (non-physical) parameter $A_L$ sets the amplitude of the CMB lensing spectrum, with $A_L=1$ corresponding to the amplitude predicted by the underlying cosmological model (here, $\Lambda$CDM with massive neutrinos). The Planck collaboration reports that the data are compatible with rather large values of $A_L$, and even slightly favors such values. For the $\Lambda$CDM model with massive neutrinos, the CMB (Planck+WP+ACT+SPT) data produces $A_L=1.31^{+0.12}_{-0.14}$ (68\% CL)~\cite{PlanckExplanatorySupplement2013}. Due to non-trivial degeneracies in parameter space, the CMB bound on $\sum m_\nu$ is significantly weaker when $A_L$ is left free. There have been extensive discussions about the fact that the marginal preference for $A_L>1$ could be a manifestation of some unknown or underestimated systematics, and that one should consider the more conservative neutrino mass bound obtained with a free $A_L$.

In previous analyses presented in this paper, we always assumed $A_L=1$. We now allow this parameter to vary in the fit,  and study its impact on our results (through the correlation between $A_L$, $\sigma_8$ and $\sum m_\nu$ in the CMB likelihood). Interestingly, the preferred range is $A_L=1.12\pm 0.10$ (68\%C.L.),  well compatible with 1. The Ly$\alpha$ data prefer the highest values of $\sigma_8$ compatible with  the CMB data set. Hence, they favor models with a strong lensing effect in the CMB power spectrum, removing any need for departing from standard cosmology (i.e., from $A_L=1$). A consequence of finding $A_L$ fully compatible with 1 is that the preferred values of all cosmological parameters remain  stable when $A_L$ is left free. In particular, the upper bound on $\sum m_\nu$ only increases marginally, from 0.16 to 0.21~eV (95\%C.L.).

\subsubsection{Comparison with previous results}

\begin{figure}[htb]
\begin{center}
\epsfig{figure= 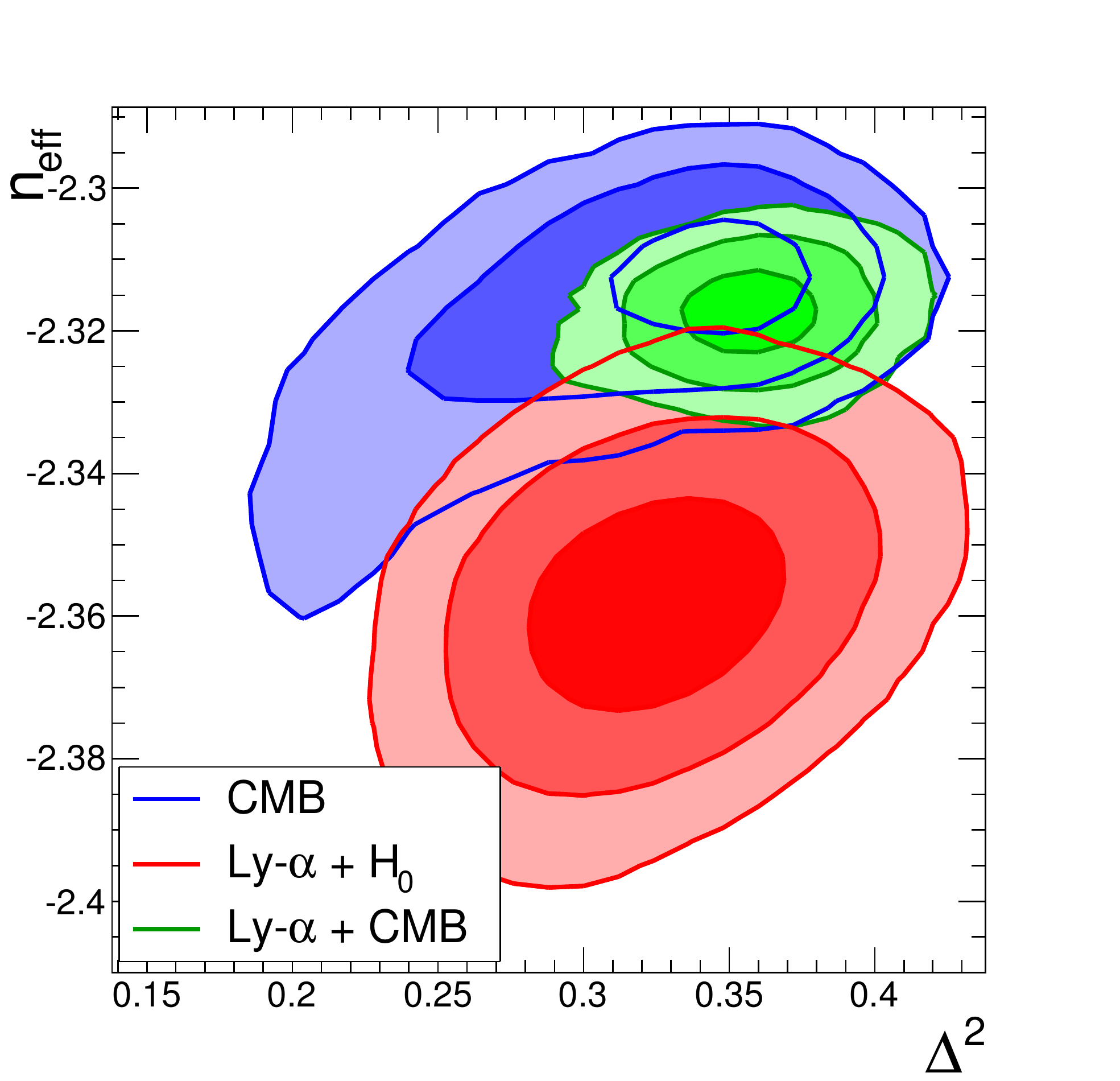,width = .8\textwidth} 
\caption{\it  Constraints on the effective slope $n_{\rm eff}$ and amplitude $\Delta^2_L$ of the linear power, measured at
$k_p=0.009 \,({\rm km/s})^{-1}$ and $z_p=3$ from Ly$\alpha$ data alone (red), or from CMB  data alone (blue contours). CMB constraints include the prior $\sum m_{\nu}>0$, which prevents the contours from entering into the higher $\Delta^2$ region, and explains their  asymmetry. The green contours are the constraints from the combined CMB+Lyman-$\alpha$ data sets.}
\label{fig:anze}
\end{center}
\end{figure}

A different parameterization has been used in some previously published results (e.g.~\cite{McDonald2005}): the dimensionless amplitude $\Delta^2_L(k,z)\equiv k^3 P_L(k,z)/2\pi^2 $ and the logarithmic slope $n_{\rm eff}(k,z) \equiv d\ln P_L/d\ln k$ of the linear power spectrum $P_L$, both evaluated at a pivot redshift $z_p$ and pivot wavenumber $k_p$. In figure~\ref{fig:anze}, we illustrate our constraints using this parameterization, with $k_p=0.009 \,{\rm s~km^{-1}}$ and $z_p=3$, a central position in the medium-resolution SDSS Lyman-$\alpha$ data. The conversion between wavenumbers expressed in ${\rm s~km^{-1}}$ and in $h~{\rm Mpc}^{-1}$ is redshift dependent and given by the factor $H(z)/(1+z)$. At redshift $z=3$ and for the central cosmology used in our simulation grid, it equals 113~$\rm km~s^{-1}~{Mpc}^{-1}$.

Our results correspond to an amplitude  $\Delta^2_L = 0.32\pm 0.03$ and an effective slope $n_{\rm eff} = -2.36 \pm 0.01$, in  agreement within 1.5~$\sigma$ with previous results from \cite{Viel2005}and \cite{McDonald2005} (we measure a slightly lower amplitude and steeper slope).  The Ly$\alpha$ results are illustrated as the red contours in Fig.~\ref{fig:anze}.  We also illustrate (blue contours) the constraints derived from the {CMB} chains on the same parameters. 
The amplitudes from {\rm Planck} and Ly$\alpha$ are consistent within  less than $1\sigma$, and the effective slopes are about $2\sigma$ apart, as expected from the similar tension observed for $n_s$ (cf. sections~\ref{sec:FreqCombined} and \ref{sec:combinedbayesian}). 
The combined CMB+Ly$\alpha$ best-fit region is illustrated in green. Recall that the CMB constraints include a $\sum m_\nu>0$ prior, preventing a combination ``by eye'' of the posterior contours shown in the figure. The combined constraint is obtained by proper addition of the CMB and Ly$\alpha$ likelihoods, on which the $\sum m_\nu>0$ prior is then applied. 

\subsubsection{Conclusions on combined constraints}

In conclusion of section~\ref{sec:combined},   the different statistical methods used in this analysis, as well as  the various  data combinations we considered,  all yield a bound of the order of $\sum m_\nu <0.15$ (95\% C.L.). The bound depends of course on the assumed underlying cosmology (flat $\Lambda$CDM). We leave the investigation of extended models for future work. 

Including additional cosmological ingredients could be motivated by the  tension that is observed on $n_s$ between Ly$\alpha$ and CMB+BAO data. However, one can check in figure 21 of the Planck result paper \cite{PlanckCollaboration2013} that it is not easy to accommodate smaller values of $n_s$ with CMB+BAO data, at least for the most obvious extensions of the $\Lambda$CDM model. Promoting $N_\mathrm{eff}$ or $Y_P$ as free parameters results in accommodating {\it larger} values. Introducing spatial curvature or a tensor-to-scalar ratio has a negligible impact on $n_s$ bounds. However, models with phantom dark energy ($w<-1$) or negative running $[d \, n_s / d \ln k]$ could  reconcile the different data sets and  deserve further investigation. 
Moreover, to reconcile large simulation volumes with good mass resolutions, we used a splicing technique in our simulation procedure, which affects the value of $n_s$. We have modeled the $k$-mode dependence due to splicing approximation with a simple linear fit, but intend to refine this correction in the  future with more detailed simulations.  We will also measure the splicing correction using a large simulation with higher resolution than what we currently have available. We thus expect to improve our knowledge of the impact of the splicing procedure on the power spectrum, and thus to better model it. For instance,  removing the splicing correction  increases the value of $n_s$ to 0.935 using Ly$\alpha$ data only in the configuration of the last column of table~\ref{tab:fit_Lya}. This compares to the value $n_s=0.928$ that is currently obtained when letting $\alpha_{splicing}$ free to vary. Removing this correction would thus reduce the tension on $n_s$, with negligible impact, however, on our constraint on $\sum m_\nu$ because of the weak correlation between   $n_s$ and  $\sum m_\nu$. A quantitative study of the systematic effect of the splicing is discussed in section~\ref{sec:systsimus}.


\section{Discussion}
\label{sec:discussion}
\subsection{Sensitivity to simulation parameters and interpretation} 
\label{sec:sensitivity}

The central element of our neutrino mass
constraints is our simulation-based
Taylor-expansion model for the dependence of the
\lya\ flux power spectrum on cosmological and
astrophysical parameters.  Here, we
illustrate the response of the $P(k)$ predictions
to isolated variations in individual parameters 
from the best-fit \lya+CMB model, at levels close
to the parameter errors derived from this joint
fit.  We show results at redshifts $z=2.2$, 3.0,
and 4.0 as representative examples.

For reference, the
15th-percentile (5$^{\rm th}$ lowest value out of 35, with one power spectrum value for each $k$-mode), median, and
85th-percentile (30$^{\rm th}$ lowest out of 35) fractional errors per
$k$-bin in the measured $P(k)$ are 2\%, 4\%, and
12\% at $z=2.2$, 2\%, 2\%, and 6\% at $z=3.0$,
and 9\%, 11\%, and 15\% at $z=4.0$ (see detail in table~\ref{tab:distrib_z}), and that
off-diagonal covariances are substantially smaller
than diagonal errors (see the plots and tables of
\cite{Palanque-Delabrouille2013}).  Thus, coherent
percent-level changes in $P(k)$ across multiple
$k$ bins and redshift slices are detectable at
enormous statistical significance; the
limiting factor in cosmological constraints is the
ability to break degeneracies among the effects of
different parameters.

\begin{figure}[htp]
\begin{center}

\epsfig{figure= 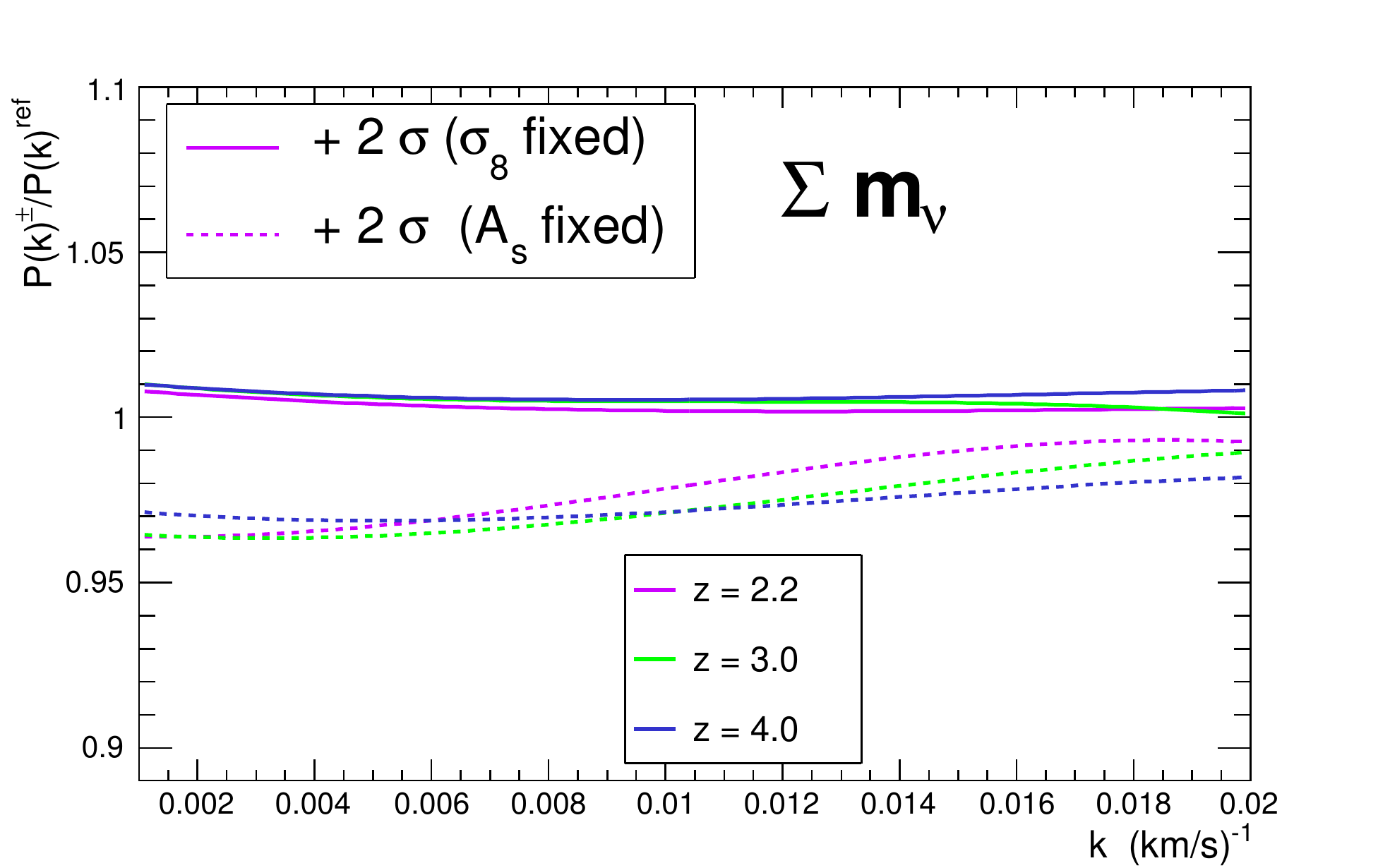,width = 8cm} \\
\epsfig{figure= 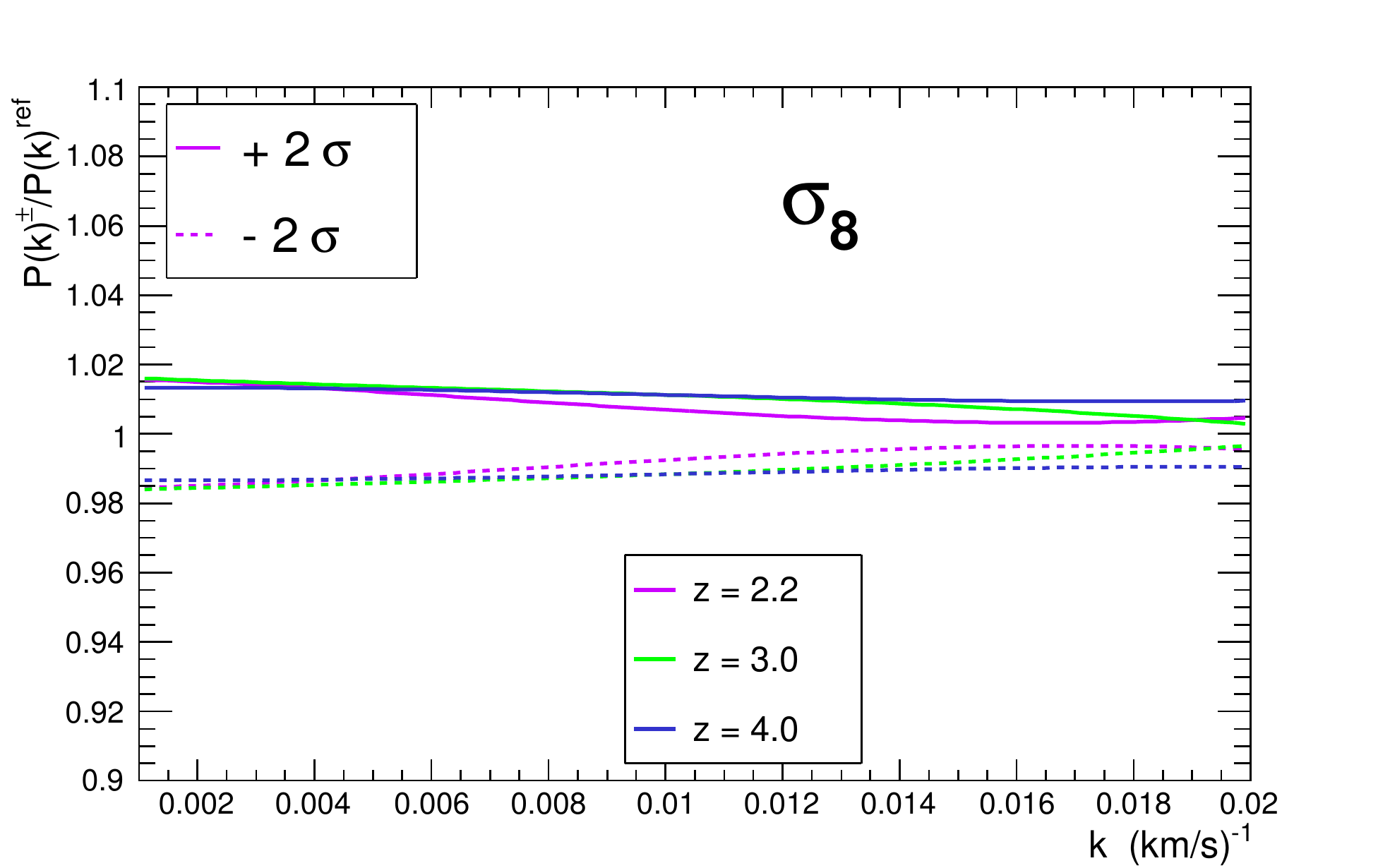,width = 8cm}\hspace{-.8cm}
\epsfig{figure= 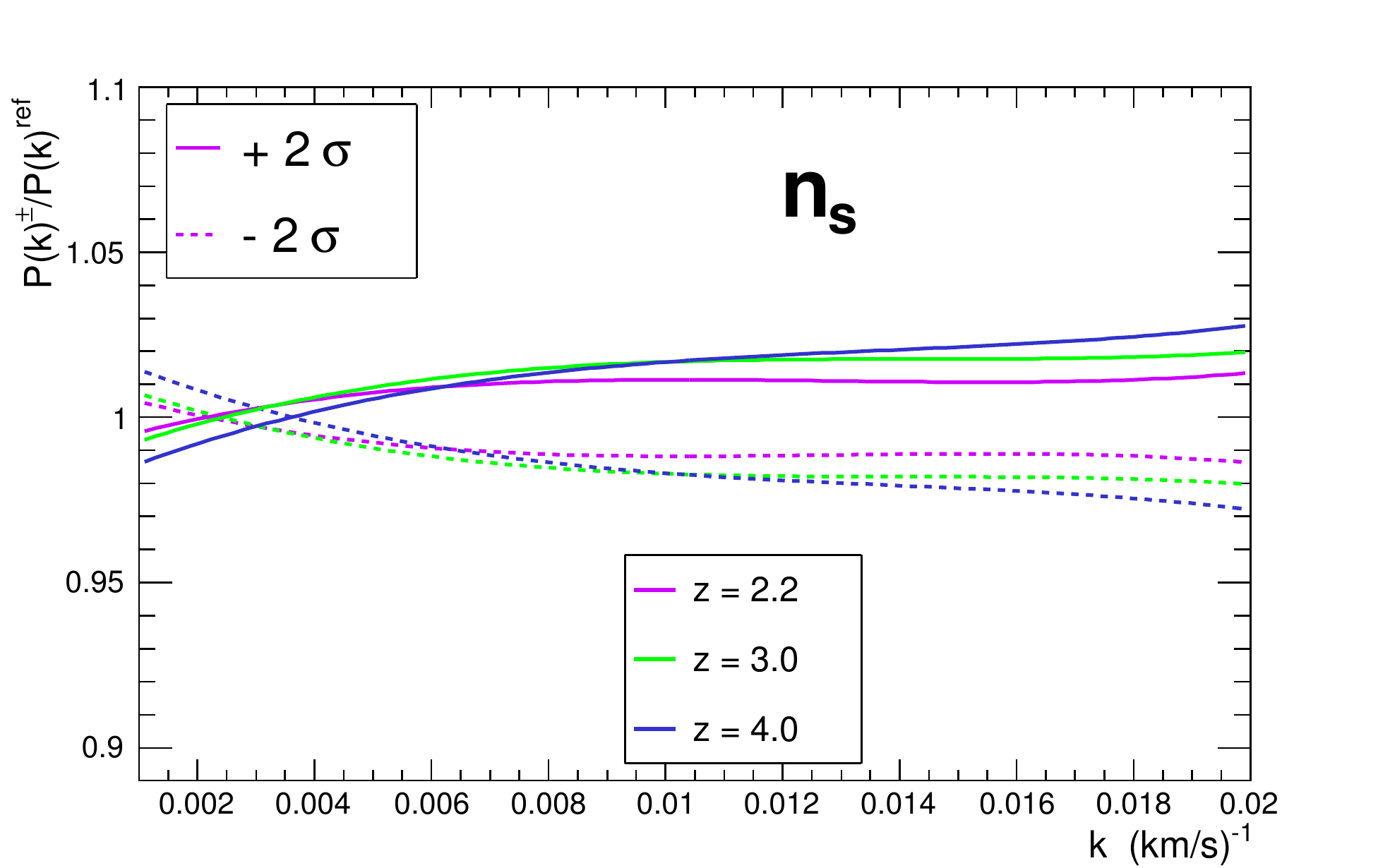,width = 8cm} 
\epsfig{figure= 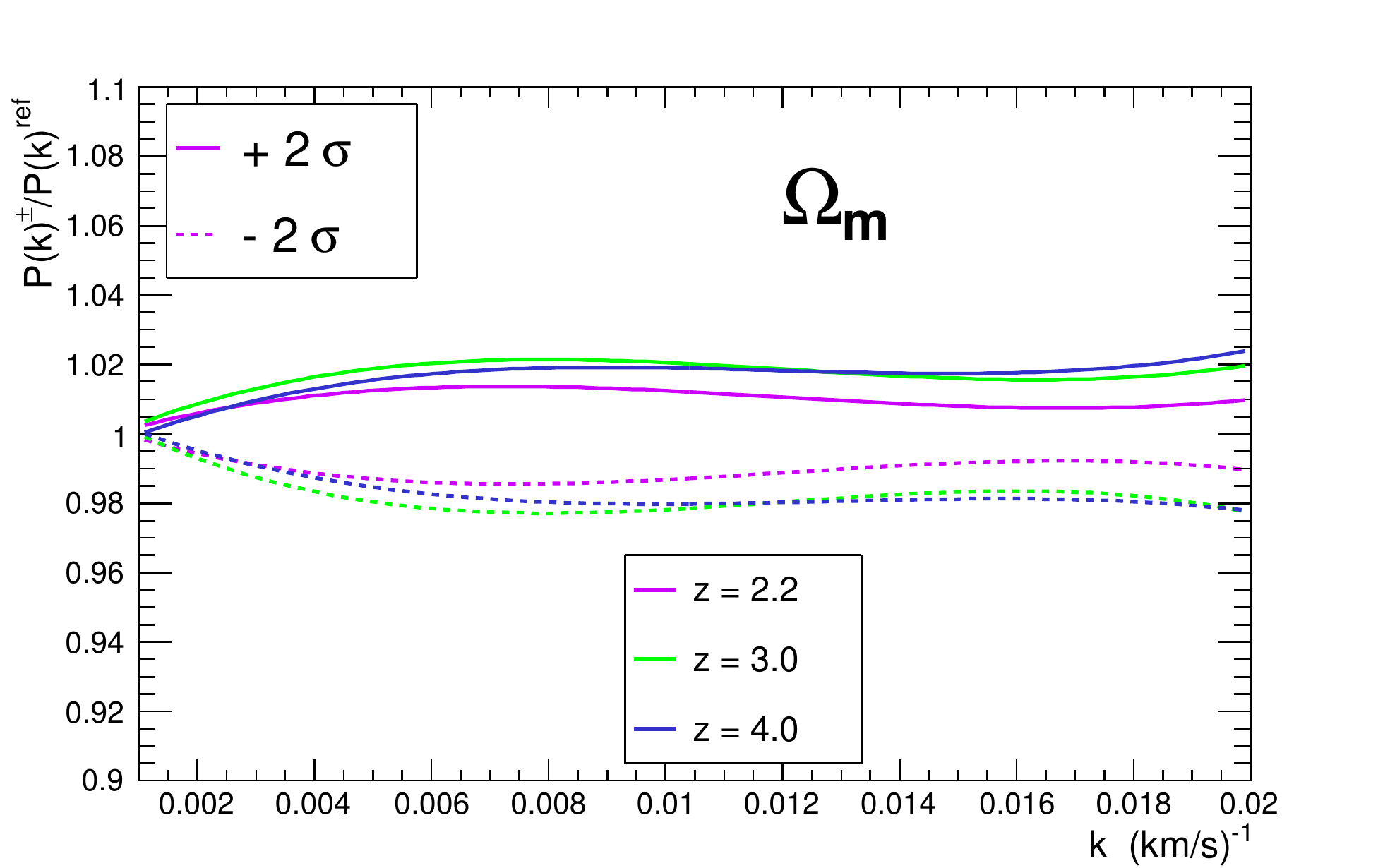,width = 8cm}\hspace{-.8cm}
\epsfig{figure= 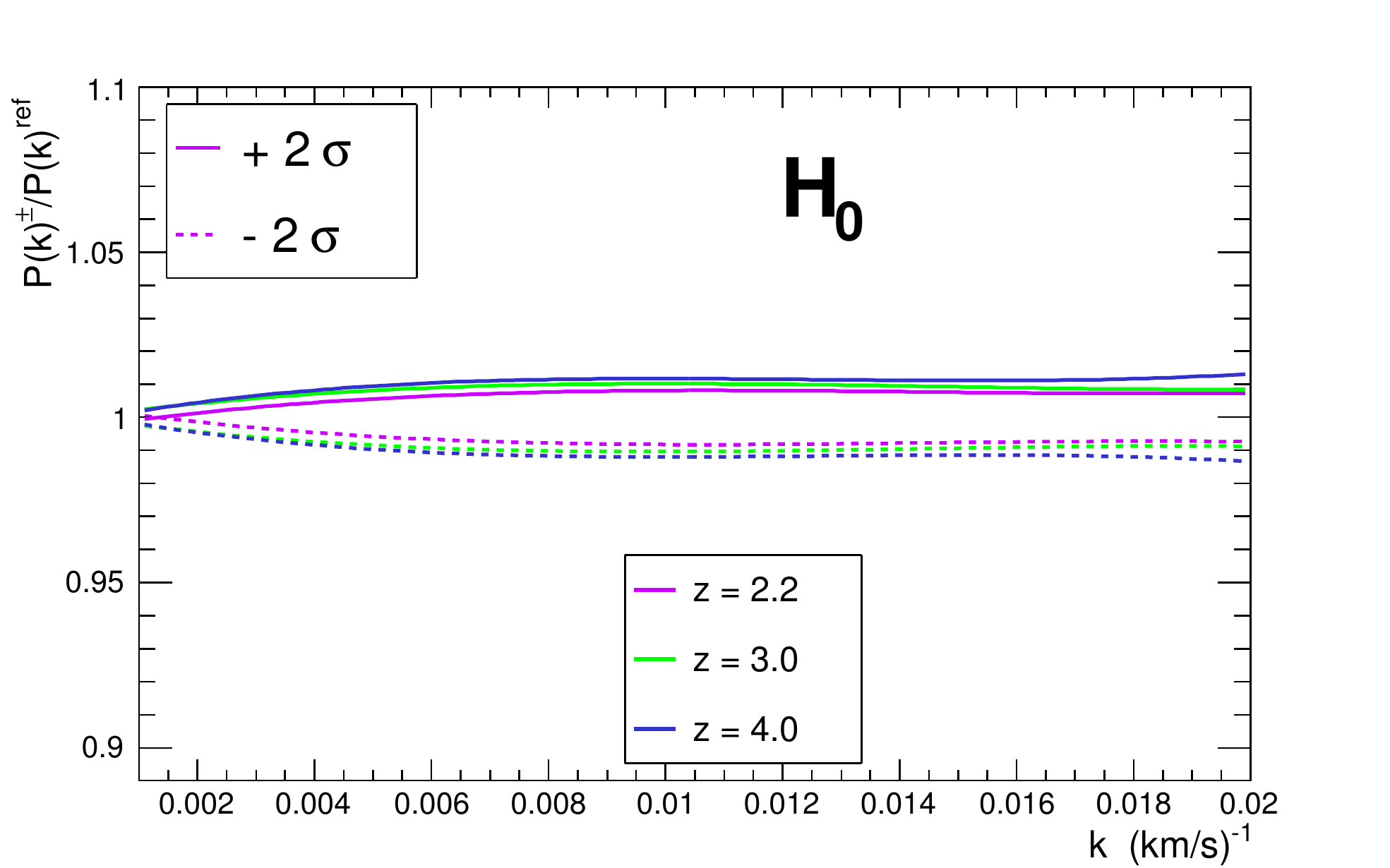,width = 8cm}
\caption{\it Sensitivity diagrams illustrating the change in the power spectrum for a $\pm 2\sigma$ variation of each parameter around its best-fit value in the Ly$\alpha$ $+$ CMB configuration,  for the cosmological parameters used in the fit. Curves are third-degree polynomial fits, for better visibility. }

\label{fig:sensitivityplotsCosmo}
\end{center}
\end{figure}

\begin{figure}[htp]
\begin{center}

\epsfig{figure= 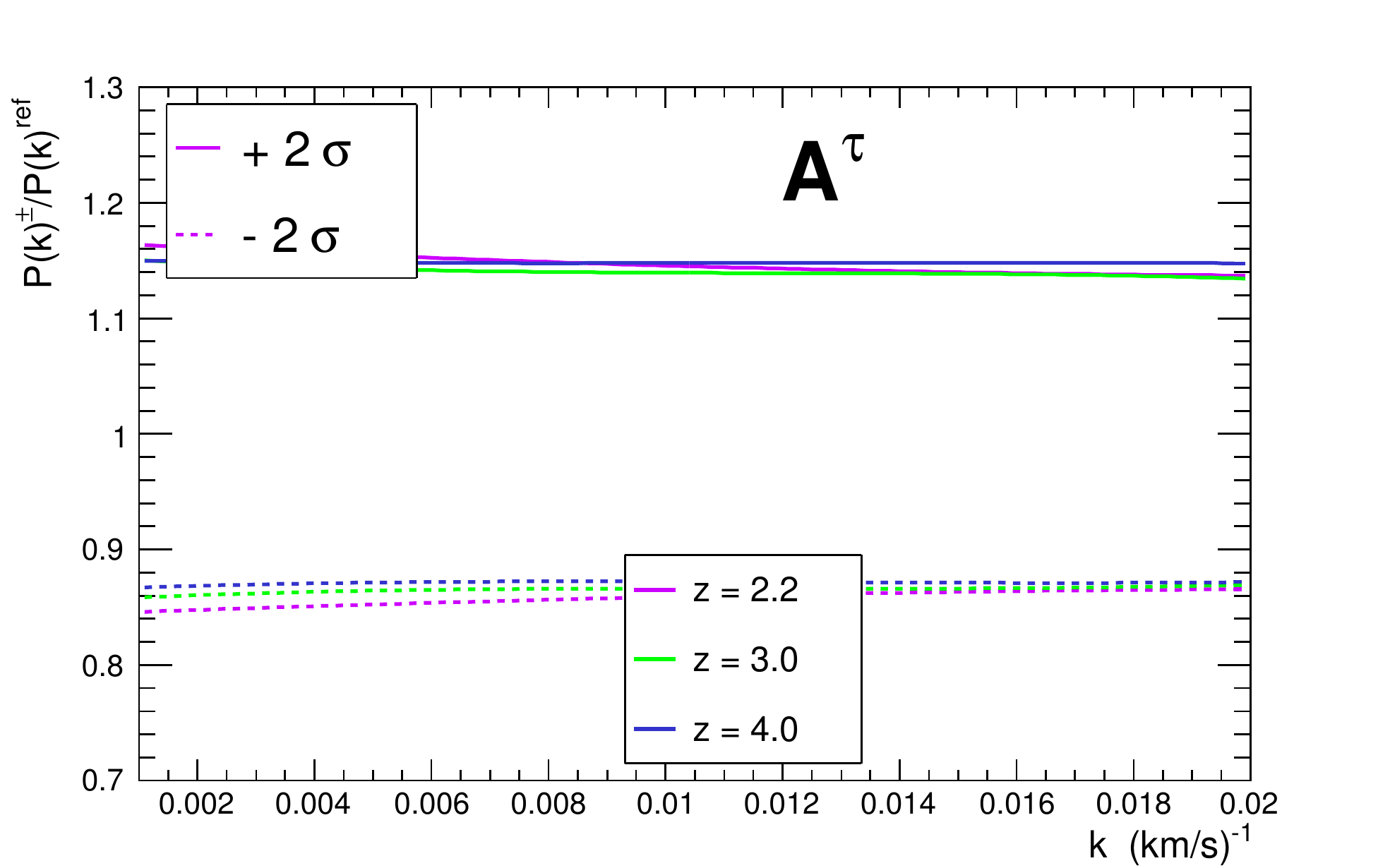,width = 8cm}\hspace{-.8cm}
\epsfig{figure= 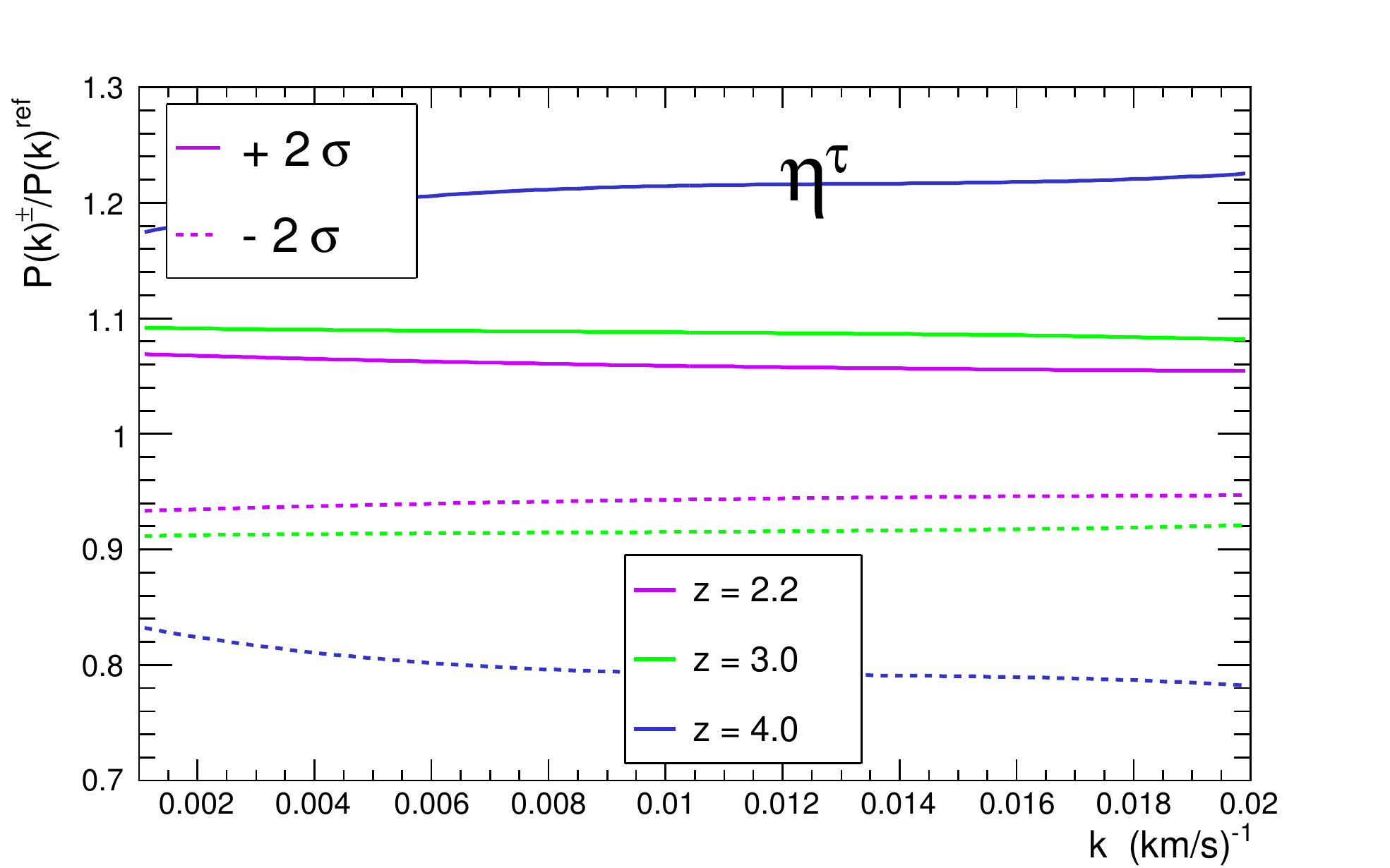,width = 8cm}
\epsfig{figure= 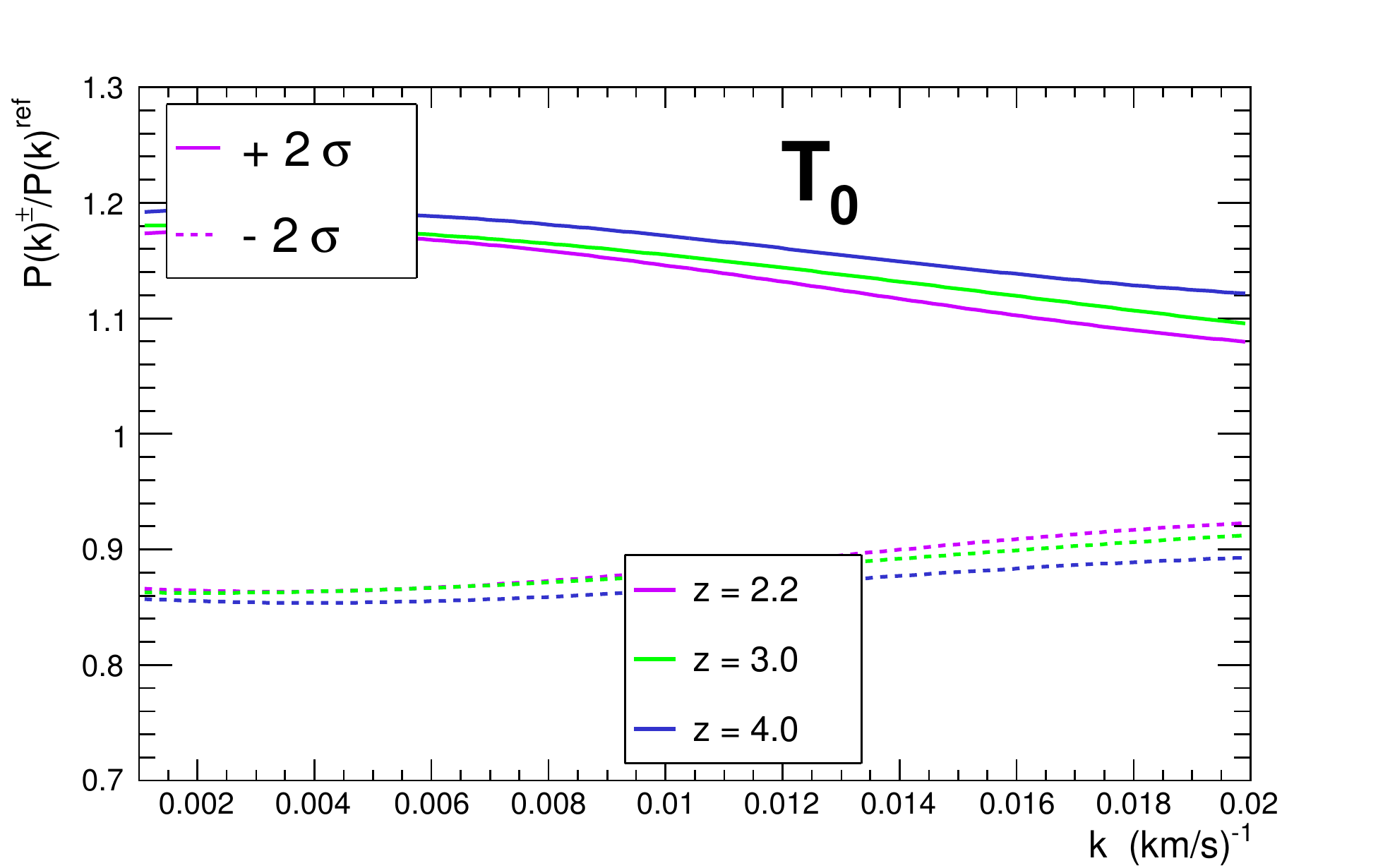,width = 8cm}\hspace{-.8cm}
\epsfig{figure= 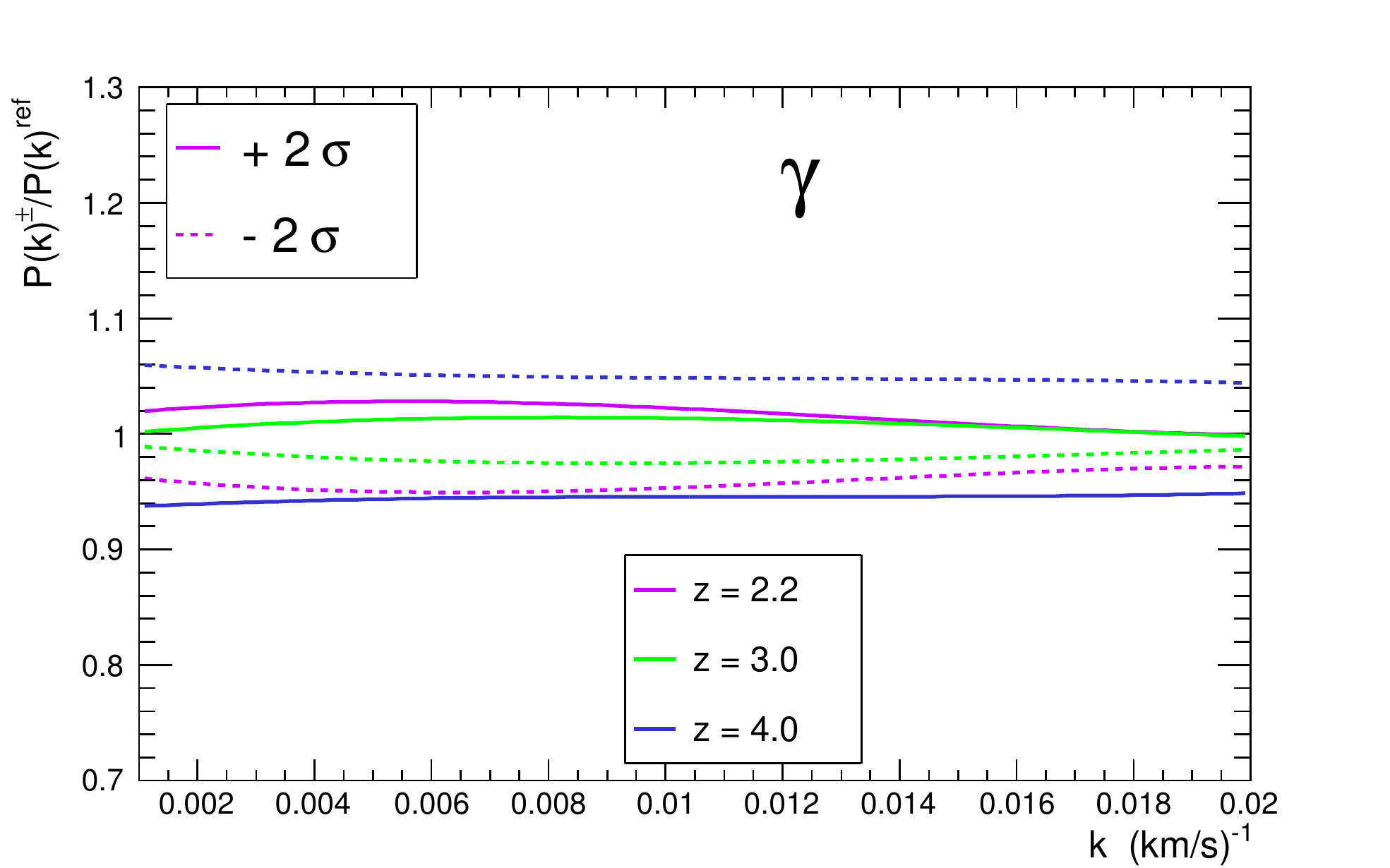,width = 8cm}

\caption{\it Same as figure~\ref{fig:sensitivityplotsCosmo}, for the astrophysical parameters used in the fit. }

\label{fig:sensitivityplotsAstro}
\end{center}
\end{figure}

The top panel of Figure~\ref{fig:sensitivityplotsCosmo} shows the
response of $P(k)$ to an increase of $\sum m_\nu$
from 0 eV to 0.15 eV, the level of our 95\%
confidence upper limit from the joint fit.  Solid
and dotted curves illustrate two different cases:
one in which $\sigma_8$ is held fixed as $\sum
m_\nu$ changes, and one in which the CMB
normalization $A_s$ is held fixed.  For fixed
$\sigma_8$, the change in $P(k)$ is a nearly
constant 1\% increase at $z=4.0$, while at $z=2.2$
the change declines from 1\% at $k=0.002$ to
nearly zero change at $k=0.02$.  Qualitatively,
the small and approximately flat response of
$P(k)$ to a $\sum m_\nu$ change at fixed
$\sigma_8$ can be understood with reference to
figure~1 of \cite{Rossi2014}, which shows that a
0.1 eV neutrino mass produces a $\sim 5\%$
suppression of the 3-d linear {\it matter} power
spectrum on the scales $k > 0.1 h\,{\rm Mpc}^{-1}$
probed by the BOSS \lya\ $P(k)$, relative to a
$\sum m_\nu=0$ model with the same large scale
power spectrum normalization.  Since this
suppression also covers the $k-$ range that
dominates contributions to $\sigma_8$, the initial
conditions of a simulation with $\sum m_\nu = 0.1$
eV are nearly the same as those of a massless
neutrino simulation with $\sigma_8$ lower by 5\%.
We therefore expect the impact of a $\sum m_\nu$
change to be approximately degenerate with that of
a $\sigma_8$ change, so that altering $\sum m_\nu$
at fixed $\sigma_8$ has no impact on $P(k)$.  The
residual 1\% boost at $k \approx 0.002$ arises
because the suppression of linear power at $z=2-4$
is slightly less than the suppression at $z=0$,
which is the reference redshift for fixed
$\sigma_8$.  The small but significant scale
dependence of the $P(k)$ response is a consequence
of non-linear evolution that can only be modeled
accurately using hydrodynamic simulations that
incorporate a neutrino component as in
\cite{Rossi2014}.

The weak and nearly flat response of $P(k)$ to a
$\sum m_\nu$ change at fixed $\sigma_8$ explains
why the \lya\ $\sigma_8$ contours in the
$\sigma_8-\sum m_\nu$ plane are nearly vertical
(Figure~\ref{fig:ContourFreqCombi}); the BOSS \lya\ $P(k)$ constrains
$\sigma_8$ with little dependence on $\sum m_\nu$.
As emphasized in section~\ref{sec:FreqCombined}, it is the
combination of these vertical contours with the
tilted $\sigma_8-\sum m_\nu$ constraints from the
CMB data that leads to a tight upper limit on
$\sum m_{\nu}$ in the joint fit.  The fact that the
$P(k)$ response at fixed $\sigma_8$ is not
perfectly flat or perfectly redshift-independent
explains why the \lya\ contours close at high
$\sum m_{\nu}$.

The amplitude of CMB anisotropies determines $A_s$
rather than $\sigma_8$, so it is the dotted curves
in this panel that are more directly relevant to
understanding joint \lya-CMB constraints.  These curves 
show that changing $\sum m_\nu$ from 0 eV to 0.15
eV suppresses the predicted \lya\ $P(k)$ by
$3-4\%$ at $k=0.002$, declining to a $1-2\%$
suppression at $k=0.02$, with a scale dependence
that is stronger at low redshift.  Except for
being a factor $\sim 2$ larger, this response is
similar to the effect of reducing $\sigma_8$ by
$2\sigma$ (from 0.832 to 0.814), shown by the dotted
lines in the second panel of
Figure~\ref{fig:sensitivityplotsCosmo}.  The physics of this
latter response was described by
\cite{McDonald2005}: an increase in $\sigma_8$
(solid lines of this panel) boosts the flux power
spectrum on linear scales where the effective bias
of the \lya\ forest is constant, but this boost is
diluted by non-linear effects at smaller
scales.

The primary obstacle to measuring the amplitude of
the matter power spectrum with the \lya\ $P(k)$ is
the degeneracy with the mean optical depth,
controlled in our models by the parameters
$A_\tau$ and $\eta_\tau$.  As shown in the top
left panel of Figure~\ref{fig:sensitivityplotsAstro}, the effect
of increasing $A_\tau$ by $2\sigma$ (from 0.0033 to
0.0037) is a 14-16\% amplification of $P(k)$ that is
nearly independent of scale and redshift.  Early
measurements of matter clustering from the \lya\
forest (e.g.,
\cite{Croft1999,McDonald2000,Croft2002}) imposed
external constraints on the mean optical depth,
but with the high-precision of SDSS
\cite{McDonald2005} and BOSS measurements it
becomes possible to break this degeneracy
internally because the impact of $\sigma_8$ is
more scale-dependent than that of $A_\tau$.
Agreement with independent measurements of the
mean optical depth then becomes a (successfully
passed) consistency test.

The scale and redshift-dependent impact of other
cosmological and astrophysical parameter changes
is illustrated in the other panels of
Figures~\ref{fig:sensitivityplotsCosmo} and~\ref{fig:sensitivityplotsAstro}.
Changes to $n_s$, $\Omega_m$, or $H_0$ alter the
slope of the linear matter power spectrum over the
scales probed by the \lya\ $P(k)$.  In the absence
of non-linear effects, these changes would tilt
the predicted $P(k)$ over the full $k$-range, but
non-linearity saturates the effects at high $k$,
with a flattening that sets in more quickly at the
lowest ($z=2.2$) redshift.  Changing $\eta_\tau$
alters the mean absorption as a function of
redshift, producing strong redshift-dependent
changes in the predicted power spectrum that are
nearly independent of $k$.  
Increasing $T_0$ with $A_\tau$ fixed (and thus a
renormalized photoionization rate $\Gamma_{\rm HI}$)
produces an amplification of $P(k)$ at large scales
that decreases towards high $k$, probably as a 
consequence of thermal broadening effects on line
saturation.  Changing the
slope $\gamma$ of the temperature-density relation
alters the effective bias of the \lya\ forest in a
manner that depends significantly on redshift but
only weakly on scale.  
The other astrophysical
nuisance parameters $f_ {\rm{Si\,II}}$ and $f_ {\rm{Si\,III}}$, not
shown in the plot, have an oscillatory 
impact that reflects the characteristic scale of
the wavelength separations.  The changes in the
predicted $P(k)$ associated with $2\sigma$ changes
in the astrophysical parameters
(Fig.~\ref{fig:sensitivityplotsAstro}) are typically several times
larger than those associated with $2\sigma$
changes in the cosmological parameters because
only the \lya\ forest measurements can break
degeneracies among them, while the cosmological
parameters are also constrained by the CMB data.
For example, there is a $-0.77$ correlation
coefficient between changes in $A_\tau$ and $T_0$,
so that when both are increased together the
impact on $P(k)$ is only a weak change of shape
rather than a large shift of amplitude.
Similarly, the redshift dependence induced by changing
$\eta_\tau$ can be compensated by changing $\gamma$,
which also has a redshift-dependent impact.

Our ability to place a strong upper limit on $\sum
m_{\nu}$ from the \lya+CMB combination ultimately
derives from the fact that there is no combination
of other elements in our model that can mimic the
subtle but distinctive scale- and
redshift-dependence of the dotted curves in the
first panel of Figure~\ref{fig:sensitivityplotsCosmo}.  Our limit
would be weakened by an unmodeled astrophysical
effect or an unrecognized observational systematic
that counteracts this dependence, being largest at
low $k$ and declining but not disappearing at high
$k$.  Given the high precision of the measurement,
however, the cancellation produced by this effect
would need to be highly exact to have an important
impact on the neutrino mass constraint.

\subsection{Impact of  simplifying hypotheses } \label{sec:systsimus}

Our conclusions about neutrino mass are produced by comparing model predictions to the Ly$\alpha$ and CMB data. Here we address simplifications inherent in our modeling to assess  how they could affect the neutrino mass constraints.
For this study,  we first  test the impact of various assumptions of our analysis on the best fit parameters derived from  Ly$\alpha$ data alone. We then consider 
 Ly$\alpha$ $ with $ CMB  and assume the results obtained on this combined data set as representative of the effects on all data combinations studied in this paper. We take the fit results obtained on either configuration as our reference  (second and fourth columns  of table~\ref{tab:fit_Lya_CMB} for the Ly$\alpha$ and the Ly$\alpha$ $+$ CMB configurations, respectively).  In both cases, we compute a systematic uncertainty as the difference on the best-fit values between the reference  and the  configuration modified according to the systematic effect under study.  Tables~\ref{tab:systLya} and \ref{tab:syst} summarize  the impact of   simplifying hypotheses  on the  cosmological parameters to which we have the greatest sensitivity: $\sigma_8$ and $n_s$ for both Ly$\alpha$ and Ly$\alpha$ $+$ CMB cases, as well as  $\sum \! m_\nu$  for  Ly$\alpha$ $+$ CMB.  The list of systematic effects  explored is given below. Some items are related to the hydrodynamical simulations,  others are linked to technical details in the Ly$\alpha$ data analysis. 

\begin{itemize}
\item {\bf Splicing technique in simulations:} 
As explained in section \ref{sec:likelihood}, we introduced  a corrective term with the shape ${\cal C}_{splicing}(k) = 1.01 - \alpha_{splicing} \cdot k$ in the Ly$\alpha$ likelihood, to correct the simulated power spectrum for residuals in the splicing procedure. The best-fit value of  the nuisance parameter $\alpha_{splicing}$ is within $1\sigma$ of the value of 2.5 measured on the splicing test  (see details in Sec.~\ref{sec:1Dsim}). As  expected,  the errors on the cosmological parameters inflate compared to their value with $\alpha_{splicing}$ fixed. Here, to  estimate a systematic uncertainty associated to this approach,  we apply an ad-hoc  correction to the data  of  ${\cal R}(k) = 0.99 + 4 \,k$ and maximize the likelihood again. Given the splicing residuals   measured,  0.99 is an extreme value of the offset, and since the best-fit slope is negative, we chose as an extreme case a positive slope of twice the magnitude ($+4$ instead of about $-2$).
The deviation observed on the cosmological parameters are significant when considering Ly$\alpha$ data alone, of the order of $40\%$ and $100\%$ of the statistical uncertainty for $\sigma_8$ and $n_s$, respectively. When combining Ly$\alpha$ data to other data sets, however, the deviations  
 are extremely low, at the level of one fifth of the statistical error in the worst case, i.e., for $n_s$. The impact on the neutrino mass bound at 95\% C.L. is  0.01 eV.

\item {\bf UV fluctuations:} The fluctuations in the intensity of the ionizing background due to discrete sources can change the measured power spectrum of Ly$\alpha$ transmission. Using the  analytical framework  presented in~\cite{Satya2014}, one can estimate  that  the impact of these ionizing intensity fluctuations from shot noise is of the order of 10-20\% of the total 1D power spectrum. To study the impact on the neutrino mass bound, we add  a $k$-independent offset equal to 10\% of  the measured 1D power spectrum  at the pivot wavenumber $k_p=0.009\,\rm(km/s)^{-1}$, and we refit the data. 
While the spectral index $n_s$ is significantly affected when fitting Ly$\alpha$ data alone, its variation decreases to half of its statistical uncertainty  when Ly$\alpha$ data is combined with CMB.
Therefore, the neutrino mass is only slightly affected by this additive term: we observe a $-0.03$ eV shift on the 95\% C.L. bound.  

\item {\bf AGN and SN feedback:}
We have neglected AGN or SN feedbacks in the current analysis. We here investigate the effect of adding their feedback to simulations that lack it. In  ~\cite{Viel12}, a study of the impact of  AGN and SN feedback is computed for the 1D power spectrum. In both cases, it can be modeled as ${\cal C}_{\rm feedback}(k) = \alpha(z) + \beta(z)\cdot k$, with a substantially smaller impact at higher redshift. Winds driven by SNe have an effect that is qualitatively opposite to that of AGN feedback, and with a  slightly smaller magnitude.  Over our range of interest, $0.001<k<0.02$, the slope of the power spectrum can be reduced by 6\% at most at $z<2.5$ and 2\% for $z>3.5$ when adding  AGN feedback, while it is increased by 5\% and 1\% for the same redshift bins respectively in the case of SN feedback. In addition, AGN feedback can offset the amplitude of the power spectrum by as much as -9\% at $z<2.5$, and SN feedback by up to +6\%. 
  We propagate these systematics with the same method as before, modifying the simulation power spectra according to the above corrective terms and maximizing the likelihood again. Despite an impact of the AGN feedback on $\sigma_8$ of almost 50\% of the statistical uncertainty in the case of Ly$\alpha$ data alone, the final impact on the neutrino mass bound at 95\% C.L. are    +0.02 eV and $-0.01$~eV for AGN and SN feedbacks respectively in the Ly$\alpha$ $+$ CMB configuration.

\item {\bf High-density absorbers:}
In~\cite{Palanque-Delabrouille2013}, we used the DR9 quasar catalog of BOSS~\cite{Paris2012} and we rejected all quasars that are flagged as having damped Lyman alpha (DLA)  or detectable Lyman limit systems (LLS)  in their forest. We evaluated their  contribution  by comparing the power  spectra  measured with and without  removing the flagged quasars. The ratio of the power spectra as a function of $k$ for these two configurations is in good agreement with the simulation of the effect of DLAs and LLS that was studied in~\cite{McDonald2005}. This first check demonstrates that a large fraction of the forest with high-density absorbers was indeed rejected. To  account for a possible remaining contribution, we assume that 20\% at most of the effect simulated in ~\cite{McDonald2005} (${\Delta P}(k) = 1/(15.10^{3}k-8.9)+0.018$) is still present.  Despite an impact on  $n_s$ of almost 50\% of the statistical uncertainty, the final impact on the neutrino mass bound at 95\% C.L. remains small, at the order  of  0.03 eV.

\item {\bf  PSF (point spread function) of BOSS spectra:} In~\cite{Palanque-Delabrouille2013}, we have  measured and applied a wavelength and spatial-dependent correction to the resolution (i.e., PSF) given in the BOSS spectra by the official SDSS pipeline. This correction reaches 10\% in the worst case, ie., for the reddest wavelength and thus the largest redshifts (Fig.4 of \cite{Palanque-Delabrouille2013}). To take into account a possible residual effect in our modeling of the instrument profile, we introduced a nuisance parameter (see Sec.~\ref{sec:likelihood}), identical for all  redshift bins. The best-fit value of this nuisance parameter indicates an overestimate of the resolution correction of about 2.5~$\rm km~s^{-1}$. However, as the resolution is wavelength dependent, this global correction might not fully account for all residual effect.   We here relax this simplifying hypothesis by assuming  an offset of our correction of the resolution that varies from 5~$\rm km~s^{-1}$ for the lowest redshift bin to about 9~$\rm km~s^{-1}$ for the highest one. With resolutions in the range between 70 and 85~$\rm km~s^{-1}$, this offset is of similar amplitude as the correction of the pipeline resolution, and corresponds to an extreme case. We measure no significant deviation in the cosmological results from this systematic uncertainty.

\end{itemize}

\begin{table}[htdp]
\caption{\it Impact of relaxing  simplifying hypotheses on the  central values of $\sigma_8$ and $n_s$  for  the Ly$\alpha$  configuration.}
\begin{center}
\begin{tabular}{lcc}
\hline
Simplifying hypothesis & $\sigma_8$ & $n_s$ \\
\hline 
Splicing technique  & $\pm 0.014$&$\pm 0.011$\\
UV fluctuations &$ -0.002$  & $-0.024$ \\ 
SN feedback & $+0.007$& -$0.006$ \\
AGN feedback & $-0.006$ & $+0.010$ \\
High-density absorbers & $+0.007$ & $+0.007$  \\
PSF of BOSS spectra & $\pm0.002$ & $\pm 0.0015$  \\
\hline
\end{tabular}
\end{center}
\label{tab:systLya}
\end{table}

\begin{table}[htdp]
\caption{\it Impact of relaxing simplifying hypotheses on the  central values of $\sigma_8$ and $n_s$,  and on the 95\%C.L. limit on $\sum \! m_\nu$~(eV)  for  the Ly$\alpha$ $+$ CMB configuration.}
\begin{center}
\begin{tabular}{lccc}
\hline
Simplifying hypothesis & $\sigma_8$ & $n_s$ &$\sum \! m_\nu$ limit  (95\%C.L.)    \\
\hline 
Splicing technique  & $\pm 0.001$ & $\pm 0.0018$ & $\pm 0.03$\\
UV fluctuations & $+0.001$  & $-0.0023$ & $-0.03$ \\ 
SN feedback & $+0.001$& $-0.0012$ & $-0.01$ \\
AGN feedback & $-0.001$ & $+0.0009$ & $+0.02$ \\
High-density absorbers & $0.000$ & $+0.0015$ & $+0.03$ \\
PSF of BOSS spectra & 0.000 & $\pm 0.0002$ & 0.00 \\
\hline
\end{tabular}
\end{center}
\label{tab:syst}
\end{table}

In conclusion, when considering Ly$\alpha$ data alone, none of the effects considered have an impact on $\sigma_8$ at the level of the statistical uncertainty on this parameter, while UV fluctuations and the splicing  
 technique used in the simulations are the most significant sources of   systematic uncertainty on $n_s$, with an impact of similar magnitude as the statistical uncertainty.  When combining Ly$\alpha$ with other data sets, however, the large uncertainty on $n_s$ reduces significantly, and, as shown in table~\ref{tab:syst}, it has only   a small influence on the limit we derive for $\sum \! m_\nu$.

\subsection{Implication for particle physics}

What are the implications of this work for the absolute masses $(m_1,m_2,m_3)$ of the three neutrino mass eigenstates $(\nu_1,\nu_2,\nu_2)$?  The atmospheric, solar, reactor and accelerator neutrino experiments constrain two squared mass differences, $\delta m^2$ and  $\Delta m^2$ (with $\delta m^2 \ll \Delta m^2$, following the formalism in~\cite{Capozzi:2013csa}). Assuming  one of the two mass hierarchies (NH or IH), our constraint on  $\sum m_\nu$ allows  one to derive direct constraints on the absolute masses $(m_1,m_2,m_3)$.

We can compare these constraints to  direct measurements of the mass  $m_\beta$ with single $\beta$-decays as planned by the KATRIN experiment~\cite{Osipowicz:2001sq} with tritium $\beta$-decay.  In absence of neutrino oscillations, this experiment would probe the reaction $ ^{3}{\rm H} \rightarrow \, ^{3}{\rm He} + e^- + \nu_e$. However, 
what tritium $\beta$-decay experiments really probe is an incoherent sum of the three reactions $ ^{3}{\rm H} \rightarrow \, ^{3}{\rm He} + e^- + \nu_i$, where $i$ stands for 1,2,3, and the $\nu_i$ are the three mass eigenstates. At currently achievable resolution, the measurable amplitude is therefore related to the combination  shown in Eq~\ref{eq:numassbeta} where $U_{ei}$ are the 3 coefficients of the first line of the  Pontercorvo-Maki-Nakagawa-Sakata  mixing matrix.    
Indeed, this neutrino mass,  $m_\beta$, can be directly derived from the masses $(m_1,m_2,m_3)$  through the PMNS mixing matrix $U(\theta_{12}, \theta_{13},\theta_{23})$ where $\theta_{ij}$ are the mixing angles:
\begin{equation}
m_\beta =  \left( \sum_i |U_{ei}|^2 m_i^2 \right)^\frac{1}{2}=\left( c_{13}^2  c_{12}^2 m_1^2  +  c_{13}^2  s_{12}^2 m_2^2 +  s_{13}^2  m_3^2 \right)^\frac{1}{2}
\label{eq:numassbeta}
\end{equation}
with  $c_{ij}=\cos\theta_{ij}$ and $s_{ij}=\sin\theta_{ij}$.

\begin{figure}[htbp]
\begin{center}
\epsfig{figure= 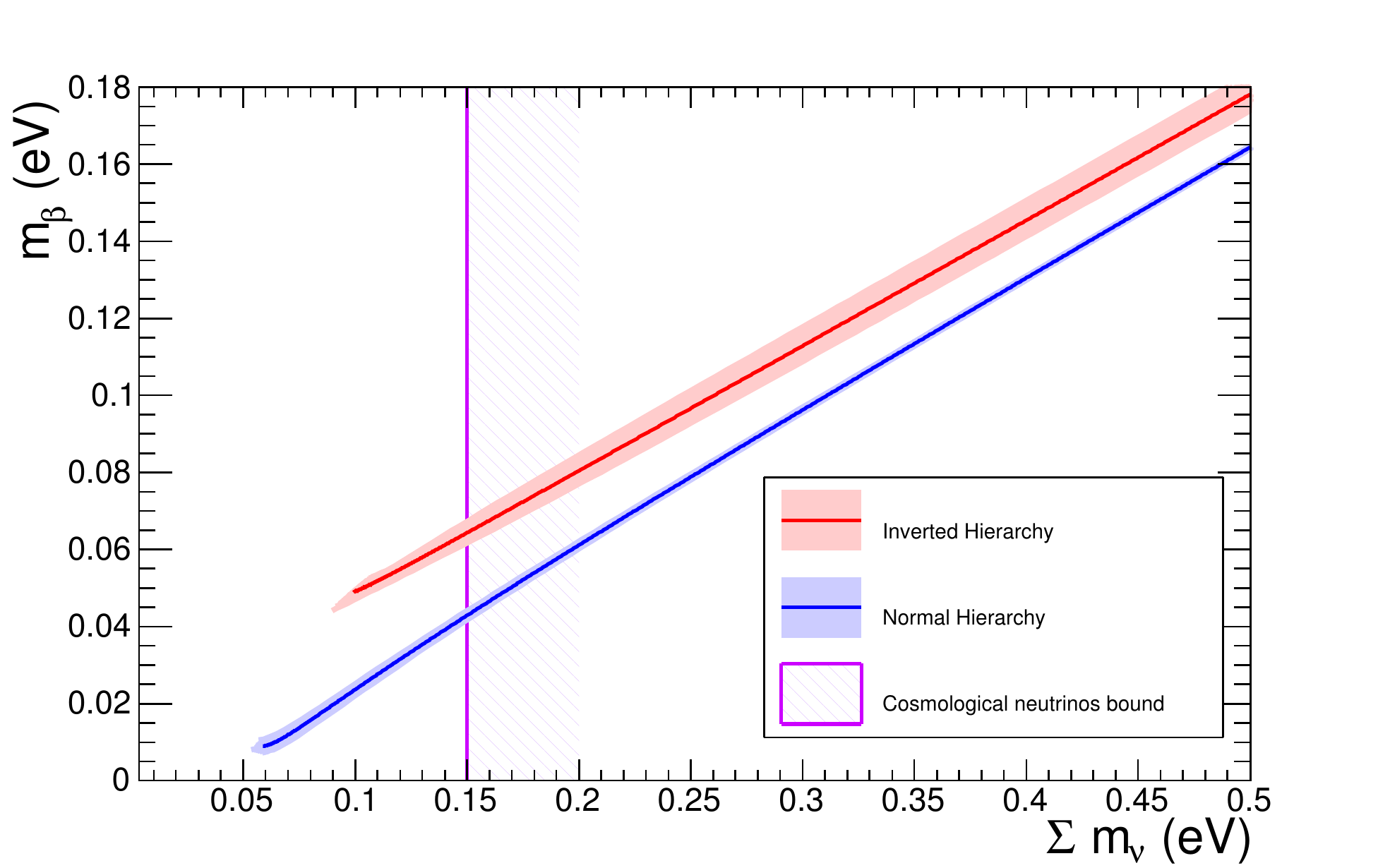,width = 15.0cm}
\caption{ Constraints on the sum of the neutrino masses and the effective electron neutrino mass, $m_\beta$. The blue and red curves correspond, respectively, to the NH scenario and the IH scenario obtained with the atmospheric, solar, reactor and accelerator neutrino experiments. The blue and red contours represent the 95\% C.L. derived from~\cite{Capozzi:2013csa} around the central value of their fit. The purple hashed region represents the 95\% C.L. bound computed in this analysis. }
\label{fig:ContourNuMass}
\end{center}
\end{figure}

The current upper limit on $m_\beta$, the ``effective electron neutrino mass", is of the order of 2 eV. The  KATRIN experiment  will improve this limit by one order of magnitude down to 0.2 eV.  Figure~\ref{fig:ContourNuMass}  shows the values of $m_\beta$ that are consistent with the bounds on   $\sum m_\nu$ given in this paper and with the constraints on $\delta m^2$, $\Delta m^2$, $s_{12}^2$ and $s_{13}^2$ derived by~\cite{Capozzi:2013csa} from  the combination of atmospheric, solar, reactor and accelerator neutrino experiments, for both hierarchy scenarios. Combined with the contours from~\cite{Capozzi:2013csa}, our results
imply $m_\beta<0.06$~eV  (respectively $m_\beta<0.04$~eV) for the inverted (resp. normal) hierachy.
A detection by Katrin of $m_\beta>0.2$~eV would thus call into question
the three-neutrino model used to interpret neutrino oscillation
experiments. 

\section{Conclusions}
\label{sec:conclusion}

In this paper, we present  the constraints on  cosmological parameters  from the 1D Ly$\alpha$ forest power spectrum, whether taken alone or  in combination with other probes.  The cosmological interpretation is obtained using a likelihood approach  to compare the Ly$\alpha$ forest power spectrum derived from SDSS-III/BOSS data~\cite{Palanque-Delabrouille2013} and   full hydrodynamical simulations produced specifically for that study~\citep{Borde2014, Rossi2014}. 

To model the Ly$\alpha$ likelihood, we  introduce parameters that describe the astrophysics of the IGM, 
the cosmological model, and  imperfections of the power spectrum measurement. In the first stage, we  perform   maximizations of the likelihood for different configurations of the astrophysical parameters and several sub-samples of  input power spectra.  The results of the various fits are extremely stable. For the rest of the paper, we  select  the configuration (redshift range $z=[2.1, 4.1]$ and no prior on the IGM temperature $T_0$) that gives the best $\chi^2$ per degree of freedom ($\chi^2 / {\rm dof} = 339/330$). The values obtained for the effective optical depth ($\tau$) and for the two parameters ($T_0$ and $\gamma$) that are related to the heating of IGM are  consistent with the ones from  previous measurements or expected from simulations.

Since we have the sensitivity to derive tight bounds on the sum $\sum m_\nu$ of the masses of active neutrinos  at a level approaching the physical limit ($\sum m_\nu=0$), we perform, in parallel, two statistical analyses: one based on a Bayesian interpretation and the other on a  frequentist interpretation. When combining Ly$\alpha$ with CMB data in the frequentist approach, we use   the central values of the parameters and the covariance matrices available in the official WMAP and {\it Planck} repositories. For the Bayesian approach, we produce new Markov chains  directly from the likelihoods. Despite  these methodological differences, the two approaches produce similar results, thus demonstrating the robustness of the analysis.  

The BOSS Ly$\alpha$ data alone produce better constraints than previous Ly$\alpha$ results, but the limits are not strong, especially for the sum of the neutrino masses where we obtain an upper bound of 1.1~eV (95\% CL), including systematics for both  data and   simulations. The combination of Ly$\alpha$ with CMB data ({\it Planck} or WMAP9, in addition to   the high-$l$ likelihoods from the ACT and SPT ground-based experiments) reduces the uncertainties  by about a factor two on $\sigma_8$, $n_s$, $\Omega_m$ and $H_0$ compared to Ly$\alpha$ alone. 

Finally, this work  contributes towards solving one of the most fundamental issues of neutrino physics:  the neutrino mass hierarchy. Indeed, the most sensitive probes of the total neutrino masses, $\sum m_\nu$, are currently cosmological observations. We present several bounds derived from  different data combinations: Ly$\alpha$ alone,  Ly$\alpha$ plus different sets of CMB data among {\it Planck}, WMAP9, ACT and SPT. When we consider the CMB data set consisting of  {\it Planck}  + polarization from WMAP9 + ACT + SPT,   combined with the one dimensional Ly$\alpha$ forest power spectrum from BOSS, we achieve a limit of the order of $\sum m_\nu <0.15$ at 95\% C.L., a factor of two  tighter than the previous bound obtained with  CMB and BAO data. Adding BAO data  tightens this constraint  further to $\sum m_\nu < 0.14$~eV at 95\% CL. While this value does not exclude the Inverted Hierarchy scenario, the new bound  favors the Normal Hierarchy scenario, and suggests interesting complementarity between  such a  measurement of the  absolute neutrino mass  and  future direct measurements of the ``effective electron neutrino mass"  as foreseen by the KATRIN experiment.

\acknowledgments

We acknowledge PRACE (Partnership for Advanced Computing in Europe) for awarding us access to resource curie-thin nodes based in France at TGCC, under allocation number 2012071264.\\
This work was also granted access to the resources of CCRT under the allocation 2013-t2013047004 made by
GENCI (Grand Equipement National de Calcul Intensif).\\
A.B., N.P.-D., G.R. and Ch.Y.  acknowledge  support from grant ANR-11-JS04-011-01 of Agence Nationale de la Recherche.\\
M.V. is supported by ERC-StG "CosmoIGM".\\
JSB acknowledges the support of a Royal Society University
Research Fellowship.\\
The work of G.R. is also supported by the faculty research fund of Sejong University in 2014, and by the National Research Foundation of Korea through NRF-SGER 2014055950.\\
We thank Volker Springel for making \texttt{GADGET-3} available to our team.

Funding for SDSS-III has been provided by the Alfred P. Sloan Foundation, the Participating Institutions, the National Science Foundation, and the U.S. Department of Energy Office of Science. The SDSS-III web site is http://www.sdss3.org/.

SDSS-III is managed by the Astrophysical Research Consortium for the Participating Institutions of the SDSS-III Collaboration including the University of Arizona, the Brazilian Participation Group, Brookhaven National Laboratory, Carnegie Mellon University, University of Florida, the French Participation Group, the German Participation Group, Harvard University, the Instituto de Astrofisica de Canarias, the Michigan State/Notre Dame/JINA Participation Group, Johns Hopkins University, Lawrence Berkeley National Laboratory, Max Planck Institute for Astrophysics, Max Planck Institute for Extraterrestrial Physics, New Mexico State University, New York University, Ohio State University, Pennsylvania State University, University of Portsmouth, Princeton University, the Spanish Participation Group, University of Tokyo, University of Utah, Vanderbilt University, University of Virginia, University of Washington, and Yale University.


\bibliographystyle{unsrtnat_arxiv}
\bibliography{biblio}

\end{document}